\documentclass[twocolumn,aps,showpacs,prb,amsmath,amssymb,floatfix,superscriptaddress]{revtex4-2}

\usepackage{amssymb,color}
\usepackage{flafter}
\usepackage{bm} 
\usepackage{color}
\usepackage{graphicx}
\usepackage{physics}
\usepackage{amsthm}
\usepackage{amsmath}
\usepackage{amssymb}
\usepackage{enumerate}
\usepackage{placeins}
\usepackage{booktabs}
\usepackage{dsfont}
\usepackage{yfonts}
\usepackage{hyperref} 
\usepackage[color=green!60,textsize=scriptsize]{todonotes}
\usepackage{comment}

\DeclareSymbolFont{bbold}{U}{bbold}{m}{n}
\DeclareSymbolFontAlphabet{\mathbbold}{bbold}

\newcommand{\ex}[1]{\langle #1 \rangle}

\newcommand{\Dlt}{\delta}
\newcommand{\sgn}{\text{sgn}}
\newcommand{\sinefreesame}[3]{\sin{(#3\frac{\pi}{L}(#1-#2))}}

\newcommand{\sineintsame}[4]{\sin{(\frac{\pi}{L}(#3(#1-#2)+i#4))}}

\usepackage{xspace}
\newcommand{\TRDM}{2\nobreakdash-\hspace{0pt}RDM\xspace}
\newcommand{\ORDM}{1\nobreakdash-\hspace{0pt}RDM\xspace}
\newcommand{\FF}{\rm{FF}}	

\newcommand{\h}[5]{h_{#5}(#2\!-\!#1,#4\!-\!#3)}
\newcommand{\hr}[1]{h_{#1,-}}
\newcommand{\hf}[1]{h_{#1,+}}
\newcommand{\hd}[1]{h_{#1,\hspace{1pt}0\hspace{1pt}}}

\begin{document}

\title{The two-particle density matrix of a Luttinger liquid}

\author{Harini Radhakrishnan}
\affiliation{Department of Physics and Astronomy, University of Tennessee, Knoxville, TN 37996, USA}
\affiliation{Department of Physics, Drexel University, Philadelphia, PA 19104, USA}
\affiliation{Department of Materials Science and Engineering, Drexel University, Philadelphia, PA 19104, USA\looseness=-1}

\author{Matthias Thamm}
\affiliation{Institut f\"{u}r Theoretische Physik, Universit\"{a}t Leipzig,  Br\"{u}derstra{\ss}e 16, 04103 Leipzig, Germany}

\author{Hatem Barghathi}
\affiliation{Department of Physics and Astronomy, University of Tennessee, Knoxville, TN 37996, USA}
\affiliation{Institute for Advanced Materials and Manufacturing, University of Tennessee, Knoxville, Tennessee 37996, USA\looseness=-1}

\author{Bernd Rosenow}
\affiliation{Institut f\"{u}r Theoretische Physik, Universit\"{a}t Leipzig,  Br\"{u}derstra{\ss}e 16, 04103 Leipzig, Germany}

\author{Adrian Del Maestro}
\affiliation{Department of Physics and Astronomy, University of Tennessee, Knoxville, TN 37996, USA}
\affiliation{Institute for Advanced Materials and Manufacturing, University of Tennessee, Knoxville, Tennessee 37996, USA\looseness=-1}
\affiliation{Min H. Kao Department of Electrical Engineering and Computer Science, University of Tennessee, Knoxville, TN 37996, USA\looseness=-1}
	
\date{\today}

\begin{abstract} 
Two-particle coherence is the first level of the reduced-density-matrix hierarchy that contains correlations inaccessible to single-particle observables, yet analytic two-particle density matrices are rare even in one dimension. We derive a closed, finite-size expression for the equal-time two-particle reduced density matrix of spinless fermions in a Tomonaga–Luttinger liquid using constructive bosonization with an explicit ultraviolet cutoff. In addition to the familiar Luttinger parameter $K$-dependent exponent $\gamma^2=(K+K^{-1}-2)/2$ which governs the spatial decay of matrix elements, the result exposes a second exponent, $\lambda=(K^{-1}-K)/2$, which encodes correlations between opposite chiralities and controls the off-diagonal structure. The diagonal limit of the two-particle reduced density matrix yields density correlations and the static structure factor, while its coherences resolve algebraic $2k_F$ charge-density-wave correlations for repulsion and odd-parity $p$-wave pairing correlations for attraction.  After fixing the cutoff from the one-particle density matrix, the analytic result quantitatively reproduces density matrix renormalization group calculations of the interacting $J$-$V$ chain within the Luttinger liquid regime.  The result connects universal Luttinger liquid scaling with observables in finite microscopic systems.
\end{abstract}
 
\maketitle

\section{Introduction}

Reduced density matrices provide a compact description of the correlations in a many-body quantum state.  The $n$-body reduced density matrix, $\rho_n$, contains all information needed to compute any observable involving at most $n$ particles \cite{coleman1963structure}.  Thus $\rho_1$ determines one-body quantities such as the density, momentum distribution, and single-particle coherence, while $\rho_2$ is sensitive to genuine two-particle correlations.  Its diagonal elements give pair-density correlations and determine quantities such as the static structure factor, which can be measured in scattering and Bragg spectroscopy experiments \cite{Altman2004,Boll2016,Yang2018,Kuhnle2010}.  Its off-diagonal elements contain pair coherences, and therefore distinguish states with similar density correlations but different phase structure.

In quantum chemistry, reduced density matrices are central to the $N$-representability problem and to approaches that determine energies without reconstructing the full many-body wavefunction \cite{Mazziotti2012,Mazzioti2012pra,Mazziotti1998,Liebert:2025nn}.  In nuclear and electronic many-body systems, reduced density matrices provide measures of single-particle coherence, exchange, and correlation beyond a Slater determinant \cite{Bulgac:2022}.  In cold atoms and quantum simulators, related few-body correlation functions can be accessed directly, and are used to characterize fluctuations, entanglement, and dynamical response \cite{Naldesi2022,Polkovnikov:2006js,Gritsev:2006tn,Moitra2023}. 
One- and two-particle density matrices have also been studied in attractively interacting one-dimensional Fermi gases, where the \TRDM resolves pairing structure \cite{Rammelmueller:2017al}, and in the one-dimensional extended Hubbard model, where its spectrum, cumulant, and coherence diagnose correlated phases \cite{Ferreira:2022qq}.
Reduced density matrices also encode the entanglement between different groups of indistinguishable particles 
\cite{Zozulya:2007fe,Zozulya:2008bg,Haque:2007mu,Haque:2009zi, Barghathi:2017ab, Radhakrishnan, Herdman:2014jy, Herdman:2015xa} with unique sensitivity to interactions and quantum statistics not present in mode entanglement.
More recently, nonlinear resonant inelastic X-ray scattering has been proposed as a way to access connected four-fermion correlations, which form the essential part of the cumulant two-particle density matrix \cite{Liu:2025rixs}.  

\begin{figure}[t]
 \centering
 \includegraphics[width=0.70\linewidth]{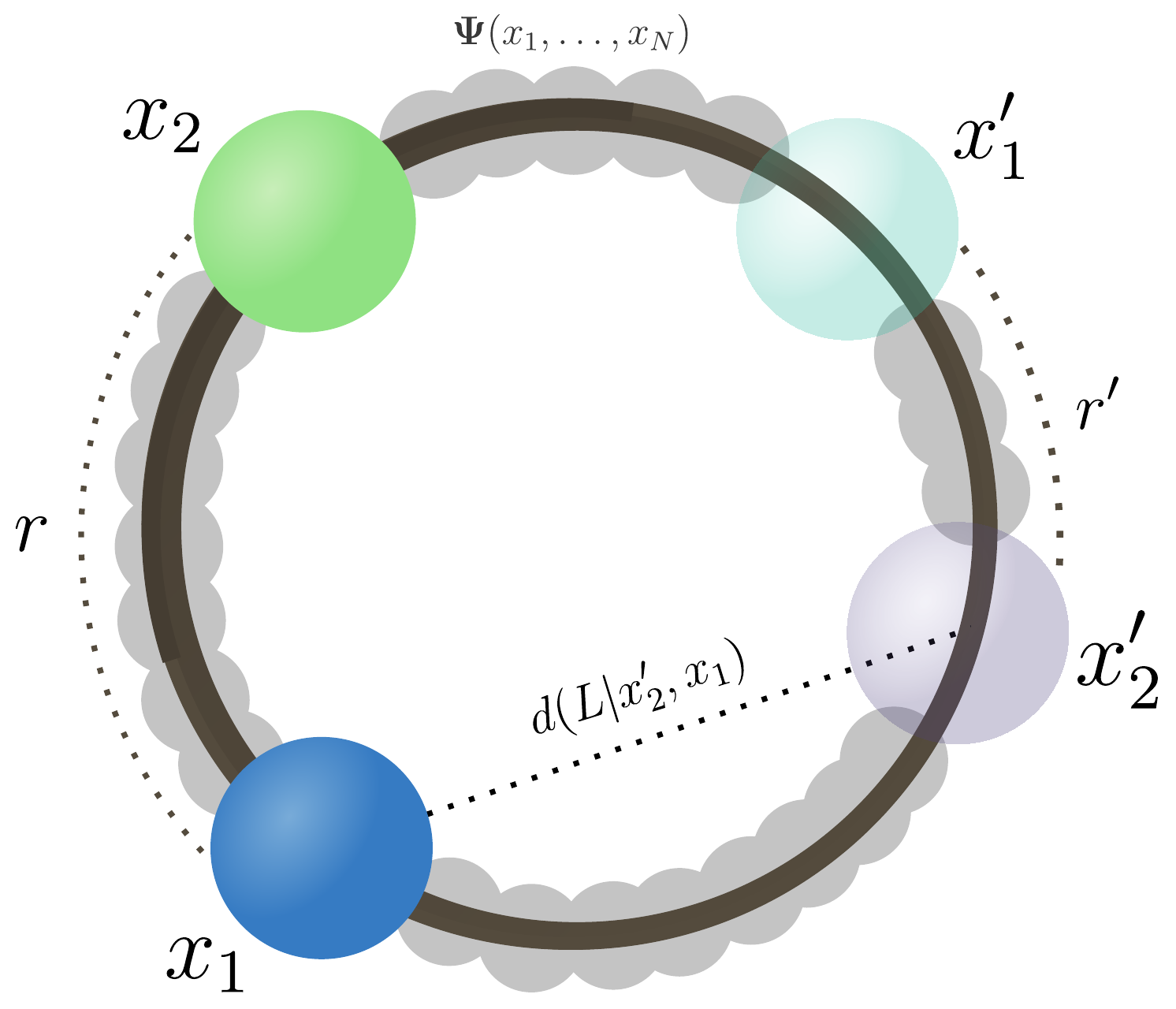}
 \caption[Graphical depiction of the two-body density matrix]{Graphical
 depiction of the two-body density of a one dimensional $N$-body system of
 length $L$ with wavefunction $\Psi(x_1,\dots x_N)$. The quantity
$\rho_2(x_2',x_1';x_1,x_2)$ denotes the matrix element obtained when two
fermions are annihilated at sites $x_1$ and $x_2$ while two are created at
coordinates $x_1'$ and $x_2'$. For periodic boundary conditions,
$d(L|x_2',x_1)$ is the chord length across the ring.}
 \label{fig:01_Configuration}
\end{figure}

These developments make it useful to have explicit, analytically controlled results for $\rho_2$ in strongly correlated systems.  Even for two particles, the matrix element depends on four spatial coordinates, and for general $n$ the number of independent elements grows rapidly with system size.  One must also keep track of statistics, interactions, boundary conditions, and short-distance regularization.  Here, we make progress by focusing on the one-dimensional Tomonaga-Luttinger liquid (TLL), where low-energy fermionic correlations can be computed by bosonization.  We consider spinless fermions on a ring and study the equal-time two-body matrix with elements
\begin{equation}
    \rho_2(x_2',x_1';x_1,x_2)
    =
    \expval{\Psi^\dagger(x_2')\Psi^\dagger(x_1')\Psi(x_1)\Psi(x_2)} \, .
    \label{eq:2RDMintro}
\end{equation}
This object is the amplitude for removing two fermions at $x_1,x_2$ and inserting them at $x_1',x_2'$.  Its diagonal limit gives the pair distribution function and density-density correlations, while its off-diagonal elements describe two-particle coherence.  The geometry of this matrix element is shown schematically in Fig.~\ref{fig:01_Configuration}.

Conformal field theory gives the universal scaling form of such correlation functions in one dimension \cite{Tsvelik_2003,Cazalilla_2004,Cardy1996}, determining the power laws for the spatial decay of matrix elements fixed by the Luttinger parameter $K$, but not the nonuniversal amplitudes and short-distance structure. Finite-size bosonization expressions for multipoint fermionic correlators have also been obtained for Luttinger liquids with normal or Andreev boundaries, mainly in the context of transport and proximity effects \cite{CauxLopezSuppa2003}. 
In this paper, we use constructive bosonization \cite{Haldane:1981eh,Giamarchi:2004bk,Gogolin1998,vonDelft:1998ae},  keeping an explicit ultraviolet cutoff and finite-size boundary-condition sensitive distances, allowing the continuum expression to be matched quantitatively to lattice data.  
Thus, our TLL result retains both the universal long-distance structure and the nonuniversal information needed to compute observables in a finite system described by a microscopic Hamiltonian.

 \begin{figure}[t]
  \centering
  \includegraphics[width=\linewidth]{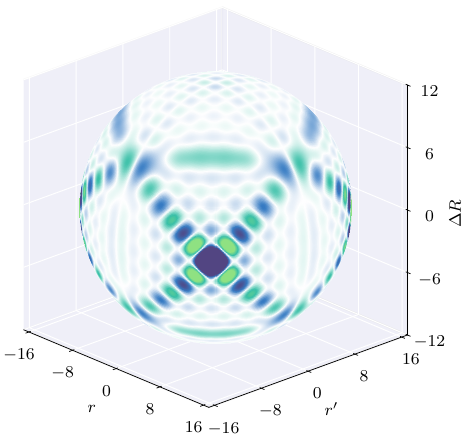}
  \caption[The two-body reduced density matrix plotted in a spherical geometry as a function of relative coordinates]{The 2-RDM $\rho_2(x_2',x_1';x_1,x_2)$ plotted on the surface of a sphere with $\Delta R^2 + {r^2}/{2}+{{r'}^2}/{2}=144$, $L=40$, $K=8/5$ (attractive), $n_0=1/2$, and $\epsilon=1$ (see Section~\ref{sec:2RDMAnalysis} for parameter details). The colormap corresponds to a range from $-0.03$ (blue) to $0.03$ (green). Alternating blue-green regions signal the sign change from fermion antisymmetry, while the vanishing of $\rho_2$ along $r=0,r'=0$ is due to the Pauli exclusion principle.}
  \label{fig:sphericalheatmap}
\end{figure}

The resulting expression for $\rho_2$ contains six chiral contributions, corresponding to the possible right- and left-moving fermion combinations, which resolve into three real terms.  A complicated interaction-dependent oscillatory structure has spatial decay controlled by both the familiar exponent
\begin{equation*}
    \gamma^2 = \frac{K+K^{-1}-2}{2},
\end{equation*}
which also appears in the one-particle density matrix \cite{DzyaloshinskiiLarkin1974,LutherPeschel1975,Barghathi:2017ab,Cazalilla:2006zw,Thamm:2022ja}, as well as a second exponent
\begin{equation*}
    \lambda = \frac{K^{-1}-K}{2}\, 
\end{equation*}
arising from anomalous correlations between fermions of opposite chirality.  $\lambda$ is absent from the spatial dependence of the 1-RDM and, unlike $\gamma^2$,  can be negative in the attractive regime, $K>1$.  The appearance of $\lambda$ is one of the main new features of the two-particle problem.  It reflects the fact that $\rho_2$ is not simply a product of one-particle coherences once interactions are present.  The dependence of $\gamma^2$ and $\lambda$ on $K$ is shown in Fig.~\ref{fig:02_Exponents}.

The four-coordinate structure of Eq.~\eqref{eq:2RDMintro} can inhibit interpretation, but
it has a simple geometry.  Translational invariance removes one coordinate,
leaving the relative separations $r=x_2-x_1$, $r'=x_2'-x_1'$ and the difference of center-of-mass coordinates $\Delta R = (x_1'+x_2'-x_1-x_2)/{2}.$ In these variables, the dominant features of $\rho_2$ lie near four pairwise-intersecting hyperplanes corresponding to $x_1'=x_1$, $x_2'=x_2$, $x_1'=x_2$, and $x_2'=x_1$.  The first two give the diagonal density correlations; the other two give their exchange-related partners.  This geometric view provides a useful way to analyze both the density-density limit and the off-diagonal coherences.  Figure~\ref{fig:sphericalheatmap} visualizes the same three-coordinate function on a fixed-radius spherical surface in $(r,r',\Delta R)$ space for an attractive interaction.

The off-diagonal elements reveal correlations not visible in the density-density limit: in the repulsive regime they show signatures of $2k_F$ charge-density-wave order, while in the attractive regime they display the antisymmetric structure expected for a $p$-wave pairing instability \cite{Baldelli23,Aase2022,kanesliding,Kane2017,nayak2000density}.

As an application, we test the analytic result for $\rho_2$ against density matrix renormalization group (DMRG) \cite{White1992DMRG,Schollwock2011DMRG} calculations for the $J$-$V$ chain of interacting spinless fermions, where the low-energy sector is a TLL for $|V/J|<2$.  At half filling, Bethe ansatz gives the Luttinger parameter $K$ as a function of the microscopic interaction.  We compute the ground-state density matrices using density matrix renormalization group calculations and compare them with the bosonization expression.  The only nonuniversal input is the cutoff $\epsilon$, which is fixed from the one-particle density matrix and then used without further adjustment in the two-particle result \cite{Thamm:2022ja}.  With this matching, the analytic expression reproduces the numerical \TRDM throughout the TLL regime.

The diagonal elements of the \TRDM give a finite-size expression for the density-density correlation function, from which we compute the static structure factor and recover the expected small-momentum behavior controlled by $K$.  We also use the same matrix elements to compute the two-body contribution to the lattice ground-state energy.  

The rest of the paper is organized as follows. In Sec.~II we review the definitions and basic properties of $n$-body reduced density matrices, including their relation to observables and entanglement measures. In Sec.~III we derive the two-body density matrix of a spinless TLL using constructive bosonization. In Sec.~IV we analyze its coordinate structure, diagonal limit, and interaction-induced correlations.  We also discuss coherence diagnostics of algebraic charge density wave ordering and $p$-wave pairing correlations. In Sec.~V we compare the result with DMRG calculations for the $J$-$V$ chain and use it to compute the static structure factor, the lattice energy.  Section~VI summarizes the results and discusses extensions to higher density matrices.

\section{The $n$-particle reduced density matrix}
\label{sec:nrdm}

The utility of $n$-particle reduced density matrices ($n$-RDMs) is that they retain the information needed to compute all observables involving at most $n$ particles, without requiring direct access to the full many-body wave function. This is a substantial compression of the many-body problem. For example, in a finite one-particle basis with $M$ orbitals, a general $N$-fermion wave function lives in a Hilbert space of dimension $\binom{M}{N}$, while the $n$-RDM acts only on the $n$-particle Hilbert space, whose dimension is $\binom{M}{n}$. Thus, for fixed $n$, the reduced description grows polynomially with the one-particle basis size, while the full many-body wave function grows combinatorially. The price is that low-order RDMs do not uniquely determine an arbitrary many-body state. 
The gain is that for a system with fixed particle number, the $n$-RDM contains the information required to compute all particle number conserving observables of rank at most $n$, where the quantities of direct physical interest are often low-order observables.

For example, the one-particle reduced density matrix (1-RDM) is fundamental for understanding single-particle coherence and the nature of quasiparticle excitations. It directly determines one-body observables such as the density, momentum distribution, and the one-particle contribution to the energy \cite{DelMaestro:2021ja,Thamm:2022ja}. The \ORDM cannot capture genuine two-particle correlations, which first appear in the \TRDM. The \TRDM thus reveals the essential pairwise correlations that underpin many-body phenomena such as superconductivity and fermionic superfluidity \cite{yang1962concept}. 

We define the $n$-RDM as the correlation function 
\begin{align}
    \rho_n(x_n',...,&x_1';x_1,...,x_n) 
    \notag\\  
    &=  \expval{ {\Psi}^{\dagger}(x_n') \cdots {\Psi}^{\dagger}(x_1') {\Psi}(x_1) \cdots{\Psi}(x_n) }\, 
    \label{eq:nrdm_definition}
\end{align}
where $\Tr \rho_n = N!/(N-n)!$ and note that there exist different conventions for the $n$-RDM in the literature, which differ by combinatorial factors but are all equivalent for computing $n$-body observables \cite{Lowdin1955,Carlson:1961dr, Ando1963,coleman1963structure}. Up to such factors, the $n$-RDM can equivalently be obtained by tracing out $N-n$ particle coordinates from the full $N$-body density matrix.

That the expectation value of any $n$-body observable can be determined from the $n$-RDM becomes clear by considering a general Hermitian $n$-body operator in second quantization
\begin{align}
    \hat{O}_n &= \frac{1}{n!} \int \dd{X_n}\dd{X_n'} o_n(X_n';X_n) \nonumber \\
              &\qquad \times {\Psi}^{\dagger}(x_n') \dots {\Psi}^{\dagger}(x_1') {\Psi}(x_1) \dots {\Psi}(x_n)\\ 
    \expval{\hat{O}_n}  &= \frac{1}{n!} \int \dd{X_n}\dd{X_n'} o_n(X_n';X_n) \rho_n(X_n'; X_n) \ ,
    \label{eq:nbody_observable_rdm}
\end{align}
where $o_n(x_n',...,x_1';x_1,...,x_n) = o_n^*(x_n,...,x_1;x_1',...,x_n')$ is the corresponding first-quantized $n$-particle kernel. Here, we introduced the abbreviations $X_n = (x_1,...,x_n)$ and $\dd{X_n} = \prod_i \dd{x_i}$, and analogously $X_n' = (x'_n,...,x'_1)$ and $\dd{X'_n} = \prod_i \dd{x'_i}$.

Several useful examples follow, including that the local density $n(x)$ is determined by the diagonal elements of the \ORDM:
\begin{equation}
n(x) \equiv \expval{\hat{n}(x)} =  
\expval{{\Psi}^{\dagger}(x){\Psi}(x)} = \rho_1(x;x),
    \label{eq:density_from_ordm}
\end{equation}
with $\hat{n}(x) = \Psi^\dagger(x) \Psi(x)$ and $\int \dd{x}n(x)=N$.  For a system of length $L$ with periodic boundary conditions, the momentum distribution is the Fourier transform of the \ORDM:
\begin{equation}
    n(q) = \expval{\hat{c}_q^{\dagger}\hat{c}_q^{\phantom\dagger}}
    = \frac{1}{L} \int_0^L \dd{x}\dd{x'} e^{iq(x'-x)} \rho_1(x';x),
    \label{eq:momentum_distribution}
\end{equation}
where
\begin{equation}
    \hat{c}_q = \frac{1}{\sqrt{L}} \int_0^L \dd{x} e^{-iqx}\hat{\Psi}(x).
\end{equation}

For a continuum Hamiltonian (in units where $\hbar=1$) with a one-body potential $U(x)$ and a two-body interaction $V(x-x')$,
\begin{align}
    \hat{H} &= \int \dd{x} {\Psi}^{\dagger}(x)
\bqty{-\frac{1}{2m}\dv[2]{x} + U(x) } {\Psi}(x) \nonumber \\
            &\quad + \frac{1}{2} \int \dd{x}\dd{x'} V(x-x')
    {\Psi}^{\dagger}(x) {\Psi}^{\dagger}(x') {\Psi}(x') {\Psi}(x),
    \label{eq:spinless_fermion_hamiltonian}
\end{align}
the energy depends only on the $1$- and $2$-RDMs:
\begin{align}
    E &= \int \dd{x} \bqty{ \left. -\frac{1}{2m} \frac{\partial^2}{\partial x^2}
        \rho_1(x';x) \right \rvert_{x'=x} +  U(x)\rho_1(x;x)} \nonumber \\
      &\quad + \frac{1}{2}\int \dd{x}\dd{x'} V(x-x') \rho_2(x,x';x',x).
    \label{eq:energy_from_rdms}
\end{align}

The pair distribution function $g_2$ is also fixed by the \TRDM:
\begin{align}
    g_2(x,x') &= \frac{\expval{{\Psi}^{\dagger}(x) {\Psi}^{\dagger}(x')
    {\Psi}(x') {\Psi}(x)}}{ n(x)n(x') } \notag \\
              & = \frac{ \rho_2(x,x';x',x)}{\rho_1(x;x)\rho_1(x';x')}\, .
    \label{eq:g2_from_2rdm}
\end{align}
The ordinary density-density correlation function contains an additional contact term:
\begin{equation}
    \expval{\hat{n}(x)\hat{n}(x')} = 
     \rho_2(x,x';x',x) + \delta(x-x')\rho_1(x;x)\, .
    \label{eq:density_density_contact}
\end{equation}
Away from coincident points, or when normal ordering is used, the contact term is absent. For spinless fermions, antisymmetry also enforces a Pauli hole in the short-distance pair correlations, so that $g_2(x,x)=0$ for regular continuum wave functions. 

For a translationally invariant system, where $\langle\hat{n}(x)\rangle=n_0$ and $r=x-x'$, Eq.~\eqref{eq:density_density_contact} simplifies to
\begin{align}
\langle\hat{n}(r)\hat{n}(0)\rangle=n_0^2g_2(r)+n_0\delta(r)\, ,
\label{Eq:DensDensCorr}
\end{align}
where $g_2(r)\equiv g_2(r,0)$.  Furthermore, considering the elements of the connected density-density correlation matrix and using translational invariance: 
\begin{align}
    \langle\hat{n}(x_2)\hat{n}(x_1)\rangle_c & =\langle\hat{n}(x_2)\hat{n}(x_1)\rangle-\langle\hat{n}(x_2)\rangle\langle\hat{n}(x_1)\rangle \nonumber \\
&=n_0^2\left(g_2(r)-1\right)+n_0\delta(r),
\label{Eq:CDensDensCorr}
\end{align}
where the eigenvalues of the circulant matrix $\langle\hat{n}(r)\hat{n}(0)\rangle_c/n_0$ define the static structure factor
\begin{align}
s(q)=1+n_0\int_0^L \dd{r} \left[g_2(r)-1\right]e^{irq}\, ,
\label{Eq:StructureFactor}
\end{align}
measurable in scattering experiments. 

% ========================================================================================
\section{Two-body density matrix from bosonization} 
\label{Sec:2RDMBosonizationAnalytical}
% ========================================================================================

In this section we provide details on the derivation of the two-body reduced density matrix for the ground state of a Tomonaga-Luttinger liquid (TLL) model using constructive bosonization \cite{Giamarchi:2004bk,vonDelft:1998ae}.  The starting point is the interacting 1D Hamiltonian for spinless fermions confined to a finite system of length $L$ with periodic boundary conditions as defined above in Eq.~\eqref{eq:spinless_fermion_hamiltonian}. 
% %
% \begin{align}
%     H = &-\frac{1}{2M}\int_0^L {\rm d} x\Psi^\dagger(x)\nabla_x^2\Psi(x) 
%      + \mu\int_0^L dx \hat{\varrho}(x) 
%      \notag\\  
%      &+ \int_0^L {\rm d}x'\int_0^L {\rm d}x \hat{\varrho}(x')V(x'-x)\hat{\varrho}(x) \ ,
%     \label{Eq: continuumHam}
% \end{align} 
% %
% for particles of mass $M$, short-range pair-interaction potential $V$, chemical potential $\mu$, and density operator $\hat{\varrho}(x)=\Psi^\dagger(x)\Psi(x)$. 
We focus on the low energy sector and linearize the dispersion close to the Fermi points $\pm k_F$ with $k_F= {\pi N}/{L}=\pi n_0$,  giving rise to a decomposition in terms of left$(-)$ and right$(+)$ moving fields 
\begin{equation}
    \Psi(x)=e^{-ik_{F}x}\Psi_{+}(x)+e^{ik_{F}x}\Psi_{-}(x).
\end{equation}
Within the bosonization scheme, the fermionic field operators are expressed as
\begin{align}
    \Psi_\alpha(x)&=\frac{\chi_\alpha}{\sqrt{2\pi\eta}}e^{i(\varphi_{0,\alpha}+\alpha\frac{2\pi x}{L}N_\alpha)}e^{-i\phi_\alpha(x)}\ ,
    \label{eq:fermionfield}
\end{align}
where $\alpha=+1$ ($-1$) for right (left) movers. Here, $\chi_\alpha=e^{i\alpha\frac{\pi}{2}N_{-\alpha}}$ is a Klein factor, satisfying 
$\chi_\alpha^\dagger\chi_\alpha=1$. $N_\alpha$ is the particle number operator and $\varphi_{0,\alpha}$ is the zero mode operator which satisfy canonical commutation relations $[N_\alpha,\varphi_{0,\alpha}]=i$ \cite{Eggert.2009,Giamarchi:2004bk}. The bosonic fields can be expressed in the momentum representation as
\begin{align}
    \phi_\alpha(x)&=-\sum_{q>0}\sqrt{\frac{2\pi}{qL}}e^{-q\eta/2}[e^{i\alpha q x}b_{\alpha q}+e^{-i\alpha qx}b^\dagger_{\alpha q}]\ ,
    \label{Eq:bosonicfield}
\end{align}
where $\eta$ is a short-distance cutoff measured in units of the lattice spacing $a_0$. The plasmon operators $b_{q}$ satisfy $[b_{q\phantom{'}},b^\dagger_{q'}] = \delta_{q,q'}$.

Substituting the bosonized density operators in Eq.~\eqref{eq:spinless_fermion_hamiltonian} and keeping only quadratic terms yields the effective Hamiltonian
\begin{equation}
  H = \sum_{q\neq 0}[\omega_0(q)+m(q)]b_q^\dagger b_q +\frac{1}{2}\sum_{q\neq 0}g_2(q)(b_qb_{-q}+b^\dagger_qb^\dagger_{-q})\, .
  \label{Eq:bosonizedhamiltonian}
\end{equation}  
To capture the low energy properties of Eq.~\eqref{Eq:bosonizedhamiltonian} for short range interactions $V$, only momenta close to the Fermi surface are relevant, allowing us to take %where we take
$\omega_0(q)=v_F|q|$, $m(q)=g_4 |q|$, and $g_2(q)=g_2 |q|$ for all momenta $q_n = {2\pi n}/{L},\, n \in \mathbb{Z}\!\setminus\!\{0\}$ together with an interaction regularization scheme (details below). Here, $v_F$ is the Fermi velocity, $g_4$ contains the interaction $V$ from forward scattering while $g_2$ corresponds to the interactions from forward dispersion and backward scattering.  

Equation~\eqref{Eq:bosonizedhamiltonian} can be diagonalized by means of a Bogoliubov transformation
\begin{align}
\begin{split}
    a_q &= \cosh{(\theta_q)}b_q+\sinh{(\theta_q)}b_{-q}^\dagger   \\
    a_{-q}^\dagger &= \sinh{(\theta_q)}b_q+\cosh{(\theta_q)}b_{-q}^\dagger \, . 
\end{split}\label{Eq:diagonalizedbosonops} 
\end{align}
where $\tanh(2\theta_q) = g_2(q) / (\omega_0(q) + m(q) )$ \cite{Cazalilla:2006zw,Thamm:2022ja}.  Ground state expectation values obey $\langle a_q^\dagger a_{q'}\rangle=\delta_{q,q'}f_b(q)$ where $f_b(q)$ is the Bose-Einstein distribution function, which vanishes at zero temperature for $q>0$. 

This description can be related to the familiar Luttinger liquid Hamiltonian by defining two field operators 
\begin{align}
    \begin{split}
\phi(x) &= \frac{1}{2}\bqty{\phi_+(x)+\phi_{-}(x)} \\
\theta(x) &= \frac{1}{2}\bqty{\phi_+(x)-\phi_{-}(x)} \ ,
\end{split} \label{Eq:phiLR}
\end{align}
where $\phi(x)$ and $\theta(x)$ are dual bosonic fields describing long-wavelength density and phase fluctuations yielding 
\begin{align}
   H &=\frac{v}{2\pi}\int_0^L dx\left[\frac{1}{K}(\nabla\phi)^2+K(\nabla\theta)^2\right].
\end{align}
In this expression, $v$ is the interaction renormalized mode velocity of the propagating bosonic excitations (reducing to the bare Fermi velocity $v_F$ in the noninteracting limit, where $K=1$), while the Luttinger parameter $K$ quantifies the interaction strength and can be extracted from $K=\lim_{q\rightarrow 0}e^{2\theta_q}$.
We now present a high-level derivation of the two-body density matrix in the Luttinger regime (full details are provided in Appendix \ref{app:Details2RDMBosonization}): 
\begin{align}
    \rho_2(x_2',x_1';x_1,x_2) = \expval{\Psi^\dagger(x_2')\Psi^\dagger(x_1')\Psi(x_1)\Psi(x_2)} \, ,
    \label{Eq:rho2tocompute}
\end{align} 
which has the physical interpretation of the amplitude that two fermions which are annihilated at coordinates $x_1$ and $x_2$ are inserted again at coordinates $x_1'$ and $x_2'$. 

\begin{figure}[b]
 \centering
 \includegraphics[width=\linewidth]{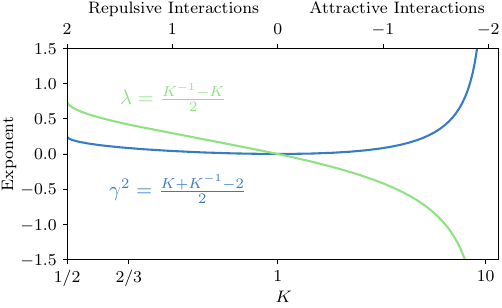}
 \caption[Dependence of $\gamma^2$ and $\lambda$ on the Luttinger parameter $K$]{Dependence of the two interaction-dependent exponents in the correlation function, $\gamma^2$ and $\lambda$, on the Luttinger parameter $K$. $\gamma^2$ is non-negative, while $\lambda$, the new exponent in this work, can attain negative values.}
 \label{fig:02_Exponents}
\end{figure}

Inserting the expansion of the fermionic field operators in terms of their right- and left-moving components, Eq.~\eqref{eq:fermionfield}, into Eq.~\eqref{Eq:rho2tocompute} and making repeated use of the Baker-Campbell-Hausdorff formula, $e^Ae^B=e^{A+B}e^{[A,B]/2}$ for c-number $[A,B]$, we obtain the non-vanishing contributions to the two-body reduced density matrix. Further using the boson cumulant formula for quadratic Hamiltonians $
\expval{e^{i(\phi_\alpha(x)-\phi_\alpha(x'))}}= e^{-\frac{1}{2}\expval{(\phi_\alpha(x)-\phi_\alpha(x'))^2}}$,  we find
\begin{widetext}
 \begin{align}
    \left\langle\Psi_\alpha^\dagger(x_2')\Psi_\alpha^\dagger(x_1')\Psi_\alpha(x_1)\Psi_\alpha(x_2)\right\rangle 
    &= 
    \frac{1}{4\pi^2\eta^2}e^{\frac{i\pi}{L}\alpha(x_2'+3x_1'-3x_2-x_1)} 
    e^{-\frac{1}{2}[\phi_\alpha(x_2'),\phi_\alpha(x_1')]}
    e^{\frac{1}{2}[\phi_\alpha(x_2'),\phi_\alpha(x_2)]}
    e^{\frac{1}{2}[\phi_\alpha(x_2'),\phi_\alpha(x_1)]}
    \notag\\&\times
    e^{\frac{1}{2}[\phi_\alpha(x_1'),\phi_\alpha(x_2)]}
    e^{\frac{1}{2}[\phi_\alpha(x_1'),\phi_\alpha(x_1)]}e^{-\frac{1}{2}[\phi_\alpha(x_2),\phi_\alpha(x_1)]}e^{-\frac{1}{2}\langle(\phi_\alpha(x_2')+\phi_\alpha(x_1')-\phi_\alpha(x_2)-\phi_\alpha(x_1))^2\rangle}
    \notag\\
    \left\langle\Psi_\alpha^\dagger(x_2')\Psi_\beta^\dagger(x_1')\Psi_\beta(x_1)\Psi_\alpha(x_2)\right\rangle &= \frac{1}{4\pi^2\eta^2}e^{\frac{i\pi}{L}(\alpha x_2'+\beta x_1'-\beta x_1 -\alpha x_2)} 
    e^{-\frac{1}{2}[\phi_\alpha(x_2'),\phi_\beta(x_1')]}
    e^{\frac{1}{2}[\phi_\alpha(x_2'),\phi_\beta(x_1)]}
    e^{\frac{1}{2}[\phi_\alpha(x_2'),\phi_\alpha(x_2)]}
    \notag\\&\times
    e^{\frac{1}{2}[\phi_\beta(x_1'),\phi_\beta(x_1)]}
    e^{\frac{1}{2}[\phi_\beta(x_1'),\phi_\alpha(x_2)]} 
    e^{-\frac{1}{2}[\phi_\beta(x_1),\phi_\alpha(x_2)]}e^{-\frac{1}{2}\langle(\phi_\alpha(x_2')+\phi_\beta(x_1')-\phi_\beta(x_1)-\phi_\alpha(x_2))^2\rangle}  
    \notag \\
    \left\langle\Psi_\alpha^\dagger(x_2')\Psi_\beta^\dagger(x_1')\Psi_\alpha(x_1)\Psi_\beta(x_2)\right\rangle &= -\left\langle\Psi_\alpha^\dagger(x_2')\Psi_\beta^\dagger(x_1')\Psi_\beta(x_2)\Psi_\alpha(x_1)\right\rangle \ ,
    \label{Eq:correlatorterms} 
\end{align}
\end{widetext} 
where $\alpha\neq \beta$.  
Zero temperature ground state expectation values of pairs of bosonic operators of the same species are given by: 
\begin{align}
\langle\phi_\alpha(x')\phi_\alpha(x)\rangle+\langle\phi_\alpha(x)\phi_\alpha(x')\rangle =\notag \\&\hspace{-4cm}\sum_{q>0}\frac{2\pi}{|q|L}e^{-\eta q}[\sinh^2(\theta_q)+\cosh^2(\theta_q)]\notag\\
    &\hspace{-4cm}\times [e^{i\alpha q(x'-x)}+e^{i\alpha q(x-x')}] \ .  
    \label{Eq:alphaalphaphicorrelator}
\end{align}  
Using the definition of the Luttinger parameter $K=\lim_{q\rightarrow 0}e^{2\theta_q}$, we introduce an interaction cutoff $\epsilon$ \cite{Cazalilla:2006zw,Thamm:2022ja}, such that
\begin{align}
\cosh^2{(\theta_q)}+\sinh^2{(\theta_q)}-1 &\approx \frac{K+K^{-1}-2}{2}e^{-\epsilon|q|}\notag\\
    &\equiv \gamma^2e^{-\epsilon|q|} \ . 
    \label{Eq:gamma2exponent}
\end{align}
There are multiple methods to regularize the momentum sum in Eq.~\eqref{Eq:alphaalphaphicorrelator}. As we will later show, $\epsilon$ can be used to make a direct connection between the continuum Luttinger model and the low energy sector of specific microscopic models. 
Equation~\eqref{Eq:gamma2exponent} also serves as the definition of the positive interaction exponent
\begin{equation}
    \gamma^2 = \frac{K + K^{-1}-2}{2}
    \label{eq:gamma2def}
\end{equation}
which is plotted as a function of the Luttinger parameter $K$ in Fig.~\ref{fig:02_Exponents}. 

For $\alpha \ne \beta$, we need to evaluate the anomalous Luttinger liquid correlator $\langle\phi_\alpha(x')\phi_\beta(x)\rangle$
\begin{align}
\langle\phi_\alpha(x')\phi_\beta(x)\rangle+\langle\phi_\alpha(x)\phi_\beta(x')\rangle &=
\notag\\&\hspace{-3cm}
2\sum_{q>0}\frac{2\pi}{|q|L}e^{-\eta q}[-\sinh{(\theta_q)}\cosh{(\theta_q)}]
\notag\\&\hspace{-3cm}\times 
[e^{i\alpha q(x'-x)}+e^{i\alpha q(x-x')}] \ .
\label{Eq:alphabetaphicorrelator}
\end{align}
from which a second interaction-dependent exponent $\lambda$ can be defined
%$\lambda$ that arises from the correlations between right- and left-moving fermions, 
%
\begin{equation}
    -2\sinh{(\theta_q)}\cosh{(\theta_q)} \approx 
    % \frac{K^{-1}-K}{2}e^{-\epsilon|q|} \!\!\!\!\! %\notag \\
    \lambda e^{-\epsilon|q|} \ ,
    \label{Eq:ExponentLambda}
\end{equation}
with
\begin{equation}
    \lambda = \frac{K^{-1}-K}{2}
\end{equation}
which arises from correlations between left- and right-moving fermions.  
The behavior of $\lambda$ as a function of $K$ is shown in Fig.~\ref{fig:02_Exponents}; unlike $\gamma^2$, it can take negative values.

Putting these results together, we can compute the exponentiated expectation values appearing in the \TRDM for $\alpha \ne \beta$
\begin{align}
    &e^{-\frac{1}{2}\langle(\phi_\alpha(x)-\phi_\alpha(x'))^2\rangle}  = 
    \frac{-i\sin{(\frac{i\pi}{L}\eta)}}{|\sin{(\frac{\pi}{L}(\alpha(x-x')+i\eta))}|}
    \notag\\&\quad\quad\times
    \left[\frac{-i\sin{(\frac{\pi}{L}i(\eta+\epsilon))}}{|\sin{(\frac{\pi}{L}(\alpha(x-x')+i(\eta+\epsilon))}|}\right]^{\gamma^2}\ ,
   \label{Eq:alphaalpha} \\
         &e^{-\frac{1}{2}\langle \phi_\alpha(x)\phi_\beta(x') \rangle} e^{-\frac{1}{2}\langle\phi_\alpha(x')\phi_\beta(x)\rangle} = 
     \notag\\&\quad\quad
         \left|e^{-\frac{\pi}{L}(\eta+\epsilon)}\sin{\left(\frac{\pi}{L}\left(\alpha(x-x')+i(\eta+\epsilon)\right)\right)} \right|^\lambda\ .
         \label{Eq:alphabeta}
    \end{align}

The final step is to compute the exponentials of the commutators of the bosonic field operators. Inserting the definition of $\phi_\alpha(x)$ given by Eq.~\eqref{Eq:bosonicfield}, straightforwardly, for $\alpha\neq \beta$, $e^{[\phi_\alpha(x'),\phi_\beta(x)]}=1$. In the case of the non-vanishing commutator, we obtain \cite{Thamm:2022ja},
\begin{align}
e^{\frac{1}{2}[\phi_\alpha(x),\phi_\alpha(x')]} &= i\sgn(\alpha(x-x'))e^{-i\alpha\frac{\pi}{L}(x-x')} 
\label{Eq:commutatorexponential}
\end{align} 
after taking the limit $\eta \rightarrow 0$.

We note that from the above expressions, the \ORDM for spinless fermions in 1D can be directly obtained as
\begin{equation}
\langle\Psi^\dagger(x')\Psi(x)\rangle = \frac{\sin\big(k_F\Delta x\big)}{L\sin\Big(\frac{\pi}{L}\Delta x\Big)} \left| \frac{\sin\big(i\pi\epsilon/L\big)}{\sin\Big(\frac{\pi}{L}(\Delta x+i\epsilon)\Big)} \right|^{\gamma^2},
\label{Eq:OBDMLL}
\end{equation}
where $\Delta x = x'-x$, in agreement with known results \cite{DzyaloshinskiiLarkin1974,LutherPeschel1975,Cazalilla:2006zw,Thamm:2022ja}.

We now have all the pieces to evaluate all terms in Eq.~\eqref{Eq:correlatorterms}. By keeping $\epsilon q$ finite, we can safely take the limits 
$\eta/x \rightarrow 0, \eta q\rightarrow 0, \eta/L \rightarrow 0$
in Eq.~\eqref{Eq:alphaalpha} and Eq.~\eqref{Eq:alphabeta} and we obtain a final constructive bosonization expression for the \TRDM:
\begin{widetext}
\begin{align}
\!\!\!\!&\langle\Psi^\dagger(x_2')\Psi^\dagger(x_1')\Psi(x_1)\Psi(x_2)\rangle = 
         \notag\\
&\qquad \frac{\cos{(k_F(x_2'\!+\!x_1'\!-\!x_2\!-\!x_1))}}{2\pi^2}\left[\frac{\h{x_1'}{x_2'}{x_1}{x_2}{0}}{\h{x_2}{x_2'}{x_1}{x_1'}{0}\h{x_1}{x_2'}{x_2}{x_1'}{0}}\right]\left|\frac{h_\epsilon(0,0)\h{x_1'}{x_2'}{x_1}{x_2}{\epsilon}}{\h{x_2}{x_2'}{x_1}{x_1'}{\epsilon}\h{x_1}{x_2'}{x_2}{x_1'}{\epsilon}}\right|^{\gamma^2}   \notag\\
    &\quad+\frac{\cos{(k_F(x_2'\!-\!x_1'\!-\!x_2\!+\!x_1))}}{2\pi^2}\left[\frac{1}{\h{x_2}{x_2'}{x_1}{x_1'}{0}}\right]\left|\frac{h_\epsilon(0,0)}{\h{x_2}{x_2'}{x_1}{x_1'}{\epsilon} }\right|^{\gamma^2}\left|\frac{\h{x_1'}{x_2'}{x_1}{x_2}{\epsilon}}{\h{x_1}{x_2'}{x_2}{x_1'}{\epsilon}}\right|^\lambda \notag\\
&\quad-\frac{\cos{(k_F(x_2'\!-\!x_1'\!+\!x_2\!-\!x_1))}}{2\pi^2}\left[\frac{1}{\h{x_1}{x_2'}{x_2}{x_1'}{0}}\right]\left|\frac{h_\epsilon(0,0)}{\h{x_1}{x_2'}{x_2}{x_1'}{\epsilon}}\right|^{\gamma^2}\left|\frac{\h{x_1'}{x_2'}{x_1}{x_2}{\epsilon}}{\h{x_2}{x_2'}{x_1}{x_1'}{\epsilon}}\right|^\lambda \ ,
\label{Eq:eq_twoRDM}
    \end{align}
\end{widetext}
where we have defined a convenient shorthand notation
\begin{align}
    h_\epsilon(x,y) &= d_\epsilon(x)d_\epsilon(y) \\
    d_\epsilon(x) &= \frac{L}{\pi}\sin\left[\frac{\pi}{L}(x+i \epsilon)\right]\, . 
\end{align} 
Here, $\vert d_0(x_2-x_1)\vert$ represents the chord length between the two coordinates $x_1$ and $x_2$ across the ring (due to the presence of periodic boundary conditions in 1D, see e.g.~Fig.~\ref{fig:01_Configuration}), and $\vert d_\epsilon(x_2-x_1)\vert$ is its generalization including the interaction cutoff $\epsilon$. 
We see that the exponent $\gamma^2$ appears in all three terms of Eq.~\eqref{Eq:eq_twoRDM}, whereas the exponent $\lambda$ only appears in the latter two due to the effect of anomalous correlators.
 
%========================================================================
\section{Analysis of the 2-RDM}
\label{sec:2RDMAnalysis}
%========================================================================

The complete expression for the two-particle density matrix given in Eq.~\eqref{Eq:eq_twoRDM} is at first glance rather unwieldy.  Without taking into account any symmetries, it is naively a function of four separate coordinates. In this section, we present an extensive analysis of its structure, focusing on its diagonal elements, and various limiting cases to build intuition before a final discussion of its power in elucidating the effects of interactions and exchange statistics in the Luttinger model.  We consider a system of $N$ fermions on a ring of size $L$ and to simplify expressions we set the unit for all lengths such that the density $n_0=N/L=1/2$ and we fix the interaction-dependent cutoff to be $\epsilon=1$ (unless otherwise stated).

% -----------------------------------------------------------------------------
\subsection{General Geometric and Symmetry Considerations}
% -----------------------------------------------------------------------------
In the limit $x_1'=x_1$, and using our previous definition of the density
operator $\hat{n}(x_1)=\Psi^\dagger(x_1)\Psi(x_1)$, we can write the \TRDM as
$\langle\Psi^\dagger(x_2')\hat{n}(x_1)\Psi(x_2)\rangle$. This makes it
physically transparent that it measures the correlated hopping of one particle
between positions $x_2$ and $x_2'$ in the presence of another particle at
$x_1$.  The limits $x_2' \to x_2$, $x_2' \to x_1$ and $x_1' \to x_2$ can be
related to $x_1' \to x_1$ by exchanging the particle coordinate labels
($x_1\leftrightarrow x_2$ and/or $x_1'\leftrightarrow x_2'$) and thus the
resulting \TRDM can be obtained from
$\langle\Psi^\dagger(x_2')\hat{n}(x_1)\Psi(x_2)\rangle$ by anticommuting the field operators.  

The intersection between hyperplanes $x_1'=x_1$ and $x_2'=x_2$ gives rise to
the diagonal elements of the \TRDM, $\rho_2(x_2,x_1;x_1,x_2)$, while the intersection between hyperplanes $x_2'=x_1$ and $x_1'=x_2$ generates a negative copy of the diagonal elements due to antisymmetrization ($x_1'\leftrightarrow x_2'$). However, for all of the
other four intersections between the hyperplanes, the \TRDM vanishes exactly as
this condition requires $x_1=x_2$ or $x_1'=x_2'$, which is forbidden by the
Pauli exclusion for spinless fermions. 

An investigation of the arguments of the $\cos(\dots)$ multiplicative prefactors
in each of the terms of the \TRDM in Eq.~\eqref{Eq:eq_twoRDM}, leads to the recognition
that the arguments define a set of four orthogonal hyperplanes: 
\begin{equation}
    \begin{split}
        P_1 & :  x_2'+x_1'-x_2-x_1 \\
        P_2 & :  x_2'-x_1'-x_2+x_1 \\
P_3 & : x_2'-x_1'+x_2-x_1 \\
P_4 & : x_2'+x_1'+x_2+x_1
\label{eq:hyperplanes}
\end{split}
\end{equation}
where the fourth is orthogonal to the set $\qty{P_1,P_2,P_3}$ as the sum of all coordinates.  To understand the origin of Eq.~\eqref{eq:hyperplanes}, and before adopting it in the description of the \TRDM, we can return to a more natural choice for describing
the coordinates of a two-body object, \emph{i.e.}, the relative $(r)$ and center
of mass $(R)$ coordinates: 
\begin{align}
   r &= x_2 - x_1 \\
   R &= \frac{1}{2}(x_1 + x_2)\, ,
\label{eq:rRdef}
\end{align}
and the respective primed versions $r'=x'_2-x'_1$ and $R'=(x'_2+x'_1)/2$.  As
we consider spinless fermions on a ring (periodic boundary conditions), the \TRDM
is invariant under a constant shift of all cartesian coordinates. In this case, $r'$ and $r$ are independent of such a shift, while for $R'$ and $R$, we can define the difference 
\begin{equation}
    \Delta R=R'-R= \frac{1}{2}(x'_2+x'_1-x_2-x_1)
\label{eq:ΔRdef}
\end{equation}
as the third independent argument from such a change. Finally, the sum $\Sigma
R=R'+R=(x'_2+x'_1+x_2+x_1)/2$ must drop out from Eq.~\eqref{Eq:eq_twoRDM} due to translation symmetry, and as a result, in the following sections we will often express quantities in terms of the coordinates $(r,r',\Delta R)$ only, utilizing the shorthand notation:
\begin{equation}
   \rho_2(r',r,\Delta R) \equiv \rho_2(x_2',x_1';x_1,x_2). 
\label{eq:rho2rrpDR}
\end{equation}

%-----------------
\subsection{Density-Density Correlation Function}

We begin our detailed analysis with the diagonal elements of the \TRDM defined by taking the limits $x'_1\to x_1$ and $x'_2\to x_2$ of Eq.~(\ref{Eq:eq_twoRDM}), which due to the translational invariance, can be captured by the single relative coordinate $r=x_2-x_1$, hence
\begin{align}
&\langle\Psi^\dagger(r)\hat{n}(0)\Psi(r)\rangle=n_0^2-\frac{1}{2L^2\sin^2\left(\frac{\pi}{L}r\right)}\notag\\&+\frac{(K-1)\cos\left(\frac{2\pi}{L}r\right)}{2L^2\abs{\sin\left(\frac{\pi}{L}(r+i\epsilon)\right)}^{2}}-\frac{(K-1)\sin^2\left(\frac{2\pi}{L}r\right)}{4L^2\abs{\sin\left(\frac{\pi}{L}(r+i\epsilon)\right)}^{4}}\notag\\
&+\frac{\abs{\sin\left(\frac{i\pi}{L}\epsilon\right)}^{2(K-1)}}{2L^2\abs{\sin\left(\frac{\pi}{L}(r+i\epsilon)\right)}^{2(K-1)}}\frac{\cos\left(2k_Fr\right)}{\sin^2\left(\frac{\pi}{L}r\right)},
\label{Eq:diagonal}
\end{align}
where, as expected, the above expression vanishes in the limit $r \to 0$. Details of taking the formal limits are rather involved and are included in Appendix~\ref{app:2RDMdetails}. Equation~\eqref{Eq:diagonal} immediately provides access to the pair correlation function as defined by Eq~(\ref{eq:g2_from_2rdm}), with the uniform fermionic density $\langle\hat{n}(x)\rangle=n_0$, 
\begin{equation*}
    g_2(r)=\frac{\expval{\Psi^\dagger(r)\hat{n}(0)\Psi(r)}}{n_0^2}\,, 
\end{equation*}
which is related to the density-density correlation function via Eq.~\eqref{Eq:DensDensCorr}.  Thus $g_2(r)$ captures $\expval{\hat{n}(r)\hat{n}(0)}$ away from $r=0$.

In the limit $L\gg r\gg 1$, Haldane's result for the low-dimensional quantum fluid \cite{Haldane:1981eh,Cazalilla_2004}\footnote{In Ref.~\cite{Haldane:1981eh}, the second term in Eq.~(7) has the wrong sign \cite{Cazalilla_2004}} is
\begin{align}
    \langle\hat{n}(r)\hat{n}(0)\rangle &\approx n_0^2\big[1-\eta(2\pi n_0 r)^{-2}\notag\\
                                       &\;+\sum_{m=1}^\infty A_m(n_0 r)^{-m^2\eta}\cos{(2\pi mn_0 r)}\big].
    \label{Eq:haldane}
\end{align}
where the exponent $\eta=2K$ and the coefficients $A_m$ are model-dependent \cite{LukyanovTerras2003}. In this limit, Eq~(\ref{Eq:diagonal}) yields (re-inserting $\epsilon$)
\begin{align}
    \expval{\hat{n}(r)\hat{n}(0)}\!\approx\! n_0^2\left[1-\frac{K}{2\pi^2(n_0r)^2}+\frac{(n_0\epsilon)^{2K-2}\cos\left(2\pi n_0 r\right)}{2\pi^2\abs{n_0 r}^{2K}}\right]
\label{Eq:latticedensity}
\end{align}
which fixes the coefficient $A_1=(n_0\epsilon)^{2K-2}/(2\pi^2)$, demonstrating the utility of the constructive bosonization approach. It is worth mentioning that in the noninteracting limit $K\to1$, all higher-order corrections $A_{m>1}$ vanish while $A_1$ approaches $1/(2\pi^2)$. Comparing Eq.~(\ref{Eq:haldane}) and Eq.~(\ref{Eq:DensDensCorrFF}) and our expression, Eq~(\ref{Eq:diagonal}),  yields the known result for free fermions (FF) when $K=1$
\begin{align}
\langle\hat{n}(r)\hat{n}(0)\rangle_{\rm{FF}}=n_0^2-\frac{\sin^2\left(k_F r\right)}{L^2\sin^2\left(\frac{\pi}{L}r \right)}+n_0\delta(r)
\label{Eq:DensDensCorrFF}\, .
\end{align}

\begin{figure}[h]
    \centering
    \includegraphics[width=\columnwidth]{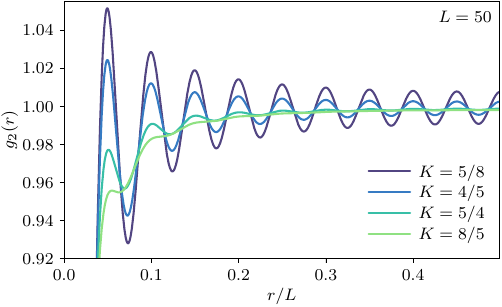}
    \caption{The density-density correlation function $g_2(r)$ at fixed $L=50$ for different values of the Luttinger parameter $K$.}
    \label{fig:04_densitydensity}
\end{figure}

For attractive interactions ($K>1$), determining the dependence of the coefficient $A_1$ on $\epsilon$ and $n_0$ influences only the sub-subleading oscillating term of the density-density correlations. However, for repulsive interactions ($K<1$) it sets the decay of oscillating terms to the leading order as can be seen in Fig.~\ref{fig:04_densitydensity}.
%%%%%%
%%%%%%

%
\subsection{Interaction Induced Two-Body Correlations}

\begin{figure}[h!]
    \centering
    \includegraphics[width=0.99\columnwidth]{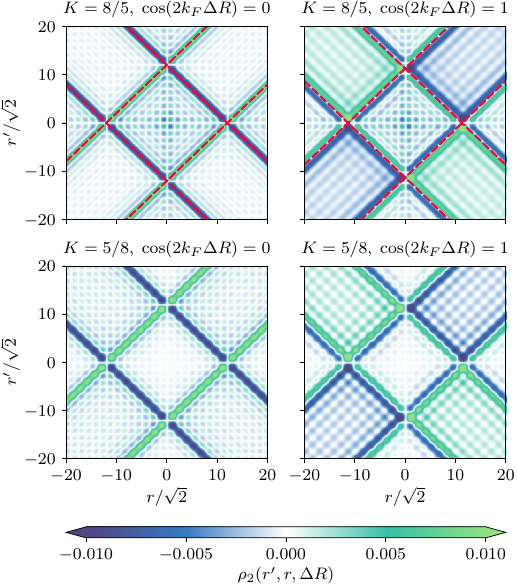}
    \caption{
        The 2-RDM of the Luttinger model on a ring for attractive $K=8/5$,
        (top row) and repulsive $K=5/8$ (bottom row) at two fixed values of
        $\Delta R = L/5 + 1/2$ (left column) and $\Delta R = L/5$ (right
        column) where $L=40$. The values of $\Delta R$ are chosen
        to suppress (enhance) the first oscillating fermionic term in
        Eq.~\eqref{Eq:eq_twoRDM} as indicated in the panel titles.  Red dashed
    lines denote the lines corresponding to the limits (hyperplanes)
    $x'_1\to x_1$, $x'_2\to x_2$, $x'_2\to x_1$, and $x'_1\to x_2$.}
    \label{fig:Drho_Fix_DR}
\end{figure}

We now turn our attention to the effects of interactions on the full structure of $\rho_2$, and
Fig.~\ref{fig:Drho_Fix_DR} shows four slices of the \TRDM as a function of the relative coordinates $r$ and $r'$. 
%%%%%%
%%%%%%
Here, we chose the values of the Luttinger parameter $K$ such that the
attractive ($K=8/5$, top row) and repulsive ($K=5/8$, bottom row) cases have
the same $\gamma^2=(K+K^{-1}-2)/2$ exponent while $\lambda=(K^{-1}-K)/2$ has
the same magnitude but opposite signs (see Fig.~\ref{fig:02_Exponents}).
$\lambda$ does not appear in the expression for the \ORDM, and to further
isolate interaction effects, we choose $\Delta R$ to neutralize the first term
in Eq.~(\ref{Eq:eq_twoRDM}) with $\cos(2k_F\Delta R)=0$ (left column) while
$\cos(2k_F\Delta R)=1$ (right column). 

In the left panels of the figure, we see a strong signal along the two pairs of
parallel lines (indicated by red dashed lines) defined by the coordinate limits
\begin{equation*}
\begin{array}{lll}
r'=r+2\Delta R &\Longleftrightarrow& x_1'=x_1,\\
r'=-r-2\Delta R &\Longleftrightarrow& x_2'=x_1,\\
r'=-r+2\Delta R &\Longleftrightarrow& x_1'=x_2,\\
r'=r-2\Delta R &\Longleftrightarrow& x_2'=x_2.
\end{array}
\end{equation*}
whereas in the right panel, the elements of the density matrix along these
lines are strongly suppressed, with a sign-change as one crosses them in the
orthogonal direction. Along these special lines, the elements are robust and
decays are not observed as $r$ and $r'$ are increased. 

The effects of changing the sign of the interaction from attractive to
repulsive are most evident in the interior of the diamond structure enclosing
the origin in Fig.~\ref{fig:Drho_Fix_DR}. For attractive $K>1$,  a rapidly
oscillating (at the scale of the short-distance cutoff) grid-like formation is
apparent that only weakly decays with increasing $r$ and $r'$. The same region
is suppressed for repulsive ($K<1$) interactions.

To further disentangle contributions arising from interparticle interactions
and fermionic exchange, we examine two additional quantities.  First, the
$2^{\rm nd}$ cumulant \cite{KutzelniggMukherjee1999}, 
\begin{multline}
    \Lambda_2(x_2',x_1';x_1,x_2) \equiv \rho_2(x_2',x_1';x_1,x_2) \\
    - \qty[\rho_1(x_1';x_1)\rho_1(x_2';x_2) - \rho_1(x_1';x_2)
    \rho_1(x_2';x_1)]\, ,
\label{eq:2cumulant}
\end{multline}
that measures the failure of the \TRDM to factorize according to Wick's theorem. It 
isolates two-particle correlations that cannot be reconstructed from
one-particle occupations and coherences alone. In particular,
Eq.~\eqref{eq:2cumulant} vanishes for a single Slater determinant, even though
the corresponding two-particle density matrix still contains nontrivial
exchange correlations. A nonzero cumulant consequently signals correlations
beyond the independent-particle description, although such correlations
need not arise exclusively from interactions in more general mixed or
degenerate states.

%%%%%%
\begin{figure}
    \centering
    \includegraphics[width=0.99\columnwidth]{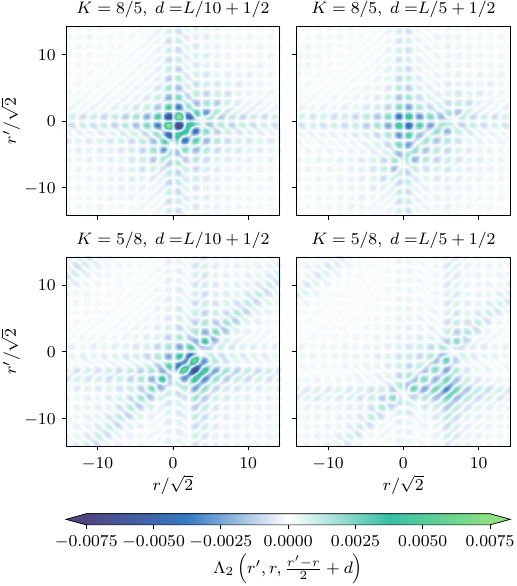}
    \caption{The second cumulant of the Luttinger model for spinless fermions on a ring for attractive $K=8/5$, (top row) and repulsive $K=5/8$ (bottom row) at two fixed values of $d = x_1'-x_1 = L/10 + 1/2$ (left column) and $d = L/5+1/2$ (right
    column) where $L=40$. The cumulant is localized and decays away from coordinate-coincidence lines.}
    \label{fig:cumulant}
\end{figure} 
%%%%%%

A distinct second quantity is the change in the two-particle density matrix relative
to a non-interacting free fermion (FF) reference state: 
\begin{multline}
\Delta \rho_2 (x_2',x_1';x_1,x_2)
\equiv \\ \rho_2 (x_2',x_1';x_1,x_2) - \rho_2^{\rm FF}
(x_2',x_1';x_1,x_2),
\label{eq:interaction_shift_2rdm}
\end{multline}
where $\rho_2^{\rm FF}$ is evaluated for the corresponding noninteracting
Hamiltonian with the same particle number, system geometry, and boundary
conditions. Unlike the cumulant, $\Delta \rho_2$ is explicitly
reference-dependent. It measures the total redistribution of two-particle
density and coherence produced by turning on interactions and
generally contains two physically different effects. The first captures changes in
the disconnected two-particle contribution resulting from interaction-induced
modifications of the one-particle density matrix characterized by the exponent
$\gamma^2$. These include changes in natural-orbital occupations,
single-particle coherence, and the associated exchange structure. The second
is the change in the irreducible two-particle correlations. 
Even for a weakly correlated interacting state, the difference between
Eqs.~\eqref{eq:2cumulant} and \eqref{eq:interaction_shift_2rdm}
may be appreciable because the interaction changes the
one-particle density matrix and hence its direct and exchange contributions.
More details on the analytic form of Eq.~\eqref{eq:interaction_shift_2rdm} are provided in Appendix~\ref{app:2RDMdetails}.

Figure~\ref{fig:cumulant} shows the second cumulant for the same interaction
values as Fig.~\ref{fig:Drho_Fix_DR} but with columns now showing two distances $d=x_1'-x_1$ small (left) and larger (right) across the interesting diagonals identified in the fixed $\Delta R$ slices of $\rho_2$.
Here, all panels show the sign oscillations observed inside the inner diagonal
and there is a clear decay as $r$ and $r'$ are increased.  Moving from
attractive to repulsive interactions suppresses elements with $r'>r$ due to the
sign choice of $d$.  The exact pair-coincidence lines at $r=0$ and $r'=0$ are
Pauli zeros. 

\begin{figure}
    \centering
    \includegraphics[width=0.99\columnwidth]{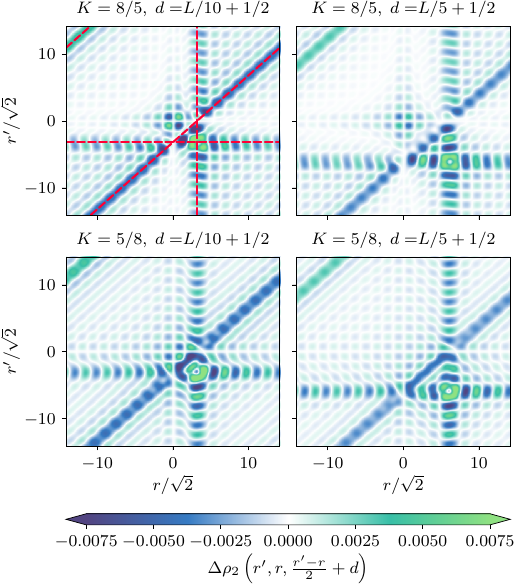}
    \caption{The effects of interactions as measured with respect to  
        the free fermion 2-RDM for attractive $K=8/5$, (top row)
        and repulsive $K=5/8$ (bottom row) at two fixed values of $d = x_1'-x_1
    = L/10 + 1/2$ (left column) and $d = L/5+1/2$ for $L=40$. Red dashed lines indicate
$x_1'=x_2$, $x_2'=x_1$, and $x_2'=x_2$.
    Subtraction of the free 2-RDM exposes interaction-induced redistribution, including attractive clustering.} 
    \label{fig:04_heatmap}
\end{figure}

Finally, Fig.~\ref{fig:04_heatmap} shows the interaction induced change in
$\rho_2$ as defined in Eq.~\eqref{eq:interaction_shift_2rdm} for the same
interaction parameters as  Fig.~\ref{fig:cumulant}.
%%%%%%
%%%%%%
Unlike the preceding figure, the subtraction here is the exact free fermions
\TRDM rather than a Wick determinant formed from the interacting \ORDM.
We exclude the hyperplane $x_1'=x_1$ by fixing $x_1'-x_1=d$
where $d$ is a constant.  We observe a star-shaped feature centered around
$r=r'=0$ in the upper row, signaling clustering of the spinless fermions when $K>1$. 
The red dashed guides in the upper-left panel identify representative cross-coordinate coincidence lines:
\begin{equation*}
\begin{array}{lll}
r=d &\Longleftrightarrow& x_1'=x_2,\\
r'=-d &\Longleftrightarrow& x_2'=x_1,\\
r'=r-d &\Longleftrightarrow& x_2'=x_2
\end{array}
\end{equation*}
where we expect to observe strong signals and antisymmetrization effects. 

In summary, we find that having analytic access to the full structure of the
\TRDM provides a detailed picture of how interactions affect diagonal and
off-diagonal elements.  Studying both $\Lambda_2$ and $\Delta\rho_2$
separates interaction-driven changes in single-particle
occupations and coherences from the emergence of correlated two-particle
structure. This distinction is particularly useful when investigating pairing:
an enhancement of a pair mode in the full two-particle density matrix can
result either from a reorganization of the occupied one-particle states, or
from the development of an intrinsically correlated pair mode. In one
dimension, where pairing is commonly characterized through spatial
correlations and finite-size scaling rather than a conventional local order
parameter, this separation provides a useful diagnostic of the origin and
spatial structure of the dominant pair correlations.

\subsection{Coherences}

The off-diagonal elements, or \emph{coherences}, of the \TRDM are useful and provide insight into the phase structure of our spinless fermion system.  In the repulsive regime $1/2 \le K \le 1$ we expect charge-density-wave (CDW) correlations, and the \TRDM lets us both confirm their onset and
determine the orbital character of the algebraic order.

\subsubsection{Onset of CDW correlations}
A CDW is a condensate of particle-hole pairs at wavevector $2k_F$, and it can be diagnosed in the diagonal (density--density) sector. From the diagonal limit of the \TRDM, Eq.~\eqref{Eq:diagonal}, the $2k_F$ component of the pair correlation function decays as a power law fixed by the Luttinger parameter [cf.~Eq.~\eqref{Eq:latticedensity}],
\begin{eqnarray}
C(r) &\equiv & \big[g_2(r)-1\big]_{2k_F}
   = \frac{(n_0\epsilon)^{2K-2}}{2\pi^2}\,
     \frac{\cos\!\left(2k_F r\right)}{(n_0 r)^{2K}}
 \nonumber \\   & & \propto\; \frac{\cos\!\left(2k_F r\right)}{|r|^{2K}}\, ,
\label{eq:cdw_decay}
\end{eqnarray}
so that smaller $K$ (stronger repulsion) gives slower decay -- the hallmark
of enhanced CDW correlations \cite{Giamarchi:2004bk,Voit1995,nayak2000density}.
Figure~\ref{fig:04_densitydensity} shows the full correlator $g_2(r)$ for
several values of $K$: the $2k_F$ oscillations persist to larger $r$ as $K$
decreases, and their envelope follows the power law $r^{-2K}$. In the
noninteracting limit $K\to1$ this reduces to the free-fermion result,
Eq.~\eqref{Eq:DensDensCorrFF}, whose oscillating term decays fastest
($\sim r^{-2}$).

\subsubsection{Orbital character}

\begin{figure}[t]
  \centering
  \includegraphics[width=\linewidth]{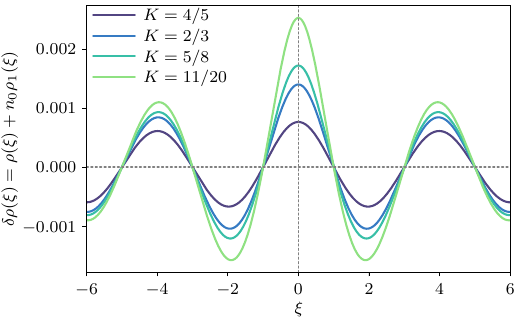}
  \caption{Orbital character of the CDW from the unnormalized
  background-subtracted coherence $\delta\rho(\xi)=\rho(\xi)+n_0\rho_1(\xi)$,
  Eq.~\eqref{eq:rho_CDW}, at $L=50$ for several repulsive $K$. Subtracting the single-particle background
  $-n_0\rho_1(\xi)$ [1-RDM, Eq.~\eqref{Eq:OBDMLL}] isolates the form factor
  $\varphi(\xi)$, the dominant eigenvector of the particle--hole block of
  $\rho_2$ at $Q=2k_F$. The residual is even in $\xi$ and peaked at $\xi=0$
  with no node, identifying the order as a site-centred ($s$-wave)
CDW.}
  \label{fig:CDW}
\end{figure}

The density-density limit establishes that phase with algebraically enhanced CDW correlations forms, but not which one: the condensing particle-hole pair may be on-site ($s$-wave) or bond-centered ($p$-wave), and these are distinguished only by the off-diagonal sector \cite{nayak2000density,Baldelli23}. We therefore introduce the off-diagonal coherence
\begin{equation}
\rho(\xi) = \Big\langle \Psi^{\dagger}(0)\,
   \Psi^{\dagger}\!\big(\tfrac{L}{2}+\xi\big)\,
   \Psi(0)\,\Psi\!\big(\tfrac{L}{2}\big)\Big\rangle .
\label{eq:rho_CDW}
\end{equation}
Anticommuting operators at the origin yields
$\rho(\xi) = -\big\langle \hat n(0)\, \Psi^{\dagger}\!\big(\tfrac{L}{2}+\xi\big)\,\Psi\!\big(\tfrac{L}{2}\big) \big\rangle$, implying that 
 $\rho(\xi)$ is the density at $x=0$ correlated with a bond of
internal extent $\xi$ starting at $L/2$. Here, the particle--hole (bond) operator
$\Psi^{\dagger}\!\big(\tfrac{L}{2}+\xi\big)\,\Psi\!\big(\tfrac{L}{2}\big)$ carries the orbital structure of the order in its
internal coordinate $\xi$. The disconnected term
$-n_0\,\rho_1(\xi)$ in $\rho(\xi)$
is a pure single-particle coherence: it produces a peak of magnitude $n_0^2$ at
$\xi=0$, decays as $\xi^{-(K+K^{-1})/2}$ with $k_F$ 
oscillations and has the same form for
any ordering symmetry, so it carries no CDW information. The orbital
wavefunction of the order resides in the connected residual
$\delta\rho(\xi)\equiv \rho(\xi)+n_0\,\rho_1(\xi)$. In the repulsive regime,
where the $2k_F$ correlations dominate this residual, its asymptotic
behavior is
\begin{equation}
\delta\rho(\xi)
   \;\xrightarrow[\;L/2\gg|\xi|\;]{}\;
   -\,n_0^{2}\,C\!\big(\tfrac{L}{2}\big)\,\varphi(\xi)
   \qquad (K<1)\, ,
\label{eq:formfactor}
\end{equation}
where $\varphi$, normalized to $\varphi(0)=1$, is the dominant eigenvector of
the particle--hole block of $\rho_2$ at $Q=2k_F$ -- the wavefunction of the
condensed particle--hole pair \cite{coleman1963structure,GarrodRosina1969} -- and the
amplitude $-n_0^2\,C(L/2)$ is the CDW onset [Eq.~\eqref{eq:cdw_decay}]
evaluated at separation $L/2$. In one dimension the associated eigenvalue is
sub-extensive (quasi-long-range order), but the eigenvector's orbital
symmetry remains well defined. The full block is a matrix in the two bond
coordinates; $\delta\rho(\xi)$ is the single column obtained by fixing the
reference bond at the on-site density and scanning $\xi$, so for an
effectively rank-one block it returns $\varphi$ directly; we confirm this by
checking that the extracted shape is unchanged when the reference bond is
shifted to nearest-neighbour separation. The parity of $\varphi(\xi)$ under
$\xi\to-\xi$ then classifies the order in the spirit of
Ref.~\cite{nayak2000density}: an even $\varphi$ peaked at $\xi=0$ is an
on-site ($s$-wave) CDW, while an odd $\varphi$ with a node at $\xi=0$ is a
bond-centred ($p$-wave/Peierls) density wave.

Figure~\ref{fig:CDW} shows the (unnormalized) residual $\delta\rho(\xi)$ for
several repulsive $K$. 
It is even in $\xi$ and peaked at $\xi=0$ with no node,
identifying the order as a site-centred ($s$-wave) CDW. This is the
particle--hole counterpart of the pairing analysis below
(Fig.~\ref{fig:pwavepairing}), where the internal coordinates of the
\emph{pair} expose a $p$-wave pairing wavefunction; here the internal
coordinate of the \emph{bond} exposes the density-wave wavefunction. We note
that the full angular-momentum classification of Ref.~\cite{nayak2000density}
($s$, $p$, $d,\dots$) requires a two-dimensional Fermi surface; in one
dimension only the $s$/$p$ parity of $\varphi$ survives.

\subsubsection{Pairing correlations}  
\begin{figure}[t!]
    \centering
    \includegraphics[width=0.99\linewidth]{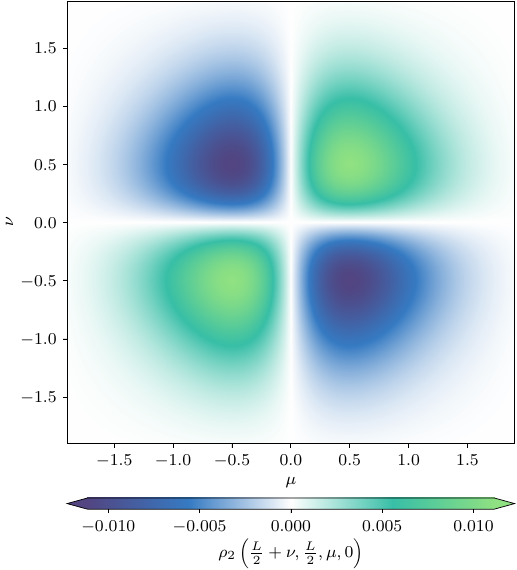}
    \caption{Heat map of \(\rho_2(L/2+\nu, L/2, \mu, 0)\) in the attractive regime (\(K > 1\)). Positive (green) and negative (blue) lobes alternate across the nodal lines \(\mu = 0\) and \(\nu = 0\), consistent with the antisymmetric \(p\)-wave character of the pair wavefunction. Here, we choose $L=40$, $K=8$, $n_0=1/2$, and $\epsilon=1$.}
    \label{fig:pwavepairing}
\end{figure}

We now examine the attractive regime, $K > 1$, where the system should exhibit a superconducting instability with $p$-wave pairing \cite{Giamarchi:2004bk}. Although true long-range order is suppressed in one dimension due to strong phase fluctuations, bosonization predicts an algebraic decay of the pair correlator, $\rho_2 \propto |x|^{-2/K}$, consistent with quasi-long-range order \cite{LutherPeschel1975}. We verify this by considering the off-diagonal element $\rho_2(L/2+\nu, L/2, \mu, 0)$, plotted in Fig.~\ref{fig:pwavepairing} as a function of the relative displacements \(\mu\) and $\nu$ for $K=8$. Unlike the particle--hole (CDW) channel, no background subtraction is required here: the pair operator $\Psi\Psi$ changes fermion number by two and has vanishing expectation value in a number-conserving ground state.  The only residual contribution is the particle--hole exchange product of two $1$-RDMs, which decays as $R^{-(K+K^{-1})}$ and is negligible against the pairing quasi-condensate ($\sim R^{-2/K}$) at the pair separation $R=L/2$, overwhelmingly so for $K=8$, so the raw element returns the pair wavefunction $\psi(\mu)\,\psi^{*}(\nu)$ directly, with $\psi$ the dominant eigenvector of the particle--particle block of $\rho_2$ (the pairing counterpart of $\varphi$) \cite{yang1962concept}. A key feature of $p$-wave symmetry is its odd parity, meaning the pair wavefunction changes sign under spatial inversion. We find that the correlator vanishes along the lines $\mu=0$ and $\nu=0$ due to Pauli exclusion and forms distinct lobes that alternate in sign across these lines, consistent with antisymmetric pairing.

\section{Application: Spinless Fermions on a Lattice}

The \TRDM provides a complete description of all two-body observables, allowing for the computation of key quantities such as interaction energies, pair correlations, and static two-body observables in lattice models describable by TLL theory. In principle, given an appropriate mapping between the continuum and lattice descriptions, the \TRDM formalism can be applied to any microscopic model, making it a versatile tool for studying strongly correlated systems. We demonstrate this explicitly in the $J$-$V$ model of spinless fermions, where we show that the ultraviolet cutoff $\epsilon$ introduced in the analytic formulation can be directly linked to the microscopic parameters of the lattice model, ensuring consistency between the low-energy field theory and numerical simulations.
\begin{figure}[t]
    \centering 
    \includegraphics[width=\linewidth]{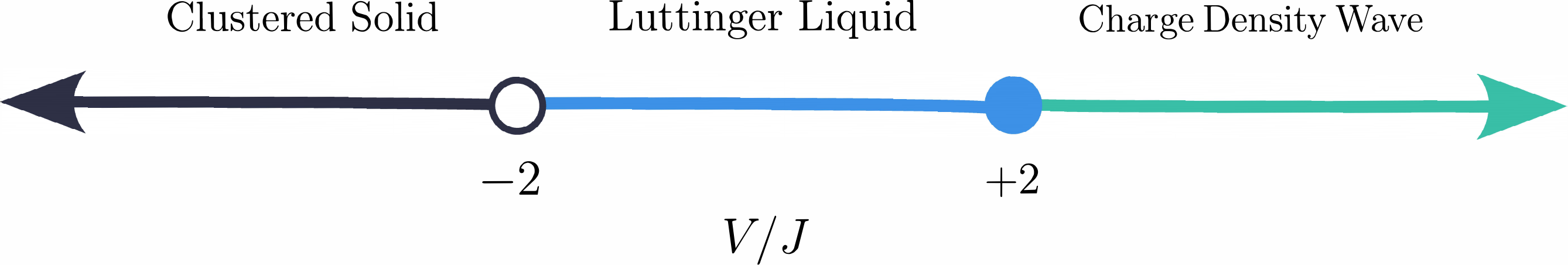}
    \caption{Illustration of the phase diagram for the $J$-$V$ model of spinless, interacting fermions on a lattice. When the ratio of the nearest neighbor interaction to the hopping parameter is within the range $-2<V/J<2$, the system can be modelled by Luttinger liquid theory. The system undergoes a first order phase transition to the clustered solid phase at $V/J=-2$ and an infinite order phase transition to the charge density wave phase at $V/J=2$.}
    \label{fig:03_JVPhaseDiagram}
\end{figure}

\subsection{$J$-$V$ Model}
We compare the TLL \TRDM with numerical results obtained from Density Matrix Renormalization Group (DMRG) techniques for the $J$-$V$ model of spinless fermions on a 1D lattice with $L$ sites at half-filling $L=2N$.  The Hamiltonian for this model is given by \begin{align}
    H &= -J \sum_{i=1}^L(c_{i+1}^\dagger c_i+c_i^\dagger c_{i+1}) + V\sum_{i=1}^L n_in_{i+1}. \label{Eq:JVModel}
\end{align} 
Here, $J$ denotes the hopping amplitude, $V$ is the nearest-neighbor interaction strength, and $n_i=c_i^\dagger c_i$ is the occupation number operator at site $i$. To ensure ground state non-degeneracy, we use periodic boundary conditions for an odd number of particles $N$ and anti-periodic boundary conditions for even $N$. The $J$-$V$ model can be modelled by TLL theory in the parameter range $-2 < V/J < 2$. The TLL interaction parameter $K$ can be exactly determined by relating the analytic model to the $J$-$V$ model at half filling using Bethe ansatz results from mapping to the spin-1/2 XXZ chain \cite{DesCloizeaux:1966},\cite{Yang:1966yc},\cite{Giamarchi:2004bk}. 
\begin{align}
    K &\equiv \sqrt{\frac{v_F+g_4+g_2}{v_F+g_4-g_2}}=  \frac{\pi}{2\cos^{-1}(-{V}/{2J})}
    \label{Eq:mapJVtoTLL}
\end{align} 

As the ratio $V/J$ increases, the system undergoes an infinite order phase transition to a charge-density wave phase at $V/J=2$ or $K=1/2$. Repulsive interactions ($V/J > 0$) have the TLL parameter $K<1$. As $V/J$ decreases, there is a first-order phase transition to the clustered solid phase occurring at $V/J=-2$, $K \rightarrow\infty$. Consequently, attractive interactions ($V/J < 0$) are associated with $K>1$. The schematic phase diagram is shown in Fig.~\ref{fig:03_JVPhaseDiagram}.

\subsection{DMRG Simulation}
\begin{figure}[t]
    \centering
    \includegraphics[width=8.6cm]{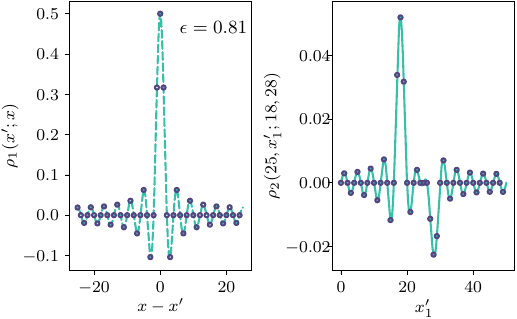}
    \caption{We plot the one-body reduced density matrix $\rho_1$ as a function of the relative coordinate $x_2-x_1$ at fixed interaction $V/J=-0.5$ and system size $L=50$. From fitting DMRG results to our expression for the 1-RDM, we obtain a value for the interaction cutoff $\epsilon=0.81$. Blue dots represent the DMRG data points, while dashed and solid lines are analytic predictions from bosonization for the 1-RDM and a cut of the 2-RDM with $x_2'=25$, $x_1=18$, and $x_2=28$, respectively. On the right panel, we use the same value of $\epsilon$ for our computation of the 2-RDM as a function of the coordinate $x_1'$ and find similarly strong agreement with the numerical results.}
    \label{fig:05_oneandtwordm}
\end{figure}

To test the utility of our analytic TLL \TRDM result, we compare to numerical DMRG calculations of the $J$-$V$~model, Eq.~\ref{Eq:JVModel}, in the Luttinger phase $|V/J|<2$. To perform DMRG computations, we use the \texttt{ITensors.jl} library \cite{itensor1,itensor2}. By carefully choosing initial states and projecting an orthogonal subspace to the desired ground state \cite{Thamm:2022ja}, we can reach system sizes of $N>50$ fermions on $L=2N$ lattice sites even for (anti-)periodic boundary conditions. For the DMRG calculations, we use a truncation cutoff of $10^{-12}$ keeping at maximum $6000$ states. 

We compute the \TRDM via 
\begin{align}
        \rho_2^{(i,j),(n,m)} &= \bra{\Psi_0} c_i^\dagger c_j^\dagger c_n c_m\ket{\Psi_0} 
\end{align}
from the ground state  $\ket{\Psi_0}$ obtained with DMRG.  The key step to make the calculation of the \TRDM numerically feasible for large systems is to use all symmetries of the $J$-$V$ Hamiltonian and general properties of the \TRDM, which allows us to drastically reduce the number of computed expectation values \cite{Radhakrishnan}. In particular, we use (i) the translational symmetry $(i,j,n,m)\to (i+1,j+1,n+1,m+1)$, (ii) the reflection symmetry $(i,j,n,m)\to (L-i,L-j,L-n,L-m)$, and (iii) the particle-hole symmetry $(i,j,n,m)\to(m,n,j,i)$. Here, a phase factor of $-1$ may occur in the case of antiperiodic boundary conditions. In addition, we can further reduce the number of computed entries in \TRDM by using fermionic anti-commutation relations that result in $ \rho_2^{(i,j),(n,m)}=-\rho_2^{(j,i),(n,m)}=-\rho_2^{(i,j),(m,n)}=\rho_2^{(j,i),(m,n)}$.

At the discrete positions of the lattice sites, we can then compare the analytic TLL expression Eq.~\eqref{Eq:eq_twoRDM} to the DMRG results as seen in Fig.~\ref{fig:05_oneandtwordm}. 
To make this comparison, we need to fix the interaction cutoff $\epsilon$ in the analytic expression. We obtain $\epsilon$ from a fit of the TLL \ORDM  result $\rho_1(x';x)$ to DMRG simulations for $\rho_1^{(i,j)}=\bra{\Psi_0}c_i^\dagger c_j\ket{\Psi_0}$ [left panel], showing excellent agreement for $\epsilon(V/J=-0.5,L=50)=0.81$. Using this value of the interaction cutoff $\epsilon$, we find similar agreement for the \TRDM [right panel]. For visual clarity, we focus on a cut through $\rho_2(x_2',x_1';x_1,x_2)$ for fixed $x_2', x_1, x_2$. 

\begin{figure}%[t!]
    \centering
    \includegraphics[width=8.6cm]{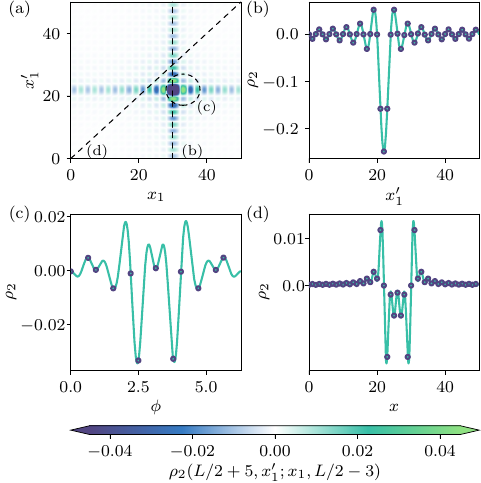}
    \caption{Panel (a) shows a heat map of the full correlation function $\rho_2(L/2+5,x_1';x_1,L/2 - 3)$ with two fixed coordinates for the interaction strength $V/J=-0.5$ and system size $L=50$. Dashed lines indicate the cuts that are plotted in the remaining panels. Panel (b) shows agreement between the DMRG data and analytic result where only the coordinate $x_1'$ is free; $\rho_2(L/2+5,x_1';L/2+5,L/2-3)$. Panel (c) shows the analytic and numerical results for a circular cut of the 2-RDM $\rho_2(L/2+5,x_1'(\phi),x_1(\phi),L/2-3)$ as a function of the angle $\phi$ where $x_1'(\phi)=L/2-3+5\sin(\phi)$, and 
    $x_1(\phi)=L/2+8+5\cos(\phi)$, and the other two coordinates are fixed. Panel (d) shows the agreement between DMRG and the analytic result when $x_1'=x_1\equiv x$, i.e.\@ for the cut $\rho_2(L/2+5,x;x,L/2-3)$. For panels (b-d) we use the value $\epsilon=0.81$ found in Fig.~\ref{fig:05_oneandtwordm}.}
    \label{fig:06_cuts_2rdm}
\end{figure}

Fixing two coordinates, $x_2'$ and $x_2$, we show a cut through the \TRDM as a function of $x_1, x_1'$ in Fig.~\ref{fig:06_cuts_2rdm}(a). To demonstrate the agreement between the analytic result and numerical simulations, we show various cuts in panels (b)-(d), along the dashed black lines in panel (a). 
The agreement between DMRG (points) and bosonization predictions (lines) is essentially exact using only a single parameter $\epsilon$ extracted as described above.

\subsection{Two-Body Observables}

\begin{figure}[t]
    \centering  %\vspace{0.5cm}
    \includegraphics[width=0.98\linewidth]{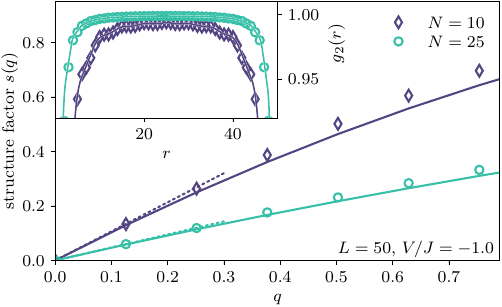}
    \caption{Structure factor $s(q)$ and density-density correlations $g_2(r)$ (inset) for filling fractions $1/5$ (purple diamonds) and $1/2$ (green circles) obtained from DMRG simulations of the lattice model with size $L=50$ and $V/J=-1.0$. The dashed lines show the analytically obtained $q\to0$ limiting behavior, $s(q\to0)$, a linear function with slope $K/(2k_F)$. The solid lines depict the analytical result in very good agreement with the numerical data.}
    \label{fig:structure_factor}
\end{figure}

\subsubsection{Static Structure Factor}
In this section, we consider the static structure factor, which is an experimentally accessible quantity defined via the density-density correlations $g_2$ as \cite{Giuliani2005} 
\begin{align}
    g_2(i-j) &= \frac{\ex{n_i n_j}}{n_0^2} -\frac{\delta_{ij}}{n_0} \ ,\\
s(q) &= 1 +  n_0 \sum_{j = 0}^{L-1} \left[g_2(j) - 1\right] e^{-i q j} \ .
\end{align}
Here, the momentum takes values $q  = 2\pi n/L$ with $n\in\mathbb{N}$.
An important property of $s(q)$ for experiments, which can be described by the Luttinger theory, is its small momentum behavior. In the canonical case of a fixed particle number $N$ it is directly proportional to the Luttinger parameter, $s(q\to0)\sim K {|q|}/{2k_F}$ \cite{Giamarchi:2004bk}, which provides a route to extract $K$ and thus to connect to the theory. 

We compute the structure factor for filling fractions $1/2$ and $1/5$ from the analytical expression, Eq.~\eqref{Eq:eq_twoRDM}, by discretizing it on the $L$ lattice sites using that $g_2(r)=\rho_2(r,0;0,r)/n_0^2$  
for $r>0$. We extract the interaction cutoff $\epsilon$ from a fit to $g_2$. 
Here, we numerically computed the Luttinger parameter at 1/5 filling for $V/J=-1.0$ as described in Appendix F of Ref.~\cite{Giamarchi:2004bk} and find $K_{1/5}(V/J=-1)=1.34135(1)$.

In the absence of a fixed particle number, the zero momentum structure factor is determined by the variance of the particle number $s(q=0)=\Delta N^2/N$. Related to this, the trace of the discretized expression, $\sum_j g_2(x_j)$, shows small deviations from the value $(N-1)/n_0 $, obtained from $g_2$ for the canonical lattice model.    Therefore, to apply the analytical Luttinger liquid result to the canonical numerical lattice simulation, we correct for the trace offset by shifting the value of $g_2(x_j)$ at sites $j=1$ and $j=L-1$ in a symmetric manner, i.e.\@ $g_2(x_j) \to g_2(x_j)  - \frac{1}{2}(\delta_{j,1}+\delta_{j,L-1}) \left(\sum_j g_2(x_j) - \frac{N-1}{n_0} \right)$. By correcting the trace, we ensure that $\Delta N^2 = \ex{\hat{N}^2}-N^2 = 0$. 

We find that the numerical structure factor obtained from DMRG simulations of the lattice model for both filling fractions, as shown in Fig.~\ref{fig:structure_factor}, is in excellent agreement with the analytical results (solid line) for $s(q)$ obtained from $g_2$ (inset) in both values of $n_0$. As expected, the structure factor for small momenta $q$ follows the linear relation with slope $K/2k_F$ (dashed line). 

\subsubsection{Ground State Lattice Energy }

\begin{figure}[t]
    \centering
    \includegraphics[width=8.6cm]{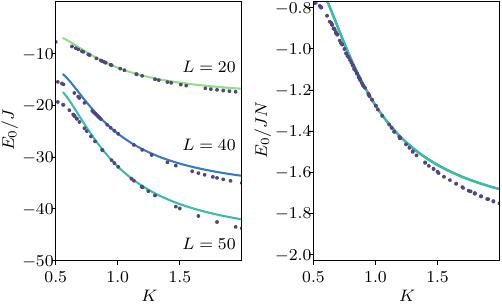}
    \caption{The ground state energy $E_0$ divided by the hopping parameter $J$ as a function of the Luttinger parameter $K$ for different system sizes. Solid lines show the analytic result while dark blue circles are the numerical results from DMRG. The righthand panel shows the energy  per particle. Strong agreement is shown for values of $K$ below 1.5.}
    \label{fig:09_dmrg_energies}
\end{figure}
To demonstrate the utility of the constructive bosonization approach taken
here, we show that continuum expressions for the \ORDM and \TRDM can be
leveraged (through the interaction cutoff $\epsilon$) to provide access to even short-range observables in the context of a microscopic model.   We compute the two-body lattice ground state energy given by \begin{align}
E_0 &= -J\sum_i\ex{c_i^\dagger c_{i+1}+c^\dagger_{i+1}c_i}+V\sum_i
\expval{n_i n_{i+1}} 
\end{align}
where the kinetic energy only depends on the one-body density matrix  $\rho_1(x_i,x_{i+1})= \langle c^\dagger_ic_{i+1} \rangle$, and the potential energy depends on $g_2(x_i,x_{i+1}) = \ex{c_i^\dagger c_ic_{i+1}^\dagger c_{i+1}}/n_0^2$, where the number operator is $n_i=c_i^\dagger c_i$. We can thus express the energy for the lattice model as 
\begin{align}
    E_0 = &-J\sum_i(\rho_1(x_i;x_{i+1})+\rho_1(x_{i+1};x_i))
    \notag \\&
    +V\sum_ig_2(x_i,x_{i+1}) n_0^2 \ . 
    \label{Eq:groundstateenergy}
\end{align}
Using translational invariance to rewrite in terms of relative coordinates, this becomes $E_0=-JL[\rho_1(1;0)+\rho_1(-1;0)]+VLn_0^2 g_2(1)$  
Fig.~\ref{fig:09_dmrg_energies}  shows the ground state energy, measured in units of the 
hopping parameter $J$, at different system sizes, comparing numerical DMRG results (blue circles) to the analytical expression, Eq.~\eqref{Eq:groundstateenergy} (solid lines).
We again obtain the short-distance cutoff by fits to the 1-RDM and correct the trace of $g_2$ as described for the structure factor above. Even though the ground state energy is a purely local observable, only involving elements of the 1-RDM and 2-RDM at distance of the lattice spacing, we find excellent agreement deep in the Luttinger phase. Deviations appear only close to the boundaries of the Luttinger liquid regime. For large $K$, the system approaches the clustered solid phase where the Luttinger liquid theory has no short range features, while the Umklapp term becomes weakly irrelevant for $K \to 1/2^{+}$, so its finite-size and short-distance corrections become important. For $K\to 1/2$ they drive the onset of charge density wave order.  
Deviations between DMRG and the TLL prediction for the energy per particle near the phase transitions do not depend on the system size $L$ (right panel), demonstrating %an absence of 
that finite-size effects are negligible for the system sizes considered.

\section{Conclusions} 
In this paper, we derived a closed finite-size expression for the two-body density matrix (\TRDM) of interacting, spinless fermions within the cutoff regularized Tomonaga-Luttinger liquid framework via a constructive bosonization approach. A notable outcome is the appearance of an interaction-dependent exponent $\lambda=(K^{-1}\!-\!K)/2$, which enters only when correlations between right- and left-moving sectors are relevant for $n>1$ density matrices. In the $n=2$ case, we obtained a finite-size, cutoff-regularized expression for the density–density correlation function that reduces to the familiar continuum form when the separations are larger than the lattice scale, yet small compared with the system size $L$.  Off-diagonal elements further indicate $p$-wave pairing in the attractive regime and signatures of $2k_{F}$ charge density wave order in the repulsive regime. The same matrix elements can be used to determine a wide class of observables, for example, interference fringe amplitudes in condensate experiments \cite{Polkovnikov:2006js}.

We benchmarked the analytic predictions against density matrix renormalization group data for the $J$-$V$ chain at half-filling. Using an ultraviolet cutoff, obtained through a self-consistent fitting procedure of the \ORDM, the analytic expressions agree with the lattice results in both attractive ($K>1$) and repulsive ($K<1$) Luttinger liquid regimes. This demonstrates that lattice effects, encoded by the cutoff, are necessary for comparison with the microscopics, and highlights the utility of our constructive bosonization approach. 

The present results invite several natural extensions. First, the method used here should generalize to any $n$, giving closed forms for higher $n$-body density matrices and thus opening a route to systematic calculations of the $n$-particle entanglement. In particular, the sub-leading $1/N$ term observed numerically in R{\'e}nyi entropies for $n>1$ \cite{Carlen:2016fv,Radhakrishnan} can be traced to the structure of the \TRDM, particularly the exponent $\lambda$. Second, a similar analysis could be applied to the time-dependent version of this quantity \cite{Protopopov2011},  allowing one to track the growth of correlations and entanglement after an interaction quantum quench. 
The closed-form \TRDM derived here is not only a complete description of equal-time two-body physics but also, for a fixed-$N$, nondegenerate ground state of a Hamiltonian with at most two-body interactions, \cite{Rosina1968}  a formally complete fingerprint of the many-body state, and hence of every higher reduced density matrix. This makes the Tomonage-Luttinger liquid a concrete setting in which to confront the still nontrivial problem of explicit and stable reduced density matrix reconstruction \cite{Mazziotti1998,Mazziotti2000Reconstruction,Massaccesi2026}.

\section{Data and Code Availability}
All code and data \cite{repo} needed to reproduce the results of this study are available online.

\acknowledgments
A.D. acknowledges support from the U.S. Department of Energy, Office of Science, Office of Basic Energy Sciences, under Award Number DE-SC0024333. 

\appendix

\onecolumngrid

% ==============================================================================
\section{Details on the bosonization calculation} 
\label{app:Details2RDMBosonization}
% ==============================================================================

In this appendix, we present more details on the derivation of the exact two-body density matrix in the Luttinger regime of main text Sec.~\ref{Sec:2RDMBosonizationAnalytical}. We expand the fermionic field operators in terms of their right- and left- moving components: 
\begin{align}
    \expval{\Psi^\dagger(x_2')\Psi^\dagger(x_1')\Psi(x_1)\Psi(x_2)}
    &= \left \langle \qty[e^{-ik_Fx_2'}\Psi_{-}^\dagger(x_2')+e^{ik_Fx_2'}\Psi_{+}^\dagger(x_2')]
    \qty[e^{-ik_Fx_1'}\Psi_{-}^\dagger(x_1')+e^{ik_Fx_1'}\Psi_{+}^\dagger(x_1')] \right.
    \notag\\&\quad\quad\quad\times \left.
    \qty[e^{ik_Fx_1}\Psi_{-}(x_1)+e^{-ik_Fx_1}\Psi_{+}(x_1)]
    \qty[e^{ik_Fx_2}\Psi_{-}(x_2)+e^{-ik_Fx_2}\Psi_{+}(x_2)]\right \rangle \, . 
    \label{Eq:eq_two_point_correlator_rl}
\end{align}%
Carrying out the multiplication, we obtain sixteen total terms; however, the majority will vanish. In order to satisfy the condition $\langle 0|\Psi^\dagger(x_2')\Psi^\dagger(x_1')\Psi(x_1)\Psi(x_2)|0\rangle \neq 0$, if a right or left-moving fermion is created, one must also be destroyed. The non-vanishing terms are of the form:
\begin{align*}
&\langle\Psi^\dagger_\alpha(x_2')\Psi_\alpha^\dagger(x_1')\Psi_\alpha(x_1)\Psi_\alpha(x_2)\rangle, \quad 
\langle\Psi^\dagger_\alpha(x_2')\Psi_\beta^\dagger(x_1')\Psi_\beta(x_1)\Psi_\alpha(x_2)\rangle,  \quad
\langle\Psi^\dagger_\alpha(x_2')\Psi_\beta^\dagger(x_1')\Psi_\alpha(x_1)\Psi_\beta(x_2)\rangle.  
\end{align*}
Then, for example, $\langle\Psi^\dagger_\alpha(x_2')\Psi_\beta^\dagger(x_1')\Psi_\beta(x_1)\Psi_\alpha(x_2)\rangle$ corresponds to the terms $\langle\Psi^\dagger_R(x_2')\Psi_L^\dagger(x_1')\Psi_L(x_1)\Psi_R(x_2)\rangle$ and $\langle\Psi^\dagger_L(x_2')\Psi_R^\dagger(x_1')\Psi_R(x_1)\Psi_L(x_2)\rangle$. We  now insert the expressions for $\Psi_\alpha(x)$ given by Eq.~\eqref{eq:fermionfield} and make repeated use of the Baker-Campbell-Hausdorff formula, $e^Ae^B=e^{A+B}e^{[A,B]/2}$, to obtain expressions for the non-zero terms in Eq.~\eqref{Eq:eq_two_point_correlator_rl}.   
Using the boson cumulant formula and the commutator $[N_\alpha,\varphi_{0,\alpha}]=i$, we arrive at main text Eq.~\eqref{Eq:correlatorterms}.

We work in the diagonal basis of the Hamiltonian, so we can apply Eq.~\eqref{Eq:diagonalizedbosonops} and get the bosonic field operator in terms of the operators $a_q$, in which the Hamiltonian is diagonal, 
\begin{align}
    \phi_\alpha(x)&= -\sum_{q>0}\sqrt{\frac{2\pi}{qL}}e^{-q\eta/2} 
     \big[e^{i\alpha qx}(\cosh{(\theta_q)}a_q-\sinh{(\theta_q)}a_{-q}^\dagger)  +e^{-i\alpha qx}(\cosh{(\theta_q)}a_q^\dagger-\sinh{(\theta_q)}a_{-q})\big] \ .
    \label{Eq:diagonalizedbosonicfield}
\end{align}
We start with the expectation values of pairs of bosonic operators of the same species $\langle\phi_\alpha(x)\phi_\alpha(x')\rangle$, and compute the sum 
\begin{align}
\langle\phi_\alpha(x)\phi_\alpha(x')\rangle+\langle\phi_\alpha(x')\phi_\alpha(x)\rangle = 
    \sum_{q>0}\frac{2\pi}{|q|L}e^{-\eta q}\big[e^{i\alpha q(x-x')}\cosh^2(\theta_q)
+e^{i\alpha q(x'-x)}\sinh^2(\theta_q)
\notag\\&\hspace{-6.2cm}
+e^{i\alpha q(x'-x)}\cosh^2(\theta_q)+e^{i\alpha q(x-x')}\sinh^2(\theta_q)\big] \ .
\end{align}
By factoring out the trigonometric functions and adding a zero, we arrive at main text Eq.~\eqref{Eq:alphaalphaphicorrelator} and identify the interaction exponent $\gamma^2$, Eq.~\eqref{Eq:gamma2exponent}. 
Similarly, we obtain Eq.~\eqref{Eq:alphabetaphicorrelator} for the anomalous correlator with $\alpha=-\beta$ and define the exponent $\lambda$, Eq.~\eqref{Eq:ExponentLambda}. Together these correlators yield the exponential terms Eq.~\eqref{Eq:alphaalpha} and Eq.~\eqref{Eq:alphabeta}.

Now, we compute the exponentials of the commutators of the bosonic field operators. Inserting the definition of $\phi_\alpha(x)$ given by Eq.~\eqref{Eq:bosonicfield}, sending $\beta \rightarrow -\alpha$ we get 
\begin{align}
    [\phi_\alpha(x),\phi_\beta(x')] &= \sum_{q>0}\sum_{q'>0}\frac{2\pi}{L}\frac{1}{\sqrt{qq'}}e^{-\frac{\eta}{2}(q+q')}\Big(e^{i\alpha(qx-q'x')}[b_{\alpha q},b_{-\alpha q'}]\notag+e^{-i\alpha(qx+q'x')}[b^\dagger_{\alpha q},b_{-\alpha q'}]
    \notag\\& \hspace{4.5cm}
    +e^{i\alpha(qx+q'x')}[b_{\alpha q},b^\dagger_{-\alpha q'}] 
    +e^{i\alpha(-qx+q'x')}[b_{\alpha q}^\dagger,b^\dagger_{-\alpha q'}] \Big) \notag\\ 
    &= 0 \\ 
    [\phi_\beta(x),\phi_\alpha(x')] &= 0  \\
     [\phi_\alpha(x),\phi_\alpha(x')] &= \sum_{q>0}\sum_{q'>0}\frac{2\pi}{L}\frac{1}{\sqrt{qq'}}e^{-\frac{\eta}{2}(q+q')} \Bigl(e^{i\alpha(qx+q'x')}[b_{\alpha q},b_{\alpha q'}]  + e^{-i\alpha(qx+q'x')}
     [b^\dagger_{\alpha q},b^\dagger_{\alpha q'}] 
     \notag \\ &\hspace{4.5cm} + e^{i\alpha(qx-q'x')}[b_{\alpha q},b^\dagger_{\alpha q'}] + e^{i\alpha(-qx+q'x')}[b_{\alpha q}^\dagger,b_{\alpha q'}] \Bigr)
     \notag\\
    &= \sum_{q>0}\frac{2\pi}{Lq}e^{-\eta q}[e^{i\alpha q(x-x')}-e^{-i\alpha q(x-x')}] \ ,
    \label{Eq:alphaalphacorr}
\end{align}
%\vspace{-\belowdisplayskip}
%\end{widetext}
%
where we have used $[b_q,b_{q'}^\dagger]=\delta_{q,q'}$.
We thus arrive at Eq.~\eqref{Eq:commutatorexponential}.
 
 We now have all the pieces to compute the terms in Eq.~\eqref{Eq:correlatorterms}. As stated, by keeping $\epsilon q$ finite, we can take the limit $\eta \rightarrow 0$. Using Eq.~\eqref{Eq:alphaalpha} and Eq.~\eqref{Eq:alphabeta}, with limit $\eta/x \rightarrow 0,\, \eta q\rightarrow 0,\, \eta/L \rightarrow 0$ and the substitutions $\sin{\frac{i\pi\eta}{L}}/\eta \rightarrow i\pi/L$, $i\sin{(ix)}=-|\sin{(ix)}|$, we obtain
%
%\begin{widetext}

%
\begin{align}
\langle\Psi_\alpha^\dagger(x_2')\Psi_\alpha^\dagger(x_1')\Psi_\alpha(x_1)\Psi_\alpha(x_2)\rangle &= -\frac{1}{4L^2}\frac{1}{\sin\left(\frac{\pi}{L}(x_2'-x_2)\right)\sin\left(\frac{\pi}{L}(x_1'-x_1)\right)}   \frac{\sin\left(\frac{\pi}{L}(x_2'-x_1')\right)}{\sin\left(\frac{\pi}{L}(x_2'-x_1)\right)}\frac{\sin\left(\frac{\pi}{L}(x_2-x_1)\right)}{\sin\left(\frac{\pi}{L}(x_1'-x_2)\right)}  \notag\\
&\qquad \times \frac{|\sin\left(\frac{i\pi}{L}\epsilon\right)|^{2\gamma^2}}{|\sin\left(\frac{\pi}{L}((x_2'-x_2)+i\epsilon)\right)|^{\gamma^2}|\sin\left(\frac{\pi}{L}((x_1'-x_1)+i\epsilon)\right)|^{\gamma^2}} \notag\\
&\qquad \times \frac{|\sin\left(\frac{\pi}{L}((x_2'-x_1')+i\epsilon)\right)|^{\gamma^2}|\sin\left(\frac{\pi}{L}((x_2-x_1)+i\epsilon)\right)|^{\gamma^2}}{|\sin\left(\frac{\pi}{L}((x_2'-x_1)+i\epsilon)\right)|^{\gamma^2}|\sin\left(\frac{\pi}{L}((x_1'-x_2)+i\epsilon)\right)|^{\gamma^2}} \\
\langle\Psi_\alpha^\dagger(x_2')\Psi_\beta^\dagger(x_1')\Psi_\beta(x_1)\Psi_\alpha(x_2)\rangle &= \frac{1}{4L^2} \frac{1}{\sinefreesame{x_2'}{x_1}{}\sinefreesame{x_1'}{x_2}{}}\notag\\ & \qquad \times \frac{|\sineintsame{x_2'}{x_1'}{}{\epsilon}|^{\lambda}|\sineintsame{x_2}{x_1}{}{\epsilon}|^{\lambda}}{|\sineintsame{x_2'}{x_2}{}{\epsilon}|^{\lambda}|\sineintsame{x_1'}{x_1}{}{\epsilon}|^{\lambda}} \notag\\
&\qquad \times \frac{|\sin{\frac{i\pi}{L}\epsilon}|^{2\gamma^2}}{|\sineintsame{x_2'}{x_1}{}{\epsilon}|^{\gamma^2}|\sineintsame{x_1'}{x_2}{}{\epsilon}|^{\gamma^2} }\\
\langle\Psi^\dagger_\alpha(x_2')\Psi^\dagger_\beta(x_1')\Psi_\alpha(x_1)\Psi_\beta(x_2)\rangle &= -\langle\Psi^\dagger_\alpha(x_2')\Psi^\dagger_\beta(x_1')\Psi_\beta(x_2)\Psi_\alpha(x_1)\rangle
\label{Eq:finalterms}
\end{align}
%
%\end{widetext}
where we used $\alpha,\beta=+1$ for right-movers and $-1$ for left-movers. 

Combining all non-zero terms (e.g.~$\langle\Psi_R^\dagger(x_2')\Psi_R^\dagger(x_1')\Psi_R(x_1)\Psi_R(x_2)\rangle$ and $\langle\Psi_L^\dagger(x_2')\Psi_L^\dagger(x_1')\Psi_L(x_1)\Psi_L(x_2)\rangle$), we obtain a final expression for the \TRDM, Eq.~\eqref{Eq:eq_twoRDM}.

% ==============================================================================
\section{Details on Structure of Full 2-RDM Expression}
\label{app:2RDMdetails}
% ==============================================================================
To set the stage for the promised analyses, we observe that in the expression Eq.~\ref{Eq:eq_twoRDM}, $h_\epsilon(x,y)$ appears with only three sets of arguments. Using this observation, we define the following.
\begin{align}
    \begin{split}
    \hr{\epsilon}&= \h{x_2}{x_2'}{x_1}{x_1'}{\epsilon} \\   \hd{\epsilon}&= \h{x_1'}{x_2'}{x_1}{x_2}{\epsilon}\\   \hf{\epsilon}&=\h{x_1}{x_2'}{x_2}{x_1'}{\epsilon},
    \end{split}
\label{Eq:h_def_x}
\end{align}
and the constant $h_\epsilon(0,0)=h_\epsilon=(\frac{L}{\pi})^2\sin^2(\frac{\pi}{L}i\epsilon)$. Therefore, we can write
\begin{align}
\langle\Psi^\dagger(x_2')\Psi^\dagger(x_1')\Psi(x_1)\Psi(x_2)\rangle &= \quad \frac{\cos{(k_F(x_2'\!+\!x_1'\!-\!x_2\!-\!x_1))}}{2\pi^2}\left[\frac{\hd{0}}{\hr{0}\hf{0}}\right]\left|\frac{h_\epsilon\hd{\epsilon}}{\hr{\epsilon}\hf{\epsilon}}\right|^{\gamma^2}   \notag\\ &\quad +\frac{\cos{(k_F(x_2'\!-\!x_1'\!-\!x_2\!+\!x_1))}}{2\pi^2}\left[\frac{1}{\hr{0}}\right]\left|\frac{h_\epsilon}{\hr{\epsilon} }\right|^{\gamma^2}\left|\frac{\hd{\epsilon}}{\hf{\epsilon}}\right|^\lambda \notag\\ &\quad -\frac{\cos{(k_F(x_2'\!-\!x_1'\!+\!x_2\!-\!x_1))}}{2\pi^2}\left[\frac{1}{\hf{0}}\right]\left|\frac{h_\epsilon}{\hf{\epsilon}}\right|^{\gamma^2}\left|\frac{\hd{\epsilon}}{\hr{\epsilon}}\right|^\lambda \,
\label{Eq:eq_twoRDM_x}
\end{align}
where the choice of the subscripts $+$, $-$ and $0$ will be explained later. The intricate functional form of \TRDM  lives in the four-dimensional hyperspace defined by the cartesian coordinates $x_1$, $x_2$, $x'_1$, and $x'_2$. Finally, to connect with the set of orthogonal hyperplanes we mentioned earlier in section \ref{sec:2RDMAnalysis}, we define $r_{\pm}=(r'\pm r)/2=(x'_2-x'_1\pm x_2\mp x_1)/2$. 

The appearance of at least one of the functions $\hr{0}$ and $\hf{0}$ in the denominator in each of the three terms brings our attention to the zeros of these functions, where $\hr{0}$ vanishes in planes $\Delta R=\pm r_-$ ($x_1'=x_1$ and $x_2'=x_2$), while  $\hf{0}=0$ at $\Delta R=\pm r_+$ ($x_1'=x_2$ and $x_2'=x_1$), where the evaluation of \TRDM under these conditions requires a proper limit evaluation (see Appendix:\ref{app:limits} for details).

Now, we write the \TRDM using the new set of coordinates as
\begin{align}
    {{\rho}}_2(r_+,r_-,\Delta R) &=\quad \frac{\cos{(2k_F\Delta R)}}{2\pi^2}\left[\frac{\hd{0}}{\hr{0}\hf{0}}\right]\left|\frac{h_\epsilon\hd{\epsilon}}{\hr{\epsilon}\hf{\epsilon}}\right|^{\gamma^2}   \notag\\ &\quad+\frac{\cos{(2k_Fr_-)}}{2\pi^2}\left[\frac{1}{\hr{0}}\right]\left|\frac{h_\epsilon}{\hr{\epsilon} }\right|^{\gamma^2}\left|\frac{\hd{\epsilon}}{\hf{\epsilon}}\right|^\lambda \notag\\ &\quad-\frac{\cos{(2k_Fr_+)}}{2\pi^2}\left[\frac{1}{\hf{0}}\right]\left|\frac{h_\epsilon}{\hf{\epsilon}}\right|^{\gamma^2}\left|\frac{\hd{\epsilon}}{\hr{\epsilon}}\right|^\lambda \,
\label{Eq:eq_twoRDM_r}
\end{align}
where now,
\begin{align}
    \begin{split}
    \hr{\epsilon}&= h_\epsilon(\Delta R+r_{-},\Delta R-r_{-}) \\   \hd{\epsilon}&= h_\epsilon(r_{+}+r_{-},r_{+}-r_{-})\\   \hf{\epsilon}&=h_\epsilon(\Delta R+r_{+},\Delta R-r_{+}),
    \end{split}
\label{Eq:h_def_r}
\end{align}
Here, the subscripts $+$ and $-$ in $h_{\epsilon,\pm}$ signal the explicit dependence on the coordinates $r_+$ or $r_-$, respectively, while the $0$ in $h_{\epsilon,0}$ is to indicate the dependence of the function on both $r_+$ and $r_-$. Clearly, the hyperplane $x_2'=x_2$, $x_1'=x_1$, $x_2'=x_1$ and $x_1'=x_2$ can be identified by $\Delta R=-r_{-}$, $\Delta R=r_{-}$, $\Delta R=-r_{+}$ and $\Delta R=r_{+}$, respectively.

In the case of free fermions, we have $\gamma^2=\lambda=0$, and thus
\begin{equation}
    {{\rho}}_{2,\FF}(r_+,r_-,\Delta R) =\frac{\cos{(2k_F\Delta R)}}{2\pi^2}\left[\frac{\hd{0}}{\hr{0}\hf{0}}\right]  +\frac{\cos{(2k_Fr_-)}}{2\pi^2}\left[\frac{1}{\hr{0}}\right] -\frac{\cos{(2k_Fr_+)}}{2\pi^2}\left[\frac{1}{\hf{0}}\right] \,
\label{Eq:eq_twoRDM_r_0}.
\end{equation}
This shows that ${{\rho}}_{2,\FF}$ is a combination of three terms, each led by an oscillation in the directions of  $r_-$ and $r_+$, $\Delta R$. The amplitude of oscillations depends on the functions $\hd{0}$, $\hr{0}$, and $\hf{0}$. The same holds for ${{\rho}}_2$  but with the additional modification of the amplitude of the oscillating terms by an interaction-dependent positive factors, and thus we can write
\begin{align}
    \Delta {{\rho}}_2(r_+,r_-,\Delta R) &=\quad \frac{\cos{(2k_F\Delta R)\hd{0}}}{2\pi^2\hr{0}\hf{0}}\left[\left|\frac{h_\epsilon\hd{\epsilon}}{\hr{\epsilon}\hf{\epsilon}}\right|^{\gamma^2}\!\!\!\!\!\!-\!\!1\!\right]   \notag\\ &\quad+\frac{\cos{(2k_Fr_-)}}{2\pi^2\hr{0}}\left[\left|\frac{h_\epsilon}{\hr{\epsilon} }\right|^{\gamma^2}\!\!\left|\frac{\hd{\epsilon}}{\hf{\epsilon}}\right|^\lambda\!\!\!\!-\!\!1\!\right]\notag\\  &\quad-\frac{\cos{(2k_Fr_+)}}{2\pi^2\hf{0}}\left[\left|\frac{h_\epsilon}{\hf{\epsilon}}\right|^{\gamma^2}\!\!\left|\frac{\hd{\epsilon}}{\hr{\epsilon}}\right|^\lambda \!\!\!\!-\!\!1\!\right] \,
\label{Eq:eq_twoRDM_r_D}.
\end{align}

To gain some understanding of this structure in a simplified way, let's consider a large ring in the regime $L\gg\vert r_+\vert\gg \vert\Delta R_0\vert$, $\epsilon$ and set $r_-=\Delta R=\Delta R_0$. In this limit we get 
\begin{align}
    \Delta {\rho}_2\approx B_1+\frac{\cos{(2k_Fr_+)}}{2\pi^2r_+^2}\left[B_2\left(\frac{\epsilon^2}{r_+^2}\right)^{K-1}\!\! \!\!\!\!-\!\!1\!\right] \,
\label{Eq:eq_twoRDM_DR0},
\end{align}
where $B_1=\frac{k_F\sin(2k_F\Delta R_0)}{2\pi^2\Delta R_0}\left[\left(4\Delta R_0^2/\epsilon^2+1\right)^{-\gamma^2/2}-1\right]$ and  $B_2=\left(4\Delta R_0^2/\epsilon^2+1\right)^{-\lambda}$.
This shows that on the line defined by the intersection of the two planes, $\Delta R=\Delta R_0$ and $r_-=\Delta R_0$, $\Delta {{\rho}}_2$ approaches the constant value $B_1$, as $r_+$ is increased, where the sub-leading term decays, asymptotically, as $r_+^{-2\min(1, K)}$. This indicates that the effects of the oscillating term will be more visible in the repulsive case ($K<1$), as evidenced in Fig.~\ref{fig:Drho_Fix_DR}, which can be attributed to the fermions' tendency to avoid each other in the presence of repulsive interactions. It is worth noting that, in such conditions,  the difference in $\Delta {\rho}_2$ persists at large $r_+$.

To explore the $\Delta R$ dependence of the central oscillating region in Fig.~\ref{fig:Drho_Fix_DR}, 
% we consider calculating $\Delta{\rho}_2$ on the surface of a  cylinder with a radius $r^2_0=r_+^2+r_-^2=(r'^2+r^2)/2$ encircling the $\Delta R$ axis, as represented in Fig.~\ref{fig:Vertical-cylinder}.
% %%%%%
%  \begin{figure}
%      \centering
%      \includegraphics[width=8.6cm]{Figures/VerticalCylinder.pdf}   \caption{ $\Delta {{\rho}}_2$ at $K=8/5$, evaluated on a cylinder of radius $r_0=3.8$ with $L=40$. }
%      \label{fig:Vertical-cylinder}
%  \end{figure}
% %%%%%
%The Figure shows that the signals weakly decay with increasing $\Delta R$. Also, the lack of oscillation concerning increasing $\Delta R$ is clear. We can understand this by revisiting our large ring with the conditions 
we consider $\vert\Delta R\vert\gg\vert r_+\vert+\vert r_-\vert+\epsilon$, where we can write
\begin{align}
    \Delta {{\rho}}_2\approx \frac{rr'\cos{(2k_F\Delta R)}}{2\pi^2\Delta R^4}\! + \! 
    \frac{\sin{(k_Fr')}\sin{(k_Fr)}}{\pi^2\Delta R^{2/K}}\mathcal{F}_K\,
\label{Eq:eq_Drho_LargeDR},
\end{align}
where,
\begin{align}
\mathcal{F}_K=\epsilon^{2\gamma^2}\left(r'^2+\epsilon^2\right)^{\lambda/2}\left(r^2+\epsilon^2\right)^{\lambda/2}-\Delta R^{\frac{2}{K}-2}
\end{align}
and we used the substitution $\cos{(2k_Fr_-)}-\cos{(2k_Fr_+)}=2\sin{(k_Fr')}\sin{(k_Fr)}$.
For $K<1$, 
\begin{align}
    \Delta {\rho}_2(K<1)\approx
    -\frac{\sin{(k_Fr')}\sin{(k_Fr)}}{\pi^2\Delta R^{2}}\,
\label{Eq:eq_Drho_LargeDR_KL1},
\end{align}
where the leading term in the expansion of $\Delta {\rho}_2$ has a power-law decay in $\Delta R$ as $1/ \Delta R^2$ and oscillating pattern with respect to $r'$ and $r$. On the contrary, when $K>1$,  $\Delta {\rho}_2$ show a slower decay in $\Delta R$ as $1/ \Delta R^{2/K}$($K>1$), where   
\begin{align}
    \Delta {\rho}_2(K>1)\approx
    \frac{\epsilon^{2\gamma^2}\sin{(k_Fr')}\sin{(k_Fr)}}{\pi^2\Delta R^{\frac{2}{K}}\left(r'^2+\epsilon^2\right)^{\abs{\lambda}/2}\left(r^2+\epsilon^2\right)^{\abs{\lambda}/2}}
\label{Eq:eq_Drho_LargeDR_KG1}.
\end{align} 
Also, having $K>1$ results in a negative $\lambda=(1/K-K)/2<0$ exponent, which, according to the last equation, weakens $\Delta {\rho}_2$ with increasing $r'$ or $r$, in agreement with \ref{fig:Drho_Fix_DR}. The change in the decay power is based on the competition between the decay of ${\rho}_{2}$ and ${\rho}_{2,\FF}$. In this limit, repulsive correlations weaken faster than attractive ones with increasing $\Delta R$ as $\Delta R^{-2/K}$. The oscillations due to increasing $\Delta R$ are concealed by attenuation with a factor $1/\Delta R^4$.
Overall, this behavior suggests that, for large $K$, a pair of fermions separated by a small distance $\vert r\vert$ at some position $R$, exhibit high correlations concerning the existence of fermionic pair with small $\vert r'\vert$ at a different position $R+\Delta R$ with large $\Delta R$, which echoes fermions' tendency to cluster.

\section{Computation of Limits} 
\label{app:limits}
Starting from the correlation function
\begin{align}
    \langle\Psi^\dagger(x_2')\Psi^\dagger(x_1')\Psi(x_1)\Psi(x_2)\rangle &=
\frac{\cos{(k_F(x_2'\!+\!x_1'\!-\!x_2\!-\!x_1))}}{2\pi^2}\left[\frac{\hd{0}}{\hr{0}\hf{0}}\right]\left|\frac{h_\epsilon\hd{\epsilon}}{\hr{\epsilon}\hf{\epsilon}}\right|^{\gamma^2}   \notag\\ &\qquad +\frac{\cos{(k_F(x_2'\!-\!x_1'\!-\!x_2\!+\!x_1))}}{2\pi^2}\left[\frac{1}{\hr{0}}\right]\left|\frac{h_\epsilon}{\hr{\epsilon} }\right|^{\gamma^2}\left|\frac{\hd{\epsilon}}{\hf{\epsilon}}\right|^\lambda \notag\\ &\qquad-\frac{\cos{(k_F(x_2'\!-\!x_1'\!+\!x_2\!-\!x_1))}}{2\pi^2}\left[\frac{1}{\hf{0}}\right]\left|\frac{h_\epsilon}{\hf{\epsilon}}\right|^{\gamma^2}\left|\frac{\hd{\epsilon}}{\hr{\epsilon}}\right|^\lambda ,
\end{align}
we want to rewrite this in a way that separates the interaction-dependent components of the correlation function. We define a new exponent $\Dlt = \gamma^2-\lambda$ and utilize the trigonometric identity $\cos{(A-B)}=\cos{(A+B)}+2\sin{(A)}\sin{(B)}$. Simplifying and rearranging the terms, the correlation function is now 
\begin{multline}
    \langle\Psi^\dagger(x_2')\Psi^\dagger(x_1')\Psi(x_1)\Psi(x_2)\rangle =\frac{1}{2\pi^2\hf{0}\hr{0}} \left|\frac{h_{\epsilon}^{\gamma^2}h_{\epsilon,0}^\lambda}{h_{\epsilon,+}^{\gamma^2}h_{\epsilon,-}^{\gamma^2}}\right|
    \notag \\
\qquad \qquad \qquad \times
    \cos{(k_F(x_1'+x_2'-x_1-x_2))}\qty[\hd{0}|\hd{\epsilon}|^\Dlt+\hf{0}|\hf{\epsilon}|^\Dlt-\hr{0}|\hr{\epsilon}|^\Dlt]\notag\\
    +\left|\frac{\hd{\epsilon}}{\hf{\epsilon}}\right|^\lambda\langle\Psi^\dagger(x_2')\Psi(x_2)\rangle\langle\Psi^\dagger(x_1')\Psi(x_1)\rangle 
- \left|\frac{\hd{\epsilon}}{\hr{\epsilon}}\right|^\lambda\langle\Psi^\dagger(x_2')\Psi(x_1)\rangle\langle\Psi^\dagger(x_1')\Psi(x_2)\rangle.
\end{multline}
where $\langle \Psi^\dagger(x_2)\Psi(x_1)\rangle$ is the one-body reduced density matrix,  
\begin{align}
    \langle \Psi^\dagger(x_2)\Psi(x_1)\rangle&= \rho_1^0(x_1,x_2)\left|\frac{\sin{(\frac{i\pi}{L}\epsilon)}}{\sin{(\frac{\pi}{L}((x_2-x_1)+i\epsilon)}}\right|^{\gamma^2} \notag\\
    \rho_1^0(x_1,x_2) &= \frac{\sin(k_F(x_2-x_1))}{L\sin{(\frac{\pi}{L}(x_2-x_1))}}.
\end{align}
\subsection{Wick's theorem}
We simplify the expression by introducing the relative coordinates 
$r = x_2 - x_1$, $r' = x_2' - x_1'$, and center-of-mass coordinates 
$R = \frac{x_1 + x_2}{2}$, $R' = \frac{x_1' + x_2'}{2}$, 
with $\Delta R = R' - R$. The two-body density matrix becomes:
\begin{align}
\langle\Psi^\dagger(x_2')\Psi^\dagger(x_1')\Psi(x_1)\Psi(x_2)\rangle 
&= \left|\frac{h_{\epsilon}^{\gamma^2} \hd{\epsilon}^\lambda}
         {\pi^2 \hf{\epsilon}^{\gamma^2} \hr{\epsilon} ^{\gamma^2}}\right| 
   \cos\left(2k_F \Delta R\right) \notag\\
&\quad \times 
\Bigg\{ \frac{
(|\hr{\epsilon}|^\delta - |\hd{\epsilon}|^\delta)\cos\left(\frac{\pi}{L}(r - r')\right) 
+ (|\hf{\epsilon}|^\delta - |\hd{\epsilon}|^\delta)\cos\left(\frac{\pi}{L}(r + r')\right)}{
\cos\left(\frac{\pi}{L}(r - r')\right)\cos\left(\frac{2\pi}{L} \Delta R\right)
\left[
\cos\left(\frac{\pi}{L}(r + r')\right) 
- \cos\left(\frac{2\pi}{L} \Delta R\right)
\right]} \notag\\
&\qquad
+ \frac{(|\hr{\epsilon}|^\delta - |\hf{\epsilon}|^\delta)\cos\left(\frac{2\pi}{L} \Delta R\right)}{
\cos\left(\frac{\pi}{L}(r - r')\right)\cos\left(\frac{2\pi}{L} \Delta R\right)
\left[
\cos\left(\frac{\pi}{L}(r + r')\right) 
- \cos\left(\frac{2\pi}{L} \Delta R\right)
\right]}
\Bigg\} \notag\\
&\quad + \left|\frac{\hd{\epsilon}}{\hf{\epsilon}}\right|^\lambda
          \langle\Psi^\dagger(x_2')\Psi(x_2)\rangle
          \langle\Psi^\dagger(x_1')\Psi(x_1)\rangle 
- \left|\frac{\hd{\epsilon}}{\hr{\epsilon}}\right|^\lambda
          \langle\Psi^\dagger(x_2')\Psi(x_1)\rangle
          \langle\Psi^\dagger(x_1')\Psi(x_2)\rangle.
\label{Eq: wicksbeforelimit}
\end{align}
where we used the identity $\sin{(A)}\sin{(B)}=\frac{\cos(A-B)}{2}-\frac{\cos(A+B)}{2}$. 
As $K=1$, the exponent $\Dlt = \gamma^2-\lambda$ goes to zero along with $\gamma^2=\lambda=0$. In the non-interacting limit, the ratios $\left|\frac{\hd{\epsilon}}{\hf{\epsilon}}\right|^\lambda$ and $\left|\frac{\hd{\epsilon}}{\hr{\epsilon}}\right|^\lambda$ will go to one, and the first term vanishes. \eqref{Eq: wicksbeforelimit} becomes $\langle\Psi^\dagger(x_2')\Psi^\dagger(x_1')\Psi(x_1)\Psi(x_2)\rangle_{\rm FF} = \langle\Psi^\dagger(x_2')\Psi(x_2)\rangle_{\rm FF} \langle\Psi^\dagger(x_1')\Psi(x_1)\rangle_{\rm FF} -\langle\Psi^\dagger(x_2')\Psi(x_1)\rangle_{\rm FF} \langle\Psi^\dagger(x_1')\Psi(x_2)\rangle_{\rm FF}$ as expected. 

\subsection{Calculating the limit $x_1'\to x_1$, $x_2'\to x_2$ and $x_2\to x_1$  }
Starting from
\begin{align}
    \rho(x_2',x_1';x_1,x_2) &= \langle\Psi^\dagger(x_2')\Psi^\dagger(x_1')\Psi(x_1)\Psi(x_2)\rangle\notag \\
    &=  \frac{|h_\epsilon|^{\gamma^2}|\hd{\epsilon}|^\lambda}{2\pi^2\hf{0}\hr{0}|\hf{\epsilon}\hr{\epsilon}|^{\gamma^2}} \cos{(k_F(x_1'+x_2'-x_1-x_2))} [\hd{0}|\hd{\epsilon}|^\delta+\hf{0}|\hf{\epsilon}|^\delta-\hr{0}|\hr{\epsilon}|^\delta]\notag\\
    &\qquad + \frac{|\hd{\epsilon}|^\lambda}{|\hf{\epsilon}|^\lambda}\langle\Psi^\dagger(x_2')\Psi(x_2)\rangle\langle\Psi^\dagger(x_1')\Psi(x_1)\rangle 
- \frac{|\hd{\epsilon}|^\lambda}{|\hr{\epsilon}|^\lambda}\langle\Psi^\dagger(x_2')\Psi(x_1)\rangle\langle\Psi^\dagger(x_1')\Psi(x_2)\rangle
\end{align}
This expression can be written more compactly as $\rho(x_2',x_1',x_1,x_2)=F(X-Y)+W$ where \begin{align*}
    F &= \frac{|h_\epsilon|^{\gamma^2}|\hd{\epsilon}|^\lambda}{2\pi^2|\hr{\epsilon}\hf{\epsilon}|^{\gamma^2}} \cos{(k_F(x_1'+x_2'-x_1-x_2))} \\
    X &= \frac{|\hd{\epsilon}|^\delta-|\hr{\epsilon}|^\delta}{\hf{0}}\\
    Y&= \frac{|\hd{\epsilon}|^\delta-|\hf{\epsilon}|^\delta}{\hr{0}}\\
    W &= \frac{|\hd{\epsilon}|^\lambda}{|\hf{\epsilon}|^\lambda}\langle\Psi^\dagger(x_2')\Psi(x_2)\rangle\langle\Psi^\dagger(x_1')\Psi(x_1)\rangle 
    - \frac{|\hd{\epsilon}|^\lambda}{|\hr{\epsilon}|^\lambda}\langle\Psi^\dagger(x_2')\Psi(x_1)\rangle\langle\Psi^\dagger(x_1')\Psi(x_2)\rangle
\end{align*}
We use the following notation to indicate the various limits, $F_1=\lim_{x_1'\rightarrow x_1}F, F_2 = \lim_{x_2'\rightarrow x_2}F_1$, and $F_3=\lim_{x_2\rightarrow x_1} F_2$. Similarly, we can define the limits for $X$, $Y$, and $W$. Then, we can compute the $x_1'\rightarrow x_1$ limit of $\rho_2(x_2',x_1';x_1,x_2)$ as $F_1(X_1-Y_1)+W_1$. $F_2(X_2-Y_2)+W_2$ gives the diagonal elements needed for the density-density correlation function. Finally, we can check that in the limit $r\rightarrow 0$ ($x_2\rightarrow x_1$), $F_3(X_3-Y_3)+W_3$ gives zero as expected. We start with the $F$ limits which can be done through direct substitution, 
\begin{align}
    F_1 &= \lim_{x_1'\rightarrow x_1} \frac{|h_\epsilon|^{\gamma^2}|\hd{\epsilon}|^\lambda}{2\pi^2|\hr{\epsilon}\hf{\epsilon}|^{\gamma^2}}\cos{(k_F(x_1'+x_2'-x_1-x_2))} \notag \\
    &= \frac{|h_\epsilon|^{\gamma^2}|h_\epsilon(x_2'-x_1,x_2-x_1)|^\lambda\cos{(k_F(x_2'-x_2))}}{2\pi^2|h_\epsilon(x_1-x_2,x_2'-x_1)|^{\gamma^2}|h_\epsilon(x_2'-x_2,0)|^{\gamma^2}}  \\
    F_2 &= \frac{|\frac{L}{\pi}\sin{(\frac{\pi}{L}((x_2-x_1)+i\epsilon))}|^{2\lambda-2\gamma^2}}{2\pi^2} \\
    F_3 &= \frac{|h_\epsilon|^{\lambda-\gamma^2}}{2\pi^2}
\end{align}
We perform the first two limits of $X$ through direct substitution as well, \begin{align}
    X_1 &= \frac{|h_\epsilon(x_2'-x_1,x_2-x_1)|^\delta-|h_\epsilon(x_2'-x_2,0)|^\delta}{h_0(x_1-x_2,x_2'-x_1)} \\
    X_2 &= -\left(\frac{L^2}{\pi^2}\right)^{\delta-1}\frac{|\sin{((\frac{\pi}{L}((x_2-x_1)+i\epsilon)))}|^{2\delta}-|\sin{(\frac{i\pi}{L}\epsilon)}|^{2\delta}}{\sin^2{(\frac{\pi}{L}(x_2-x_1))}}
\end{align}
The third limit, $r\rightarrow 0$, is slightly more complicated, so we rewrite the renormalized chord length as $|\sin{(x+i\epsilon)}|^2=\sinh^2{(\frac{\pi}{L}\epsilon)}+\sin^2{(\frac{\pi}{L}x)}$ with the corresponding derivative $\frac{d}{dx}|\sin{(x+i\epsilon)}|^2=\frac{\pi}{L}\sin{(\frac{2\pi}{L}x)}$. Using the relative coordinate, $r=x_2-x_1$, \begin{align}
    X_3 &=- \lim_{r\rightarrow 0}\left(\frac{L^2}{\pi^2}\right)^{\delta-1} \frac{|\sin{(\frac{\pi}{L}(r+i\epsilon))}|^{2\delta}-|\sin{(\frac{i\pi}{L}\epsilon)}|^{2\delta}}{\sin^2{(\frac{\pi}{L}r)}}\notag \\&=- \lim_{r\rightarrow 0}\left(\frac{L^2}{\pi^2}\right)^{\delta-1}\frac{(\sinh^2{(\frac{\pi}{L}\epsilon)}+\sin^2{(\frac{\pi}{L}r)})^\delta-|\sin{(\frac{i\pi}{L}\epsilon)}|^{2\delta}}{\sin^2{(\frac{\pi}{L}r)}} \notag \\
    &= - \lim_{r\rightarrow 0}\left(\frac{L^2}{\pi^2}\right)^{\delta-1}\frac{\delta(\sinh^2{(\frac{\pi}{L}\epsilon)}+\sin^2{(\frac{\pi}{L}r)})^{\delta-1}\frac{d}{dx}[+\sin^2{(\frac{\pi}{L}r)}]-0}{\frac{d}{dx}[\sin^2{(\frac{\pi}{L}r)}]} \notag \\
    &= -\delta\left(\frac{L^2}{\pi^2}\right)^{\delta-1}|\sin{(\frac{i\pi}{L}\epsilon)}|^{2\delta-2}.
\end{align}
We expect the same result for the $Y_3$ limit as in the limit $r\rightarrow 0$. We deploy L'Hôpital's rule twice to obtain the first two limits. Taking the derivative of the numerator of $Y$, we have
\begin{align}
\frac{\partial}{\partial x_1'} Y_{\rm num}
&=
\left(\frac{L^2}{\pi^2}\right)^{\delta-1}\frac{\partial}{\partial x_1'}
\Bigl[
   \bigl|\sin\!\bigl(\tfrac{\pi}{L}((x_2' - x_1') + i\epsilon)\bigr)\bigr|^{\delta}
   \,\bigl|\sin\!\bigl(\tfrac{\pi}{L}((x_2 - x_1) + i\epsilon)\bigr)\bigr|^{\delta}
\nonumber\\[-4pt]
&\qquad\qquad
   -\,
   \bigl|\sin\!\bigl(\tfrac{\pi}{L}((x_1' - x_2) + i\epsilon)\bigr)\bigr|^{\delta}
   \,\bigl|\sin\!\bigl(\tfrac{\pi}{L}((x_2' - x_1) + i\epsilon)\bigr)\bigr|^{\delta}
\Bigr]
\nonumber\\[6pt]
&
=-\left(\frac{L^2}{\pi^2}\right)^{\delta-1}\,\frac{\delta}{2}\,\frac{\pi}{L}\;
   \bigl|\sin\!\bigl(\tfrac{\pi}{L}((x_2 - x_1) + i\epsilon)\bigr)\bigr|^{\delta}
   \;\bigl|\sin\!\bigl(\tfrac{\pi}{L}((x_2' - x_1') + i\epsilon)\bigr)\bigr|^{\delta - 2}
   \sin\!\bigl(\tfrac{2\pi}{L}(x_2' - x_1')\bigr)
\nonumber\\[6pt]
&\qquad
-\left(\frac{L^2}{\pi^2}\right)^{\delta-1}\,\frac{\delta}{2}\,\frac{\pi}{L}\;
   \bigl|\sin\!\bigl(\tfrac{\pi}{L}((x_2' - x_1) + i\epsilon)\bigr)\bigr|^{\delta}
   \;\bigl|\sin\!\bigl(\tfrac{\pi}{L}((x_1' - x_2) + i\epsilon)\bigr)\bigr|^{\delta - 2}
   \sin\!\bigl(\tfrac{2\pi}{L}(x_1' - x_2)\bigr).
\end{align}
The derivative of the denominator is 
\begin{equation*}
    \frac{\partial}{\partial{x_1'}}\sin{(\frac{\pi}{L}(x_1'-x_1))}\sin{(\frac{\pi}{L}(x_2'-x_2))} = \frac{\pi}{L}\cos{(\frac{\pi}{L}(x_1'-x_1))}\sin{(\frac{\pi}{L}(x_2'-x_2))}.
\end{equation*}
The expression for $Y_1$ is therefore given by 
    
%\begin{align}
%    \left(\frac{L^2}{\pi^2}\right)^{1-\delta}Y_1&= -\frac{\delta}{2}\lim_{x_1'\rightarrow x_1}\frac{\frac{\pi}{L}|\sin{(\frac{\pi}{L}((x_2-x_1)+i\epsilon))}|^\delta(|\sin{(\frac{\pi}{L}((x_2'-x_1')+i\epsilon))}|^2)^{\delta/2-1}\sin{(\frac{2\pi}{L}(x_2'-x_1'))}}{\frac{\pi}{L}\cos{(\frac{\pi}{L}(x_1'-x_1))}\sin{(\frac{\pi}{L}(x_2'-x_2))}} \notag\\
%    &+\frac{\delta}{2}\lim_{x_1'\rightarrow x_1}\frac{\frac{\pi}{L}|\sin{(\frac{\pi}{L}((x_2'-x_1)+i\epsilon))}|^\delta(|\sin{(\frac{\pi}{L}((x_1'-x_2)+i\epsilon))}|^2)^{\delta/2-1}\sin{(\frac{2\pi}{L}(x_1'-x_2))}}{\frac{\pi}{L}\cos{(\frac{\pi}{L}(x_1'-x_1))}\sin{(\frac{\pi}{L}(x_2'-x_2))}} \notag\\
%    &= -\frac{\delta}{2}(|\sin{(\frac{\pi}{L}((x_2'-x_1)+i\epsilon))}||\sin{(\frac{\pi}{L}((x_2-x_1)+i\epsilon))}|)^{\delta-2}\notag\\ &\times \frac{|\sin{(\frac{\pi}{L}((x_2-x_1)+i\epsilon))}|^2\sin{(\frac{2\pi}{L}(x_2'-x_1))}+|\sin{(\frac{\pi}{L}((x_2'-x_1)+i\epsilon))}|^2\sin{(\frac{2\pi}{L}(x_1-x_2))}}{\sin{(\frac{\pi}{L}(x_2'-x_2))}} 
%    \label{eq:Y1expression}
%\end{align}
%Therefore, we arrive at 
\begin{align}
Y_1
&= -\left(\frac{L^2}{\pi^2}\right)^{\delta-1}\frac{\delta}{2}\,
   \frac{1}{\sin\tfrac{\pi}{L}(x_2'-x_2)} 
\Bigg[
     \left|\sin\!\frac{\pi}{L}\bigl[(x_2 - x_1)+i\epsilon\bigr]\right|^{\delta}
     \left|\sin\!\frac{\pi}{L}\bigl[(x_2' - x_1)+i\epsilon\bigr]\right|^{\delta-2}
     \sin\!\tfrac{2\pi}{L}(x_2' - x_1) \notag\\
&\qquad \qquad \qquad \qquad \qquad \qquad  \qquad\quad +
     \left|\sin\!\frac{\pi}{L}\bigl[(x_2' - x_1)+i\epsilon\bigr]\right|^{\delta}
     \left|\sin\!\frac{\pi}{L}\bigl[(x_1 - x_2)+i\epsilon\bigr]\right|^{\delta-2}
     \sin\!\tfrac{2\pi}{L}(x_1 - x_2)
   \Bigg].
\label{eq:Y1_limit}
\end{align}
Now for the limit $x_2' \rightarrow x_2$, a second application of L’Hôpital  yields
%\[
%Y_1'=
%|\sin \tfrac{\pi}{L}[(x_2-%x_1)+i\epsilon]|^{2}
%\sin \tfrac{2\pi}{L}(x_2'-x_1)
%+
%|\sin \tfrac{\pi}{L}[(x_2'-x_1)+i\epsilon]|^{2}
%\sin \tfrac{2\pi}{L}(x_1-x_2).
%\]
%Differentiating with respect to $x_2'$,
%\[
%\frac{\partial Y_1'}{\partial x_2'}
%=
%\frac{2\pi}{L}
%|\sin \tfrac{\pi}{L}[(x_2-%x_1)+i\epsilon]|^{2}
%\cos \tfrac{2\pi}{L}(x_2'-x_1)
%+\frac{\pi}{L}
%\sin \tfrac{2\pi}{L}(x_1-x_2)
%\sin \tfrac{2\pi}{L}(x_2'-x_1),
%\]
%while
%\[
%\frac{\partial}{\partial %x_2'}
%\sin \tfrac{\pi}{L}(x_2'-x_2)
%=
%\frac{\pi}{L}\cos \tfrac{\pi}{L}(x_2'-x_2).
%\]
%A second application of %L’Hôpital  yields
\begin{align}
Y_2
&=-\left(\frac{L^2}{\pi^2}\right)^{\delta-1}\frac{\delta}{2}\,
|\sin \tfrac{\pi}{L}[(x_2-x_1)+i\epsilon]|^{2\delta-4}
\Bigl[
2|\sin \tfrac{\pi}{L}[(x_2-x_1)+i\epsilon]|^{2}
\cos \tfrac{2\pi}{L}(x_2-x_1) -\sin^{2}\tfrac{2\pi}{L}(x_2-x_1)
\Bigr].
\end{align}
For the limit $x_2 \rightarrow x_1$,
we get
\[
Y_3
=\lim_{x_2\to x_1} Y_2
=-\left(\frac{L^2}{\pi^2}\right)^{\delta-1}\delta\bigl|\sin \tfrac{i\pi}{L}\epsilon\bigr|^{2\delta-2},
\]
in agreement with the independent result for $X_3$ as expected.\\

Finally we evaluate the $W$ limits.  First we write,
\begin{align}
W &=
\frac{h_{0}^{\lambda}}{h_{+}^{\lambda}}\,
\langle\Psi^{\dagger}(x_2')\Psi(x_2)\rangle
\langle\Psi^{\dagger}(x_1')\Psi(x_1)\rangle
-\frac{h_{0}^{\lambda}}{h_{-}^{\lambda}}\,
\langle\Psi^{\dagger}(x_2')\Psi(x_1)\rangle
\langle\Psi^{\dagger}(x_1')\Psi(x_2)\rangle,
\end{align}
where the one-body reduced density matrix is
\begin{align}
\langle\Psi^{\dagger}(x_2)\Psi(x_1)\rangle
&=\rho_{1}^{0}(x_1,x_2)\,
  \frac{\left|\sin\!\bigl(\tfrac{i\pi}{L}\epsilon\bigr)\right|^{\gamma^{2}}}
       {\left|\sin\!\bigl(\tfrac{\pi}{L}[(x_2-x_1)+i\epsilon]\bigr)\right|^{\gamma^{2}}},\\
\rho_{1}^{0}(x_1,x_2)
&=\frac{\sin\!\bigl[k_{F}(x_2-x_1)\bigr]}
        {L\,\sin\!\bigl(\tfrac{\pi}{L}(x_2-x_1)\bigr)}.
\end{align}
\noindent
We can insert these to obtain
\begin{equation}
\begin{aligned}
W &= 
\bigl|\sin\!\bigl(\tfrac{\pi}{L}[(x_2' - x_1') + i\epsilon]\bigr)\bigr|^\lambda
\;\bigl|\sin\!\bigl(\tfrac{\pi}{L}[(x_2 - x_1) + i\epsilon]\bigr)\bigr|^\lambda
\;\bigl|\sin\!\bigl(\tfrac{i\pi}{L}\epsilon\bigr)\bigr|^{2\gamma^2}
\\[6pt]
&\quad\times
\Biggl\{
\frac{
  \sin[k_F(x_2' - x_2)]\,
  \sin[k_F(x_1' - x_1)]\,
  \bigl|\sin\!\bigl(\tfrac{\pi}{L}[(x_2' - x_2) + i\epsilon]\bigr)\bigr|^{-\gamma^2}
}{
  L^2\,
  \sin(\tfrac{\pi}{L}(x_2' - x_2))\,
  \sin(\tfrac{\pi}{L}(x_1' - x_1))\,
  \bigl|\sin\!\bigl(\tfrac{\pi}{L}[(x_1' - x_2) + i\epsilon]\bigr)\bigr|^\lambda\,
  \bigl|\sin\!\bigl(\tfrac{\pi}{L}[(x_2' - x_1) + i\epsilon]\bigr)\bigr|^\lambda
}
\\[4pt]
&\qquad\quad\times\,
\frac{1}{
  \bigl|\sin\!\bigl(\tfrac{\pi}{L}[(x_1' - x_1) + i\epsilon]\bigr)\bigr|^{\gamma^2}
}
\\[8pt]
&\qquad
-\,\frac{
  \sin[k_F(x_2' - x_1)]\,
  \sin[k_F(x_1' - x_2)]\,
  \bigl|\sin\!\bigl(\tfrac{\pi}{L}[(x_2' - x_1) + i\epsilon]\bigr)\bigr|^{-\gamma^2}
}{
  L^2\,
  \sin(\tfrac{\pi}{L}(x_2' - x_1))\,
  \sin(\tfrac{\pi}{L}(x_1' - x_2))\,
  \bigl|\sin\!\bigl(\tfrac{\pi}{L}[(x_2' - x_2) + i\epsilon]\bigr)\bigr|^\lambda\,
  \bigl|\sin\!\bigl(\tfrac{\pi}{L}[(x_1' - x_1) + i\epsilon]\bigr)\bigr|^\lambda
}
\\[4pt]
&\qquad\quad\times\,
\frac{1}{
  \bigl|\sin\!\bigl(\tfrac{\pi}{L}[(x_1' - x_2) + i\epsilon]\bigr)\bigr|^{\gamma^2}
}
\Biggr\}.
\end{aligned}
\end{equation}
Taking the first limit,
\begin{equation}
\begin{aligned}
W_1 &= 
\left|\sin\!\tfrac{\pi}{L}[(x_2' - x_1) + i\epsilon]\right|^\lambda
\left|\sin\!\tfrac{\pi}{L}[(x_2 - x_1) + i\epsilon]\right|^\lambda
\left|\sin\!\tfrac{i\pi}{L}\epsilon\right|^{2\gamma^2} \\[4pt]
&\quad\times\Biggl\{
\frac{
    \tfrac{k_F}{L\pi} \,
    \sin[k_F(x_2' - x_2)] \,
    \left|\sin\tfrac{\pi}{L}[(x_2' - x_2) + i\epsilon]\right|^{-\gamma^2}
}{
    \sin(\tfrac{\pi}{L}(x_2' - x_2)) \,
    \left|\sin\tfrac{\pi}{L}[(x_1 - x_2) + i\epsilon]\right|^\lambda \,
    \left|\sin\tfrac{\pi}{L}[(x_2' - x_1) + i\epsilon]\right|^\lambda \,
    \left|\sin\tfrac{i\pi}{L}\epsilon\right|^{\gamma^2}
}
\\[8pt]
&\qquad
- \frac{
    \sin[k_F(x_2' - x_1)] \,
    \sin[k_F(x_1 - x_2)] \,
    \left|\sin\tfrac{\pi}{L}[(x_2' - x_1) + i\epsilon]\right|^{-\gamma^2}
}{
    L^2 \,
    \sin(\tfrac{\pi}{L}(x_2' - x_1)) \,
    \sin(\tfrac{\pi}{L}(x_1 - x_2)) \,
    \left|\sin\tfrac{\pi}{L}[(x_2' - x_2) + i\epsilon]\right|^\lambda \,
    \left|\sin\tfrac{i\pi}{L}\epsilon\right|^\lambda
}
\frac{1}{
  \left|\sin\tfrac{\pi}{L}[(x_1 - x_2) + i\epsilon]\right|^{\gamma^2}
}
\Biggr\}.
\end{aligned}
\end{equation}
Next, $x_2'\!\to x_2$ yields
\begin{align}
W_{2}&=n_{0}^{2}
-\left|\sin\tfrac{i\pi}{L}\epsilon\right|^{2\gamma^{2}-2\lambda}
 \frac{\sin^{2}[k_{F}(x_2-x_1)]}
      {L^{2}\sin^{2}(\tfrac{\pi}{L}(x_2-x_1))}
 \frac{1}{\left|\sin(\tfrac{\pi}{L}[(x_2-x_1)+i\epsilon])\right|^{2\gamma^{2}-2\lambda}}.
\end{align}
Letting $x_2\!\to x_1$ gives
\[
W_{3}
=n_{0}^{2}
-\left|\sin\tfrac{i\pi}{L}\epsilon\right|^{2\gamma^{2}-2\lambda}
 \left|\sin\tfrac{i\pi}{L}\epsilon\right|^{-2\gamma^{2}+2\lambda}
 \frac{N^{2}}{L^{2}}
=0.
\]
We can now assemble the final limits. Starting with \(x_1' \rightarrow x_1\):

\begin{align}
&\lim_{x_1'\to x_1}
\langle\Psi^\dagger(x_2')\Psi^\dagger(x_1')\Psi(x_1)\Psi(x_2)\rangle
= F_{1}(X_{1}-Y_{1}) + W_{1}
\nonumber\\[-4pt]
&\quad=
\frac{\cos\!\bigl[k_{F}(x_2' - x_2)\bigr]
  \left|\sin\!\frac{i\pi}{L}\epsilon\right|^{\gamma^{2}}
}{
  2L^{2}
  \left|\sin\!\tfrac{\pi}{L}[(x_1-x_2)+i\epsilon]\right|^{\delta}
  \left|\sin\!\tfrac{\pi}{L}[(x_2'-x_1)+i\epsilon]\right|^{\delta}
  \left|\sin\!\tfrac{\pi}{L}[(x_2'-x_2)+i\epsilon]\right|^{\gamma^{2}}
}
\nonumber\\
%&\quad\quad\times
%\cos\!\bigl[k_{F}(x_2' - x_2)\bigr]
%\nonumber\\[-4pt]
&\quad\quad\times\Biggl\{
\frac{
  \left|\sin\!\tfrac{\pi}{L}[(x_2'-x_1)+i\epsilon]\right|^{\delta}
  \left|\sin\!\tfrac{\pi}{L}[(x_2-x_1)+i\epsilon]\right|^{\delta}
  -
  \left|\sin\!\tfrac{\pi}{L}[(x_2'-x_2)+i\epsilon]\right|^{\delta}
  \left|\sin\!\tfrac{i\pi}{L}\epsilon\right|^{\delta}
}{
  \sin\!\bigl(\tfrac{\pi}{L}(x_1 - x_2)\bigr)
  \sin\!\bigl(\tfrac{\pi}{L}(x_2' - x_1)\bigr)
}
\nonumber\\[-4pt]
&\qquad\quad
+ \frac{\delta}{2}
\Bigl[
  \left|\sin\!\tfrac{\pi}{L}[(x_2'-x_1)+i\epsilon]\right|
  \left|\sin\!\tfrac{\pi}{L}[(x_2-x_1)+i\epsilon]\right|
\Bigr]^{\delta - 2}
\nonumber\\[-4pt]
&\qquad\quad\quad\times
\frac{
  \left|\sin\!\tfrac{\pi}{L}[(x_2-x_1)+i\epsilon]\right|^{2}
  \sin\!\bigl(\tfrac{2\pi}{L}(x_2' - x_1)\bigr)
  +
  \left|\sin\!\tfrac{\pi}{L}[(x_2'-x_1)+i\epsilon]\right|^{2}
  \sin\!\bigl(\tfrac{2\pi}{L}(x_1 - x_2)\bigr)
}{
  \sin\!\bigl(\tfrac{\pi}{L}(x_2' - x_2)\bigr)
}
\Biggr\}
\nonumber\\[-4pt]
&\quad\quad
+
\frac{
  N
}{
  L^2
}
% \cdot
\frac{
  \sin\!\bigl[k_{F}(x_2' - x_2)\bigr]
}{
   \sin\!\bigl(\tfrac{\pi}{L}(x_2' - x_2)\bigr)
}
% \nonumber\\[-4pt]
% &\quad\quad\quad\quad\cdot
\frac{
  \left|\sin\!\tfrac{i\pi}{L}\epsilon\right|^{\gamma^{2}}
}{
  \left|\sin\!\tfrac{\pi}{L}[(x_2'-x_2)+i\epsilon]\right|^{\gamma^{2}}
}
\nonumber\\[-4pt]
&\qquad\quad
-
\frac{
  \left|\sin\!\tfrac{\pi}{L}[(x_2'-x_2)+i\epsilon]\right|^{-\lambda}
  \left|\sin\!\tfrac{i\pi}{L}\epsilon\right|^{2\gamma^{2}-\lambda}
  \sin\!\bigl[k_{F}(x_2' - x_1)\bigr]
  \sin\!\bigl[k_{F}(x_1 - x_2)\bigr]
}{
  L^{2}
  \left|\sin\!\tfrac{\pi}{L}[(x_2'-x_1)+i\epsilon]\right|^{\delta}
  \left|\sin\!\tfrac{\pi}{L}[(x_1 - x_2)+i\epsilon]\right|^{\delta}\sin\!\bigl(\tfrac{\pi}{L}(x_2' - x_1)\bigr)
  \sin\!\bigl(\tfrac{\pi}{L}(x_1 - x_2)\bigr)
}
%\nonumber\\[-4pt]
%&\quad\quad\quad\quad\times
.
\label{eq:x1plimit}
\end{align}
Next, we can obtain an expression for the density-density correlation function.
\begin{equation}
\begin{split}
&\lim_{x_2'\to x_2}\lim_{x_1'\to x_1}
\bigl\langle \Psi^\dagger(x_2')\,\Psi^\dagger(x_1')\,\Psi(x_1)\,\Psi(x_2)\bigr\rangle
= F_{2}(X_{2}-Y_{2}) + W_{2} \\[6pt]
&\quad= \frac{\bigl|\sin\!\bigl(\tfrac{\pi}{L}[(x_2 - x_1) + i\epsilon]\bigr)\bigr|^{2\lambda - 2\gamma^{2}}}
               {2\,L^{2}}
\Biggl\lbrace
-\,\frac{
      \bigl|\sin\!\bigl(\tfrac{\pi}{L}[(x_2 - x_1) + i\epsilon]\bigr)\bigr|^{2\delta}
      \;-\;
      \bigl|\sin\!\bigl(\tfrac{i\pi}{L}\,\epsilon\bigr)\bigr|^{2\delta}
     }
     {\sin^{2}\!\bigl(\tfrac{\pi}{L}\,(x_2 - x_1)\bigr)}
\\[6pt]
&\qquad\quad
+\,\frac{\delta}{2}\,
   \bigl|\sin\!\bigl(\tfrac{\pi}{L}[(x_2 - x_1) + i\epsilon]\bigr)\bigr|^{2\delta - 4}
% \\[-2pt]
% &\qquad\qquad
% \times
   \Bigl[
      2\,\bigl|\sin\!\bigl(\tfrac{\pi}{L}[(x_2 - x_1) + i\epsilon]\bigr)\bigr|^{2}\,
      \cos\!\bigl(\tfrac{2\pi}{L}\,(x_2 - x_1)\bigr)
      \;-\;
      \sin^{2}\!\bigl(\tfrac{2\pi}{L}\,(x_2 - x_1)\bigr)
   \Bigr]
\Biggr\rbrace
\\[6pt]
&\quad\quad
+\,n_{0}^{2}
\;-\;
\frac{
      \bigl|\sin\!\bigl(\tfrac{i\pi}{L}\,\epsilon\bigr)\bigr|^{2\delta}\,
      \sin^{2}\!\bigl(k_{F}\,(x_2 - x_1)\bigr)
     }
     {
      L^{2}\,
      \sin^{2}\!\bigl(\tfrac{\pi}{L}\,(x_2 - x_1)\bigr)\,
      \bigl|\sin\!\bigl(\tfrac{\pi}{L}[(x_2 - x_1) + i\epsilon]\bigr)\bigr|^{2\delta}
     }.
\end{split}
\label{eq:densitydensitylimit}
\end{equation}
Finally, in limit $x_1'\to x_1$, $x_2'\to x_2$ and $x_2\to x_1$ we have
\begin{equation}
\begin{split}
&\lim_{x_2\to x_1}\lim_{x_2'\to x_2}\lim_{x_1'\to x_1}
\bigl\langle \Psi^\dagger(x_2')\,\Psi^\dagger(x_1')\,\Psi(x_1)\,\Psi(x_2)\bigr\rangle
=\lim_{x_2\to x_1} \left[ F_{2}(X_{2}-Y_{2}) + W_{2}\right] \\[6pt]
&\quad=F_{3}(X_{3}-Y_{3}) + W_{3}=0. 
\end{split}
\end{equation}

\twocolumngrid
\bibliography{bibliography}

\begin{thebibliography}{66}%
\makeatletter
\providecommand \@ifxundefined [1]{%
 \@ifx{#1\undefined}
}%
\providecommand \@ifnum [1]{%
 \ifnum #1\expandafter \@firstoftwo
 \else \expandafter \@secondoftwo
 \fi
}%
\providecommand \@ifx [1]{%
 \ifx #1\expandafter \@firstoftwo
 \else \expandafter \@secondoftwo
 \fi
}%
\providecommand \natexlab [1]{#1}%
\providecommand \enquote  [1]{``#1''}%
\providecommand \bibnamefont  [1]{#1}%
\providecommand \bibfnamefont [1]{#1}%
\providecommand \citenamefont [1]{#1}%
\providecommand \href@noop [0]{\@secondoftwo}%
\providecommand \href [0]{\begingroup \@sanitize@url \@href}%
\providecommand \@href[1]{\@@startlink{#1}\@@href}%
\providecommand \@@href[1]{\endgroup#1\@@endlink}%
\providecommand \@sanitize@url [0]{\catcode `\\12\catcode `\$12\catcode `\&12\catcode `\#12\catcode `\^12\catcode `\_12\catcode `\%12\relax}%
\providecommand \@@startlink[1]{}%
\providecommand \@@endlink[0]{}%
\providecommand \url  [0]{\begingroup\@sanitize@url \@url }%
\providecommand \@url [1]{\endgroup\@href {#1}{\urlprefix }}%
\providecommand \urlprefix  [0]{URL }%
\providecommand \Eprint [0]{\href }%
\providecommand \doibase [0]{https://doi.org/}%
\providecommand \selectlanguage [0]{\@gobble}%
\providecommand \bibinfo  [0]{\@secondoftwo}%
\providecommand \bibfield  [0]{\@secondoftwo}%
\providecommand \translation [1]{[#1]}%
\providecommand \BibitemOpen [0]{}%
\providecommand \bibitemStop [0]{}%
\providecommand \bibitemNoStop [0]{.\EOS\space}%
\providecommand \EOS [0]{\spacefactor3000\relax}%
\providecommand \BibitemShut  [1]{\csname bibitem#1\endcsname}%
\let\auto@bib@innerbib\@empty
%</preamble>
\bibitem [{\citenamefont {Coleman}(1963)}]{coleman1963structure}%
  \BibitemOpen
  \bibfield  {author} {\bibinfo {author} {\bibfnamefont {A.~J.}\ \bibnamefont {Coleman}},\ }\bibfield  {title} {\bibinfo {title} {{S}tructure of {F}ermion {D}ensity {M}atrices},\ }\href {https://doi.org/10.1103/revmodphys.35.668} {\bibfield  {journal} {\bibinfo  {journal} {Rev. Mod. Phys.}\ }\textbf {\bibinfo {volume} {35}},\ \bibinfo {pages} {668} (\bibinfo {year} {1963})}\BibitemShut {NoStop}%
\bibitem [{\citenamefont {Altman}\ \emph {et~al.}(2004)\citenamefont {Altman}, \citenamefont {Demler},\ and\ \citenamefont {Lukin}}]{Altman2004}%
  \BibitemOpen
  \bibfield  {author} {\bibinfo {author} {\bibfnamefont {E.}~\bibnamefont {Altman}}, \bibinfo {author} {\bibfnamefont {E.}~\bibnamefont {Demler}},\ and\ \bibinfo {author} {\bibfnamefont {M.~D.}\ \bibnamefont {Lukin}},\ }\bibfield  {title} {\bibinfo {title} {Probing many-body states of ultracold atoms via noise correlations},\ }\href {https://doi.org/10.1103/PhysRevA.70.013603} {\bibfield  {journal} {\bibinfo  {journal} {Phys. Rev. A}\ }\textbf {\bibinfo {volume} {70}},\ \bibinfo {pages} {013603} (\bibinfo {year} {2004})}\BibitemShut {NoStop}%
\bibitem [{\citenamefont {Boll}\ \emph {et~al.}(2016)\citenamefont {Boll}, \citenamefont {Hilker}, \citenamefont {Salomon}, \citenamefont {Omran}, \citenamefont {Nespolo}, \citenamefont {Pollet}, \citenamefont {Bloch},\ and\ \citenamefont {Gross}}]{Boll2016}%
  \BibitemOpen
  \bibfield  {author} {\bibinfo {author} {\bibfnamefont {M.}~\bibnamefont {Boll}}, \bibinfo {author} {\bibfnamefont {T.~A.}\ \bibnamefont {Hilker}}, \bibinfo {author} {\bibfnamefont {G.}~\bibnamefont {Salomon}}, \bibinfo {author} {\bibfnamefont {A.}~\bibnamefont {Omran}}, \bibinfo {author} {\bibfnamefont {J.}~\bibnamefont {Nespolo}}, \bibinfo {author} {\bibfnamefont {L.}~\bibnamefont {Pollet}}, \bibinfo {author} {\bibfnamefont {I.}~\bibnamefont {Bloch}},\ and\ \bibinfo {author} {\bibfnamefont {C.}~\bibnamefont {Gross}},\ }\bibfield  {title} {\bibinfo {title} {Spin- and density-resolved microscopy of antiferromagnetic correlations in fermi-hubbard chains},\ }\href {https://doi.org/10.1126/science.aag1635} {\bibfield  {journal} {\bibinfo  {journal} {Science}\ }\textbf {\bibinfo {volume} {353}},\ \bibinfo {pages} {1257} (\bibinfo {year} {2016})}\BibitemShut {NoStop}%
\bibitem [{\citenamefont {Yang}\ \emph {et~al.}(2018)\citenamefont {Yang}, \citenamefont {Gri\ifmmode~\check{s}\else \v{s}\fi{}ins}, \citenamefont {Chang}, \citenamefont {Zhao}, \citenamefont {Shih}, \citenamefont {Giamarchi},\ and\ \citenamefont {Hulet}}]{Yang2018}%
  \BibitemOpen
  \bibfield  {author} {\bibinfo {author} {\bibfnamefont {T.~L.}\ \bibnamefont {Yang}}, \bibinfo {author} {\bibfnamefont {P.}~\bibnamefont {Gri\ifmmode~\check{s}\else \v{s}\fi{}ins}}, \bibinfo {author} {\bibfnamefont {Y.~T.}\ \bibnamefont {Chang}}, \bibinfo {author} {\bibfnamefont {Z.~H.}\ \bibnamefont {Zhao}}, \bibinfo {author} {\bibfnamefont {C.~Y.}\ \bibnamefont {Shih}}, \bibinfo {author} {\bibfnamefont {T.}~\bibnamefont {Giamarchi}},\ and\ \bibinfo {author} {\bibfnamefont {R.~G.}\ \bibnamefont {Hulet}},\ }\bibfield  {title} {\bibinfo {title} {Measurement of the dynamical structure factor of a 1d interacting fermi gas},\ }\href {https://doi.org/10.1103/PhysRevLett.121.103001} {\bibfield  {journal} {\bibinfo  {journal} {Phys. Rev. Lett.}\ }\textbf {\bibinfo {volume} {121}},\ \bibinfo {pages} {103001} (\bibinfo {year} {2018})}\BibitemShut {NoStop}%
\bibitem [{\citenamefont {Kuhnle}\ \emph {et~al.}(2010)\citenamefont {Kuhnle}, \citenamefont {Hu}, \citenamefont {Liu}, \citenamefont {Dyke}, \citenamefont {Mark}, \citenamefont {Drummond}, \citenamefont {Hannaford},\ and\ \citenamefont {Vale}}]{Kuhnle2010}%
  \BibitemOpen
  \bibfield  {author} {\bibinfo {author} {\bibfnamefont {E.~D.}\ \bibnamefont {Kuhnle}}, \bibinfo {author} {\bibfnamefont {H.}~\bibnamefont {Hu}}, \bibinfo {author} {\bibfnamefont {X.-J.}\ \bibnamefont {Liu}}, \bibinfo {author} {\bibfnamefont {P.}~\bibnamefont {Dyke}}, \bibinfo {author} {\bibfnamefont {M.}~\bibnamefont {Mark}}, \bibinfo {author} {\bibfnamefont {P.~D.}\ \bibnamefont {Drummond}}, \bibinfo {author} {\bibfnamefont {P.}~\bibnamefont {Hannaford}},\ and\ \bibinfo {author} {\bibfnamefont {C.~J.}\ \bibnamefont {Vale}},\ }\bibfield  {title} {\bibinfo {title} {Universal behavior of pair correlations in a strongly interacting fermi gas},\ }\href {https://doi.org/10.1103/PhysRevLett.105.070402} {\bibfield  {journal} {\bibinfo  {journal} {Phys. Rev. Lett.}\ }\textbf {\bibinfo {volume} {105}},\ \bibinfo {pages} {070402} (\bibinfo {year} {2010})}\BibitemShut {NoStop}%
\bibitem [{\citenamefont {Mazziotti}(2012{\natexlab{a}})}]{Mazziotti2012}%
  \BibitemOpen
  \bibfield  {author} {\bibinfo {author} {\bibfnamefont {D.~A.}\ \bibnamefont {Mazziotti}},\ }\bibfield  {title} {\bibinfo {title} {Structure of fermionic density matrices: Complete $n$-representability conditions},\ }\href {https://doi.org/10.1103/PhysRevLett.108.263002} {\bibfield  {journal} {\bibinfo  {journal} {Phys. Rev. Lett.}\ }\textbf {\bibinfo {volume} {108}},\ \bibinfo {pages} {263002} (\bibinfo {year} {2012}{\natexlab{a}})}\BibitemShut {NoStop}%
\bibitem [{\citenamefont {Mazziotti}(2012{\natexlab{b}})}]{Mazzioti2012pra}%
  \BibitemOpen
  \bibfield  {author} {\bibinfo {author} {\bibfnamefont {D.~A.}\ \bibnamefont {Mazziotti}},\ }\bibfield  {title} {\bibinfo {title} {Significant conditions for the two-electron reduced density matrix from the constructive solution of $n$ representability},\ }\href {https://doi.org/10.1103/PhysRevA.85.062507} {\bibfield  {journal} {\bibinfo  {journal} {Phys. Rev. A}\ }\textbf {\bibinfo {volume} {85}},\ \bibinfo {pages} {062507} (\bibinfo {year} {2012}{\natexlab{b}})}\BibitemShut {NoStop}%
\bibitem [{\citenamefont {Mazziotti}(1998)}]{Mazziotti1998}%
  \BibitemOpen
  \bibfield  {author} {\bibinfo {author} {\bibfnamefont {D.~A.}\ \bibnamefont {Mazziotti}},\ }\bibfield  {title} {\bibinfo {title} {Contracted schr\"odinger equation: Determining quantum energies and two-particle density matrices without wave functions},\ }\href {https://doi.org/10.1103/PhysRevA.57.4219} {\bibfield  {journal} {\bibinfo  {journal} {Phys. Rev. A}\ }\textbf {\bibinfo {volume} {57}},\ \bibinfo {pages} {4219} (\bibinfo {year} {1998})}\BibitemShut {NoStop}%
\bibitem [{\citenamefont {Liebert}\ \emph {et~al.}(2025)\citenamefont {Liebert}, \citenamefont {Schouten}, \citenamefont {Avdic}, \citenamefont {Schilling},\ and\ \citenamefont {Mazziotti}}]{Liebert:2025nn}%
  \BibitemOpen
  \bibfield  {author} {\bibinfo {author} {\bibfnamefont {J.}~\bibnamefont {Liebert}}, \bibinfo {author} {\bibfnamefont {A.~O.}\ \bibnamefont {Schouten}}, \bibinfo {author} {\bibfnamefont {I.}~\bibnamefont {Avdic}}, \bibinfo {author} {\bibfnamefont {C.}~\bibnamefont {Schilling}},\ and\ \bibinfo {author} {\bibfnamefont {D.~A.}\ \bibnamefont {Mazziotti}},\ }\bibfield  {title} {\bibinfo {title} {{R}efining ensemble {N}-representability of one-body density matrices from partial information},\ }\href {https://doi.org/10.1088/1367-2630/ae228e} {\bibfield  {journal} {\bibinfo  {journal} {New J. Phys.}\ }\textbf {\bibinfo {volume} {27}},\ \bibinfo {pages} {124511} (\bibinfo {year} {2025})}\BibitemShut {NoStop}%
\bibitem [{\citenamefont {Bulgac}\ \emph {et~al.}(2023)\citenamefont {Bulgac}, \citenamefont {Kafker},\ and\ \citenamefont {Abdurrahman}}]{Bulgac:2022}%
  \BibitemOpen
  \bibfield  {author} {\bibinfo {author} {\bibfnamefont {A.}~\bibnamefont {Bulgac}}, \bibinfo {author} {\bibfnamefont {M.}~\bibnamefont {Kafker}},\ and\ \bibinfo {author} {\bibfnamefont {I.}~\bibnamefont {Abdurrahman}},\ }\bibfield  {title} {\bibinfo {title} {Measures of complexity and entanglement in many-fermion systems},\ }\href {https://doi.org/10.1103/PhysRevC.107.044318} {\bibfield  {journal} {\bibinfo  {journal} {Phys. Rev. C}\ }\textbf {\bibinfo {volume} {107}},\ \bibinfo {pages} {044318} (\bibinfo {year} {2023})}\BibitemShut {NoStop}%
\bibitem [{\citenamefont {Naldesi}\ \emph {et~al.}(2023)\citenamefont {Naldesi}, \citenamefont {Elben}, \citenamefont {Minguzzi}, \citenamefont {Cl\'ement}, \citenamefont {Zoller},\ and\ \citenamefont {Vermersch}}]{Naldesi2022}%
  \BibitemOpen
  \bibfield  {author} {\bibinfo {author} {\bibfnamefont {P.}~\bibnamefont {Naldesi}}, \bibinfo {author} {\bibfnamefont {A.}~\bibnamefont {Elben}}, \bibinfo {author} {\bibfnamefont {A.}~\bibnamefont {Minguzzi}}, \bibinfo {author} {\bibfnamefont {D.}~\bibnamefont {Cl\'ement}}, \bibinfo {author} {\bibfnamefont {P.}~\bibnamefont {Zoller}},\ and\ \bibinfo {author} {\bibfnamefont {B.}~\bibnamefont {Vermersch}},\ }\bibfield  {title} {\bibinfo {title} {Fermionic correlation functions from randomized measurements in programmable atomic quantum devices},\ }\href {https://doi.org/10.1103/PhysRevLett.131.060601} {\bibfield  {journal} {\bibinfo  {journal} {Phys. Rev. Lett.}\ }\textbf {\bibinfo {volume} {131}},\ \bibinfo {pages} {060601} (\bibinfo {year} {2023})}\BibitemShut {NoStop}%
\bibitem [{\citenamefont {Polkovnikov}\ \emph {et~al.}(2006)\citenamefont {Polkovnikov}, \citenamefont {Altman},\ and\ \citenamefont {Demler}}]{Polkovnikov:2006js}%
  \BibitemOpen
  \bibfield  {author} {\bibinfo {author} {\bibfnamefont {A.}~\bibnamefont {Polkovnikov}}, \bibinfo {author} {\bibfnamefont {E.}~\bibnamefont {Altman}},\ and\ \bibinfo {author} {\bibfnamefont {E.}~\bibnamefont {Demler}},\ }\bibfield  {title} {\bibinfo {title} {{I}nterference between independent fluctuating condensates},\ }\href {https://doi.org/10.1073/pnas.0510276103} {\bibfield  {journal} {\bibinfo  {journal} {Proc. Natl. Acad. Sci. U.S.A.}\ }\textbf {\bibinfo {volume} {103}},\ \bibinfo {pages} {6125} (\bibinfo {year} {2006})}\BibitemShut {NoStop}%
\bibitem [{\citenamefont {Gritsev}\ \emph {et~al.}(2006)\citenamefont {Gritsev}, \citenamefont {Altman}, \citenamefont {Demler},\ and\ \citenamefont {Polkovnikov}}]{Gritsev:2006tn}%
  \BibitemOpen
  \bibfield  {author} {\bibinfo {author} {\bibfnamefont {V.}~\bibnamefont {Gritsev}}, \bibinfo {author} {\bibfnamefont {E.}~\bibnamefont {Altman}}, \bibinfo {author} {\bibfnamefont {E.}~\bibnamefont {Demler}},\ and\ \bibinfo {author} {\bibfnamefont {A.}~\bibnamefont {Polkovnikov}},\ }\bibfield  {title} {\bibinfo {title} {{F}ull quantum distribution of contrast in interference experiments between interacting one-dimensional {B}ose liquids},\ }\href {https://doi.org/10.1038/nphys410} {\bibfield  {journal} {\bibinfo  {journal} {Nat. Phys.}\ }\textbf {\bibinfo {volume} {2}},\ \bibinfo {pages} {705} (\bibinfo {year} {2006})}\BibitemShut {NoStop}%
\bibitem [{\citenamefont {Moitra}\ and\ \citenamefont {Sensarma}(2023)}]{Moitra2023}%
  \BibitemOpen
  \bibfield  {author} {\bibinfo {author} {\bibfnamefont {S.}~\bibnamefont {Moitra}}\ and\ \bibinfo {author} {\bibfnamefont {R.}~\bibnamefont {Sensarma}},\ }\bibfield  {title} {\bibinfo {title} {Building entanglement entropy out of correlation functions for interacting fermions},\ }\href {https://doi.org/10.1103/PhysRevB.108.174309} {\bibfield  {journal} {\bibinfo  {journal} {Phys. Rev. B}\ }\textbf {\bibinfo {volume} {108}},\ \bibinfo {pages} {174309} (\bibinfo {year} {2023})}\BibitemShut {NoStop}%
\bibitem [{\citenamefont {Rammelm\"{u}ller}\ \emph {et~al.}(2017)\citenamefont {Rammelm\"{u}ller}, \citenamefont {Porter}, \citenamefont {Braun},\ and\ \citenamefont {Drut}}]{Rammelmueller:2017al}%
  \BibitemOpen
  \bibfield  {author} {\bibinfo {author} {\bibfnamefont {L.}~\bibnamefont {Rammelm\"{u}ller}}, \bibinfo {author} {\bibfnamefont {W.~J.}\ \bibnamefont {Porter}}, \bibinfo {author} {\bibfnamefont {J.}~\bibnamefont {Braun}},\ and\ \bibinfo {author} {\bibfnamefont {J.~E.}\ \bibnamefont {Drut}},\ }\bibfield  {title} {\bibinfo {title} {Evolution from few- to many-body physics in one-dimensional fermi systems: One- and two-body density matrices and particle-partition entanglement},\ }\href {https://doi.org/10.1103/physreva.96.033635} {\bibfield  {journal} {\bibinfo  {journal} {Phys. Rev. A}\ }\textbf {\bibinfo {volume} {96}},\ \bibinfo {pages} {033635} (\bibinfo {year} {2017})}\BibitemShut {NoStop}%
\bibitem [{\citenamefont {Ferreira}\ \emph {et~al.}(2022)\citenamefont {Ferreira}, \citenamefont {Maciel}, \citenamefont {Vianna},\ and\ \citenamefont {Iemini}}]{Ferreira:2022qq}%
  \BibitemOpen
  \bibfield  {author} {\bibinfo {author} {\bibfnamefont {D.~L.~B.}\ \bibnamefont {Ferreira}}, \bibinfo {author} {\bibfnamefont {T.~O.}\ \bibnamefont {Maciel}}, \bibinfo {author} {\bibfnamefont {R.~O.}\ \bibnamefont {Vianna}},\ and\ \bibinfo {author} {\bibfnamefont {F.}~\bibnamefont {Iemini}},\ }\bibfield  {title} {\bibinfo {title} {{Q}uantum correlations, entanglement spectrum, and coherence of the two-particle reduced density matrix in the extended {H}ubbard model},\ }\href {https://doi.org/10.1103/physrevb.105.115145} {\bibfield  {journal} {\bibinfo  {journal} {Phys. Rev. B}\ }\textbf {\bibinfo {volume} {105}},\ \bibinfo {pages} {115145} (\bibinfo {year} {2022})}\BibitemShut {NoStop}%
\bibitem [{\citenamefont {Zozulya}\ \emph {et~al.}(2007)\citenamefont {Zozulya}, \citenamefont {Haque}, \citenamefont {Schoutens},\ and\ \citenamefont {Rezayi}}]{Zozulya:2007fe}%
  \BibitemOpen
  \bibfield  {author} {\bibinfo {author} {\bibfnamefont {O.~S.}\ \bibnamefont {Zozulya}}, \bibinfo {author} {\bibfnamefont {M.}~\bibnamefont {Haque}}, \bibinfo {author} {\bibfnamefont {K.}~\bibnamefont {Schoutens}},\ and\ \bibinfo {author} {\bibfnamefont {E.~H.}\ \bibnamefont {Rezayi}},\ }\bibfield  {title} {\bibinfo {title} {{B}ipartite entanglement entropy in fractional quantum {H}all states},\ }\href {https://doi.org/10.1103/physrevb.76.125310} {\bibfield  {journal} {\bibinfo  {journal} {Phys. Rev. B}\ }\textbf {\bibinfo {volume} {76}},\ \bibinfo {pages} {125310} (\bibinfo {year} {2007})}\BibitemShut {NoStop}%
\bibitem [{\citenamefont {Zozulya}\ \emph {et~al.}(2008)\citenamefont {Zozulya}, \citenamefont {Haque},\ and\ \citenamefont {Schoutens}}]{Zozulya:2008bg}%
  \BibitemOpen
  \bibfield  {author} {\bibinfo {author} {\bibfnamefont {O.~S.}\ \bibnamefont {Zozulya}}, \bibinfo {author} {\bibfnamefont {M.}~\bibnamefont {Haque}},\ and\ \bibinfo {author} {\bibfnamefont {K.}~\bibnamefont {Schoutens}},\ }\bibfield  {title} {\bibinfo {title} {{P}article partitioning entanglement in itinerant many-particle systems},\ }\href {https://doi.org/10.1103/physreva.78.042326} {\bibfield  {journal} {\bibinfo  {journal} {Phys. Rev. A}\ }\textbf {\bibinfo {volume} {78}},\ \bibinfo {pages} {042326} (\bibinfo {year} {2008})}\BibitemShut {NoStop}%
\bibitem [{\citenamefont {Haque}\ \emph {et~al.}(2007)\citenamefont {Haque}, \citenamefont {Zozulya},\ and\ \citenamefont {Schoutens}}]{Haque:2007mu}%
  \BibitemOpen
  \bibfield  {author} {\bibinfo {author} {\bibfnamefont {M.}~\bibnamefont {Haque}}, \bibinfo {author} {\bibfnamefont {O.}~\bibnamefont {Zozulya}},\ and\ \bibinfo {author} {\bibfnamefont {K.}~\bibnamefont {Schoutens}},\ }\bibfield  {title} {\bibinfo {title} {{E}ntanglement {E}ntropy in {F}ermionic {L}aughlin {S}tates},\ }\href {https://doi.org/10.1103/physrevlett.98.060401} {\bibfield  {journal} {\bibinfo  {journal} {Phys. Rev. Lett.}\ }\textbf {\bibinfo {volume} {98}},\ \bibinfo {pages} {060401} (\bibinfo {year} {2007})}\BibitemShut {NoStop}%
\bibitem [{\citenamefont {Haque}\ \emph {et~al.}(2009)\citenamefont {Haque}, \citenamefont {Zozulya},\ and\ \citenamefont {Schoutens}}]{Haque:2009zi}%
  \BibitemOpen
  \bibfield  {author} {\bibinfo {author} {\bibfnamefont {M.}~\bibnamefont {Haque}}, \bibinfo {author} {\bibfnamefont {O.~S.}\ \bibnamefont {Zozulya}},\ and\ \bibinfo {author} {\bibfnamefont {K.}~\bibnamefont {Schoutens}},\ }\bibfield  {title} {\bibinfo {title} {{E}ntanglement between particle partitions in itinerant many-particle states},\ }\href {https://doi.org/10.1088/1751-8113/42/50/504012} {\bibfield  {journal} {\bibinfo  {journal} {J. Phys. A: Math. Theor.}\ }\textbf {\bibinfo {volume} {42}},\ \bibinfo {pages} {504012} (\bibinfo {year} {2009})}\BibitemShut {NoStop}%
\bibitem [{\citenamefont {Barghathi}\ \emph {et~al.}(2017)\citenamefont {Barghathi}, \citenamefont {Casiano-Diaz},\ and\ \citenamefont {{Del Maestro}}}]{Barghathi:2017ab}%
  \BibitemOpen
  \bibfield  {author} {\bibinfo {author} {\bibfnamefont {H.}~\bibnamefont {Barghathi}}, \bibinfo {author} {\bibfnamefont {E.}~\bibnamefont {Casiano-Diaz}},\ and\ \bibinfo {author} {\bibfnamefont {A.}~\bibnamefont {{Del Maestro}}},\ }\bibfield  {title} {\bibinfo {title} {{P}article partition entanglement of one dimensional spinless fermions},\ }\href {https://doi.org/10.1088/1742-5468/aa819a} {\bibfield  {journal} {\bibinfo  {journal} {J. Stat. Mech.: Theory Exp.}\ }\textbf {\bibinfo {volume} {2017}}\bibinfo  {number} { (8)},\ \bibinfo {pages} {083108}}\BibitemShut {NoStop}%
\bibitem [{\citenamefont {Radhakrishnan}\ \emph {et~al.}(2023)\citenamefont {Radhakrishnan}, \citenamefont {Thamm}, \citenamefont {Barghathi}, \citenamefont {Rosenow},\ and\ \citenamefont {{Del Maestro}}}]{Radhakrishnan}%
  \BibitemOpen
\bibfield  {number} {  }\bibfield  {author} {\bibinfo {author} {\bibfnamefont {H.}~\bibnamefont {Radhakrishnan}}, \bibinfo {author} {\bibfnamefont {M.}~\bibnamefont {Thamm}}, \bibinfo {author} {\bibfnamefont {H.}~\bibnamefont {Barghathi}}, \bibinfo {author} {\bibfnamefont {B.}~\bibnamefont {Rosenow}},\ and\ \bibinfo {author} {\bibfnamefont {A.}~\bibnamefont {{Del Maestro}}},\ }\bibfield  {title} {\bibinfo {title} {A scaling function for the particle entanglement entropy of fermions},\ }\href {https://doi.org/10.1088/1742-5468/ace430} {\bibfield  {journal} {\bibinfo  {journal} {J. Stat. Mech.: Theory Exp.}\ }\textbf {\bibinfo {volume} {2023}}\bibinfo  {number} { (8)},\ \bibinfo {pages} {083101}}\BibitemShut {NoStop}%
\bibitem [{\citenamefont {Herdman}\ \emph {et~al.}(2014)\citenamefont {Herdman}, \citenamefont {Roy}, \citenamefont {Melko},\ and\ \citenamefont {{Del Maestro}}}]{Herdman:2014jy}%
  \BibitemOpen
\bibfield  {number} {  }\bibfield  {author} {\bibinfo {author} {\bibfnamefont {C.~M.}\ \bibnamefont {Herdman}}, \bibinfo {author} {\bibfnamefont {P.~N.}\ \bibnamefont {Roy}}, \bibinfo {author} {\bibfnamefont {R.~G.}\ \bibnamefont {Melko}},\ and\ \bibinfo {author} {\bibfnamefont {A.}~\bibnamefont {{Del Maestro}}},\ }\bibfield  {title} {\bibinfo {title} {{P}article entanglement in continuum many-body systems via quantum {M}onte {C}arlo},\ }\href {https://doi.org/10.1103/physrevb.89.140501} {\bibfield  {journal} {\bibinfo  {journal} {Phys. Rev. B}\ }\textbf {\bibinfo {volume} {89}},\ \bibinfo {pages} {140501(R)} (\bibinfo {year} {2014})}\BibitemShut {NoStop}%
\bibitem [{\citenamefont {Herdman}\ and\ \citenamefont {{Del Maestro}}(2015)}]{Herdman:2015xa}%
  \BibitemOpen
  \bibfield  {author} {\bibinfo {author} {\bibfnamefont {C.~M.}\ \bibnamefont {Herdman}}\ and\ \bibinfo {author} {\bibfnamefont {A.}~\bibnamefont {{Del Maestro}}},\ }\bibfield  {title} {\bibinfo {title} {{P}article partition entanglement of bosonic {L}uttinger liquids},\ }\href {https://doi.org/10.1103/physrevb.91.184507} {\bibfield  {journal} {\bibinfo  {journal} {Phys. Rev. B}\ }\textbf {\bibinfo {volume} {91}},\ \bibinfo {pages} {184507} (\bibinfo {year} {2015})}\BibitemShut {NoStop}%
\bibitem [{\citenamefont {Liu}\ \emph {et~al.}(2025)\citenamefont {Liu}, \citenamefont {Xu}, \citenamefont {Liu},\ and\ \citenamefont {Wang}}]{Liu:2025rixs}%
  \BibitemOpen
  \bibfield  {author} {\bibinfo {author} {\bibfnamefont {T.}~\bibnamefont {Liu}}, \bibinfo {author} {\bibfnamefont {L.}~\bibnamefont {Xu}}, \bibinfo {author} {\bibfnamefont {J.}~\bibnamefont {Liu}},\ and\ \bibinfo {author} {\bibfnamefont {Y.}~\bibnamefont {Wang}},\ }\bibfield  {title} {\bibinfo {title} {Entanglement witness for indistinguishable electrons using solid-state spectroscopy},\ }\href {https://doi.org/10.1103/PhysRevX.15.011056} {\bibfield  {journal} {\bibinfo  {journal} {Phys. Rev. X}\ }\textbf {\bibinfo {volume} {15}},\ \bibinfo {pages} {011056} (\bibinfo {year} {2025})}\BibitemShut {NoStop}%
\bibitem [{\citenamefont {Tsvelik}(2003)}]{Tsvelik_2003}%
  \BibitemOpen
  \bibfield  {author} {\bibinfo {author} {\bibfnamefont {A.~M.}\ \bibnamefont {Tsvelik}},\ }\href@noop {} {\emph {\bibinfo {title} {Quantum Field Theory in Condensed Matter Physics}}},\ \bibinfo {edition} {2nd}\ ed.\ (\bibinfo  {publisher} {Cambridge University Press},\ \bibinfo {address} {Cambridge},\ \bibinfo {year} {2003})\BibitemShut {NoStop}%
\bibitem [{\citenamefont {Cazalilla}(2004)}]{Cazalilla_2004}%
  \BibitemOpen
  \bibfield  {author} {\bibinfo {author} {\bibfnamefont {M.~A.}\ \bibnamefont {Cazalilla}},\ }\bibfield  {title} {\bibinfo {title} {Bosonizing one-dimensional cold atomic gases},\ }\href {https://doi.org/10.1088/0953-4075/37/7/051} {\bibfield  {journal} {\bibinfo  {journal} {J. Phys. B: At. Mol. Opt. Phys.}\ }\textbf {\bibinfo {volume} {37}},\ \bibinfo {pages} {S1} (\bibinfo {year} {2004})}\BibitemShut {NoStop}%
\bibitem [{\citenamefont {Cardy}(1996)}]{Cardy1996}%
  \BibitemOpen
  \bibfield  {author} {\bibinfo {author} {\bibfnamefont {J.}~\bibnamefont {Cardy}},\ }\href@noop {} {\emph {\bibinfo {title} {Scaling and Renormalization in Statistical Physics}}}\ (\bibinfo  {publisher} {Cambridge University Press},\ \bibinfo {year} {1996})\BibitemShut {NoStop}%
\bibitem [{\citenamefont {Caux}\ \emph {et~al.}(2003)\citenamefont {Caux}, \citenamefont {Lopez},\ and\ \citenamefont {Suppa}}]{CauxLopezSuppa2003}%
  \BibitemOpen
  \bibfield  {author} {\bibinfo {author} {\bibfnamefont {J.-S.}\ \bibnamefont {Caux}}, \bibinfo {author} {\bibfnamefont {A.}~\bibnamefont {Lopez}},\ and\ \bibinfo {author} {\bibfnamefont {D.}~\bibnamefont {Suppa}},\ }\bibfield  {title} {\bibinfo {title} {Currents and correlations in {Luttinger} liquids and carbon nanotubes at finite temperature and size: A bosonization study},\ }\href {https://doi.org/10.1016/S0550-3213(02)01107-0} {\bibfield  {journal} {\bibinfo  {journal} {Nucl. Phys. B}\ }\textbf {\bibinfo {volume} {651}},\ \bibinfo {pages} {413} (\bibinfo {year} {2003})}\BibitemShut {NoStop}%
\bibitem [{\citenamefont {Haldane}(1981)}]{Haldane:1981eh}%
  \BibitemOpen
  \bibfield  {author} {\bibinfo {author} {\bibfnamefont {F.~D.~M.}\ \bibnamefont {Haldane}},\ }\bibfield  {title} {\bibinfo {title} {{Effective Harmonic-Fluid Approach to Low-Energy Properties of One-Dimensional Quantum Fluids}},\ }\href {https://doi.org/10.1103/PhysRevLett.47.1840} {\bibfield  {journal} {\bibinfo  {journal} {Phys. Rev. Lett.}\ }\textbf {\bibinfo {volume} {47}},\ \bibinfo {pages} {1840} (\bibinfo {year} {1981})}\BibitemShut {NoStop}%
\bibitem [{\citenamefont {Giamarchi}(2004)}]{Giamarchi:2004bk}%
  \BibitemOpen
  \bibfield  {author} {\bibinfo {author} {\bibfnamefont {T.}~\bibnamefont {Giamarchi}},\ }\href@noop {} {\emph {\bibinfo {title} {Quantum Physics in One Dimension}}}\ (\bibinfo  {publisher} {Clarendon Press},\ \bibinfo {address} {Oxford, U.K.},\ \bibinfo {year} {2004})\BibitemShut {NoStop}%
\bibitem [{\citenamefont {Gogolin}\ \emph {et~al.}(1998)\citenamefont {Gogolin}, \citenamefont {Nersesyan},\ and\ \citenamefont {Tsvelik}}]{Gogolin1998}%
  \BibitemOpen
  \bibfield  {author} {\bibinfo {author} {\bibfnamefont {A.~O.}\ \bibnamefont {Gogolin}}, \bibinfo {author} {\bibfnamefont {A.~A.}\ \bibnamefont {Nersesyan}},\ and\ \bibinfo {author} {\bibfnamefont {A.~M.}\ \bibnamefont {Tsvelik}},\ }\href@noop {} {\emph {\bibinfo {title} {Bosonization and Strongly Correlated Systems}}}\ (\bibinfo  {publisher} {Cambridge University Press},\ \bibinfo {year} {1998})\BibitemShut {NoStop}%
\bibitem [{\citenamefont {von Delft}\ and\ \citenamefont {Schoeller}(1998)}]{vonDelft:1998ae}%
  \BibitemOpen
  \bibfield  {author} {\bibinfo {author} {\bibfnamefont {J.}~\bibnamefont {von Delft}}\ and\ \bibinfo {author} {\bibfnamefont {H.}~\bibnamefont {Schoeller}},\ }\bibfield  {title} {\bibinfo {title} {Bosonization for beginners: Refermionization for experts},\ }\href {https://doi.org/10.1002/andp.19985100401} {\bibfield  {journal} {\bibinfo  {journal} {Ann. Phys. (Berlin)}\ }\textbf {\bibinfo {volume} {510}},\ \bibinfo {pages} {225} (\bibinfo {year} {1998})}\BibitemShut {NoStop}%
\bibitem [{\citenamefont {Dzyaloshinskii}\ and\ \citenamefont {Larkin}(1974)}]{DzyaloshinskiiLarkin1974}%
  \BibitemOpen
  \bibfield  {author} {\bibinfo {author} {\bibfnamefont {I.~E.}\ \bibnamefont {Dzyaloshinskii}}\ and\ \bibinfo {author} {\bibfnamefont {A.~I.}\ \bibnamefont {Larkin}},\ }\bibfield  {title} {\bibinfo {title} {Correlation functions for a one-dimensional {Fermi} system with long-range interaction ({Tomonaga} model)},\ }\href {https://jetp.ras.ru/cgi-bin/dn/e_038_01_0202.pdf} {\bibfield  {journal} {\bibinfo  {journal} {Sov. Phys. JETP}\ }\textbf {\bibinfo {volume} {38}},\ \bibinfo {pages} {202} (\bibinfo {year} {1974})}\BibitemShut {NoStop}%
\bibitem [{\citenamefont {Luther}\ and\ \citenamefont {Peschel}(1975)}]{LutherPeschel1975}%
  \BibitemOpen
  \bibfield  {author} {\bibinfo {author} {\bibfnamefont {A.}~\bibnamefont {Luther}}\ and\ \bibinfo {author} {\bibfnamefont {I.}~\bibnamefont {Peschel}},\ }\bibfield  {title} {\bibinfo {title} {Calculation of critical exponents in two dimensions from quantum field theory in one dimension},\ }\href {https://doi.org/10.1103/PhysRevB.12.3908} {\bibfield  {journal} {\bibinfo  {journal} {Phys. Rev. B}\ }\textbf {\bibinfo {volume} {12}},\ \bibinfo {pages} {3908} (\bibinfo {year} {1975})}\BibitemShut {NoStop}%
\bibitem [{\citenamefont {Cazalilla}(2006)}]{Cazalilla:2006zw}%
  \BibitemOpen
  \bibfield  {author} {\bibinfo {author} {\bibfnamefont {M.~A.}\ \bibnamefont {Cazalilla}},\ }\bibfield  {title} {\bibinfo {title} {{E}ffect of {S}uddenly {T}urning on {I}nteractions in the {L}uttinger {M}odel},\ }\href {https://doi.org/10.1103/physrevlett.97.156403} {\bibfield  {journal} {\bibinfo  {journal} {Phys. Rev. Lett.}\ }\textbf {\bibinfo {volume} {97}},\ \bibinfo {pages} {156403} (\bibinfo {year} {2006})}\BibitemShut {NoStop}%
\bibitem [{\citenamefont {Thamm}\ \emph {et~al.}(2022)\citenamefont {Thamm}, \citenamefont {Radhakrishnan}, \citenamefont {Barghathi}, \citenamefont {Rosenow},\ and\ \citenamefont {{Del Maestro}}}]{Thamm:2022ja}%
  \BibitemOpen
  \bibfield  {author} {\bibinfo {author} {\bibfnamefont {M.}~\bibnamefont {Thamm}}, \bibinfo {author} {\bibfnamefont {H.}~\bibnamefont {Radhakrishnan}}, \bibinfo {author} {\bibfnamefont {H.}~\bibnamefont {Barghathi}}, \bibinfo {author} {\bibfnamefont {B.}~\bibnamefont {Rosenow}},\ and\ \bibinfo {author} {\bibfnamefont {A.}~\bibnamefont {{Del Maestro}}},\ }\bibfield  {title} {\bibinfo {title} {{O}ne-particle entanglement for one-dimensional spinless fermions after an interaction quantum quench},\ }\href {https://doi.org/10.1103/PhysRevB.106.165116} {\bibfield  {journal} {\bibinfo  {journal} {Phys. Rev. B}\ }\textbf {\bibinfo {volume} {106}},\ \bibinfo {pages} {165116} (\bibinfo {year} {2022})}\BibitemShut {NoStop}%
\bibitem [{\citenamefont {Baldelli}\ \emph {et~al.}(2025)\citenamefont {Baldelli}, \citenamefont {Karlsson}, \citenamefont {Kloss}, \citenamefont {Fishman},\ and\ \citenamefont {Wietek}}]{Baldelli23}%
  \BibitemOpen
  \bibfield  {author} {\bibinfo {author} {\bibfnamefont {N.}~\bibnamefont {Baldelli}}, \bibinfo {author} {\bibfnamefont {H.}~\bibnamefont {Karlsson}}, \bibinfo {author} {\bibfnamefont {B.}~\bibnamefont {Kloss}}, \bibinfo {author} {\bibfnamefont {M.}~\bibnamefont {Fishman}},\ and\ \bibinfo {author} {\bibfnamefont {A.}~\bibnamefont {Wietek}},\ }\bibfield  {title} {\bibinfo {title} {Fragmented superconductivity in the hubbard model as solitons in ginzburg--landau theory},\ }\href {https://doi.org/10.1038/s41535-024-00718-3} {\bibfield  {journal} {\bibinfo  {journal} {npj Quantum Mater.}\ }\textbf {\bibinfo {volume} {10}},\ \bibinfo {pages} {22} (\bibinfo {year} {2025})}\BibitemShut {NoStop}%
\bibitem [{\citenamefont {Aase}\ and\ \citenamefont {Sudb\o{}}(2022)}]{Aase2022}%
  \BibitemOpen
  \bibfield  {author} {\bibinfo {author} {\bibfnamefont {N.~H.}\ \bibnamefont {Aase}}\ and\ \bibinfo {author} {\bibfnamefont {A.}~\bibnamefont {Sudb\o{}}},\ }\bibfield  {title} {\bibinfo {title} {Dominant superconducting correlations in a luttinger liquid induced by spin fluctuations},\ }\href {https://doi.org/10.1103/PhysRevB.106.L241102} {\bibfield  {journal} {\bibinfo  {journal} {Phys. Rev. B}\ }\textbf {\bibinfo {volume} {106}},\ \bibinfo {pages} {L241102} (\bibinfo {year} {2022})}\BibitemShut {NoStop}%
\bibitem [{\citenamefont {Mukhopadhyay}\ \emph {et~al.}(2001)\citenamefont {Mukhopadhyay}, \citenamefont {Kane},\ and\ \citenamefont {Lubensky}}]{kanesliding}%
  \BibitemOpen
  \bibfield  {author} {\bibinfo {author} {\bibfnamefont {R.}~\bibnamefont {Mukhopadhyay}}, \bibinfo {author} {\bibfnamefont {C.~L.}\ \bibnamefont {Kane}},\ and\ \bibinfo {author} {\bibfnamefont {T.~C.}\ \bibnamefont {Lubensky}},\ }\bibfield  {title} {\bibinfo {title} {Sliding luttinger liquid phases},\ }\href {https://doi.org/10.1103/PhysRevB.64.045120} {\bibfield  {journal} {\bibinfo  {journal} {Phys. Rev. B}\ }\textbf {\bibinfo {volume} {64}},\ \bibinfo {pages} {045120} (\bibinfo {year} {2001})}\BibitemShut {NoStop}%
\bibitem [{\citenamefont {Kane}\ \emph {et~al.}(2017)\citenamefont {Kane}, \citenamefont {Stern},\ and\ \citenamefont {Halperin}}]{Kane2017}%
  \BibitemOpen
  \bibfield  {author} {\bibinfo {author} {\bibfnamefont {C.~L.}\ \bibnamefont {Kane}}, \bibinfo {author} {\bibfnamefont {A.}~\bibnamefont {Stern}},\ and\ \bibinfo {author} {\bibfnamefont {B.~I.}\ \bibnamefont {Halperin}},\ }\bibfield  {title} {\bibinfo {title} {Pairing in luttinger liquids and quantum hall states},\ }\href {https://doi.org/10.1103/PhysRevX.7.031009} {\bibfield  {journal} {\bibinfo  {journal} {Phys. Rev. X}\ }\textbf {\bibinfo {volume} {7}},\ \bibinfo {pages} {031009} (\bibinfo {year} {2017})}\BibitemShut {NoStop}%
\bibitem [{\citenamefont {Nayak}(2000)}]{nayak2000density}%
  \BibitemOpen
  \bibfield  {author} {\bibinfo {author} {\bibfnamefont {C.}~\bibnamefont {Nayak}},\ }\bibfield  {title} {\bibinfo {title} {{D}ensity-wave states of nonzero angular momentum},\ }\href {https://doi.org/10.1103/physrevb.62.4880} {\bibfield  {journal} {\bibinfo  {journal} {Phys. Rev. B}\ }\textbf {\bibinfo {volume} {62}},\ \bibinfo {pages} {4880} (\bibinfo {year} {2000})}\BibitemShut {NoStop}%
\bibitem [{\citenamefont {White}(1992)}]{White1992DMRG}%
  \BibitemOpen
  \bibfield  {author} {\bibinfo {author} {\bibfnamefont {S.~R.}\ \bibnamefont {White}},\ }\bibfield  {title} {\bibinfo {title} {Density matrix formulation for quantum renormalization groups},\ }\href {https://doi.org/10.1103/PhysRevLett.69.2863} {\bibfield  {journal} {\bibinfo  {journal} {Phys. Rev. Lett.}\ }\textbf {\bibinfo {volume} {69}},\ \bibinfo {pages} {2863} (\bibinfo {year} {1992})}\BibitemShut {NoStop}%
\bibitem [{\citenamefont {Schollw{\"o}ck}(2011)}]{Schollwock2011DMRG}%
  \BibitemOpen
  \bibfield  {author} {\bibinfo {author} {\bibfnamefont {U.}~\bibnamefont {Schollw{\"o}ck}},\ }\bibfield  {title} {\bibinfo {title} {The density-matrix renormalization group in the age of matrix product states},\ }\href {https://doi.org/10.1016/j.aop.2010.09.012} {\bibfield  {journal} {\bibinfo  {journal} {Ann. Phys. (N.Y.)}\ }\textbf {\bibinfo {volume} {326}},\ \bibinfo {pages} {96} (\bibinfo {year} {2011})}\BibitemShut {NoStop}%
\bibitem [{\citenamefont {{Del Maestro}}\ \emph {et~al.}(2021)\citenamefont {{Del Maestro}}, \citenamefont {Barghathi},\ and\ \citenamefont {Rosenow}}]{DelMaestro:2021ja}%
  \BibitemOpen
  \bibfield  {author} {\bibinfo {author} {\bibfnamefont {A.}~\bibnamefont {{Del Maestro}}}, \bibinfo {author} {\bibfnamefont {H.}~\bibnamefont {Barghathi}},\ and\ \bibinfo {author} {\bibfnamefont {B.}~\bibnamefont {Rosenow}},\ }\bibfield  {title} {\bibinfo {title} {{E}quivalence of spatial and particle entanglement growth after a quantum quench},\ }\href {https://doi.org/10.1103/physrevb.104.195101} {\bibfield  {journal} {\bibinfo  {journal} {Phys. Rev. B}\ }\textbf {\bibinfo {volume} {104}},\ \bibinfo {pages} {195101} (\bibinfo {year} {2021})}\BibitemShut {NoStop}%
\bibitem [{\citenamefont {Yang}(1962)}]{yang1962concept}%
  \BibitemOpen
  \bibfield  {author} {\bibinfo {author} {\bibfnamefont {C.~N.}\ \bibnamefont {Yang}},\ }\bibfield  {title} {\bibinfo {title} {{C}oncept of {O}ff-{D}iagonal {L}ong-{R}ange {O}rder and the {Q}uantum {P}hases of {L}iquid {H}e and of {S}uperconductors},\ }\href {https://doi.org/10.1103/revmodphys.34.694} {\bibfield  {journal} {\bibinfo  {journal} {Rev. Mod. Phys.}\ }\textbf {\bibinfo {volume} {34}},\ \bibinfo {pages} {694} (\bibinfo {year} {1962})}\BibitemShut {NoStop}%
\bibitem [{\citenamefont {L\"{o}wdin}(1955)}]{Lowdin1955}%
  \BibitemOpen
  \bibfield  {author} {\bibinfo {author} {\bibfnamefont {P.-O.}\ \bibnamefont {L\"{o}wdin}},\ }\bibfield  {title} {\bibinfo {title} {Quantum theory of many-particle systems. i. physical interpretations by means of density matrices, natural spin-orbitals, and convergence problems in the method of configurational interaction},\ }\href {https://doi.org/10.1103/physrev.97.1474} {\bibfield  {journal} {\bibinfo  {journal} {Phys. Rev.}\ }\textbf {\bibinfo {volume} {97}},\ \bibinfo {pages} {1474} (\bibinfo {year} {1955})}\BibitemShut {NoStop}%
\bibitem [{\citenamefont {Carlson}\ and\ \citenamefont {Keller}(1961)}]{Carlson:1961dr}%
  \BibitemOpen
  \bibfield  {author} {\bibinfo {author} {\bibfnamefont {B.~C.}\ \bibnamefont {Carlson}}\ and\ \bibinfo {author} {\bibfnamefont {J.~M.}\ \bibnamefont {Keller}},\ }\bibfield  {title} {\bibinfo {title} {{E}igenvalues of {D}ensity {M}atrices},\ }\href {https://doi.org/10.1103/physrev.121.659} {\bibfield  {journal} {\bibinfo  {journal} {Phys. Rev.}\ }\textbf {\bibinfo {volume} {121}},\ \bibinfo {pages} {659} (\bibinfo {year} {1961})}\BibitemShut {NoStop}%
\bibitem [{\citenamefont {Ando}(1963)}]{Ando1963}%
  \BibitemOpen
  \bibfield  {author} {\bibinfo {author} {\bibfnamefont {T.}~\bibnamefont {Ando}},\ }\bibfield  {title} {\bibinfo {title} {Properties of fermion density matrices},\ }\href {https://doi.org/10.1103/revmodphys.35.690} {\bibfield  {journal} {\bibinfo  {journal} {Rev. Mod. Phys.}\ }\textbf {\bibinfo {volume} {35}},\ \bibinfo {pages} {690} (\bibinfo {year} {1963})}\BibitemShut {NoStop}%
\bibitem [{\citenamefont {Eggert}(2009)}]{Eggert.2009}%
  \BibitemOpen
  \bibfield  {author} {\bibinfo {author} {\bibfnamefont {S.}~\bibnamefont {Eggert}},\ }\href@noop {} {\bibinfo {title} {One-dimensional quantum wires: A pedestrian approach to bosonization}} (\bibinfo {year} {2009}),\ \Eprint {https://arxiv.org/abs/0708.0003} {arXiv:0708.0003 [cond-mat.str-el]} \BibitemShut {NoStop}%
\bibitem [{Note1()}]{Note1}%
  \BibitemOpen
  \bibinfo {note} {In Ref.~\cite {Haldane:1981eh}, the second term in Eq.~(7) has the wrong sign \cite {Cazalilla_2004}}\BibitemShut {NoStop}%
\bibitem [{\citenamefont {Lukyanov}\ and\ \citenamefont {Terras}(2003)}]{LukyanovTerras2003}%
  \BibitemOpen
  \bibfield  {author} {\bibinfo {author} {\bibfnamefont {S.}~\bibnamefont {Lukyanov}}\ and\ \bibinfo {author} {\bibfnamefont {V.}~\bibnamefont {Terras}},\ }\bibfield  {title} {\bibinfo {title} {Long-distance asymptotics of spin-spin correlation functions for the {XXZ} spin chain},\ }\href {https://doi.org/10.1016/S0550-3213(02)01141-0} {\bibfield  {journal} {\bibinfo  {journal} {Nucl. Phys. B}\ }\textbf {\bibinfo {volume} {654}},\ \bibinfo {pages} {323} (\bibinfo {year} {2003})}\BibitemShut {NoStop}%
\bibitem [{\citenamefont {Kutzelnigg}\ and\ \citenamefont {Mukherjee}(1999)}]{KutzelniggMukherjee1999}%
  \BibitemOpen
  \bibfield  {author} {\bibinfo {author} {\bibfnamefont {W.}~\bibnamefont {Kutzelnigg}}\ and\ \bibinfo {author} {\bibfnamefont {D.}~\bibnamefont {Mukherjee}},\ }\bibfield  {title} {\bibinfo {title} {Cumulant expansion of the reduced density matrices},\ }\href {https://doi.org/10.1063/1.478189} {\bibfield  {journal} {\bibinfo  {journal} {J. Chem. Phys.}\ }\textbf {\bibinfo {volume} {110}},\ \bibinfo {pages} {2800} (\bibinfo {year} {1999})}\BibitemShut {NoStop}%
\bibitem [{\citenamefont {Voit}(1995)}]{Voit1995}%
  \BibitemOpen
  \bibfield  {author} {\bibinfo {author} {\bibfnamefont {J.}~\bibnamefont {Voit}},\ }\bibfield  {title} {\bibinfo {title} {One-dimensional {Fermi} liquids},\ }\href {https://doi.org/10.1088/0034-4885/58/9/002} {\bibfield  {journal} {\bibinfo  {journal} {Rep. Prog. Phys.}\ }\textbf {\bibinfo {volume} {58}},\ \bibinfo {pages} {977} (\bibinfo {year} {1995})}\BibitemShut {NoStop}%
\bibitem [{\citenamefont {Garrod}\ and\ \citenamefont {Rosina}(1969)}]{GarrodRosina1969}%
  \BibitemOpen
  \bibfield  {author} {\bibinfo {author} {\bibfnamefont {C.}~\bibnamefont {Garrod}}\ and\ \bibinfo {author} {\bibfnamefont {M.}~\bibnamefont {Rosina}},\ }\bibfield  {title} {\bibinfo {title} {Particle-hole matrix: Its connection with the symmetries and collective features of the ground state},\ }\href {https://doi.org/10.1063/1.1664770} {\bibfield  {journal} {\bibinfo  {journal} {J. Math. Phys.}\ }\textbf {\bibinfo {volume} {10}},\ \bibinfo {pages} {1855} (\bibinfo {year} {1969})}\BibitemShut {NoStop}%
\bibitem [{\citenamefont {Des~Cloizeaux}(1966)}]{DesCloizeaux:1966}%
  \BibitemOpen
  \bibfield  {author} {\bibinfo {author} {\bibfnamefont {J.}~\bibnamefont {Des~Cloizeaux}},\ }\bibfield  {title} {\bibinfo {title} {A soluble fermi-gas model. validity of transformations of the bogoliubov type},\ }\href {https://doi.org/10.1063/1.1704899} {\bibfield  {journal} {\bibinfo  {journal} {J. Math. Phys.}\ }\textbf {\bibinfo {volume} {7}},\ \bibinfo {pages} {2136} (\bibinfo {year} {1966})}\BibitemShut {NoStop}%
\bibitem [{\citenamefont {Yang}\ and\ \citenamefont {Yang}(1966)}]{Yang:1966yc}%
  \BibitemOpen
  \bibfield  {author} {\bibinfo {author} {\bibfnamefont {C.~N.}\ \bibnamefont {Yang}}\ and\ \bibinfo {author} {\bibfnamefont {C.~P.}\ \bibnamefont {Yang}},\ }\bibfield  {title} {\bibinfo {title} {{O}ne-{D}imensional {C}hain of {A}nisotropic {S}pin-{S}pin {I}nteractions. {I}. {P}roof of {B}ethe's {H}ypothesis for {G}round {S}tate in a {F}inite {S}ystem},\ }\href {https://doi.org/10.1103/physrev.150.321} {\bibfield  {journal} {\bibinfo  {journal} {Phys. Rev.}\ }\textbf {\bibinfo {volume} {150}},\ \bibinfo {pages} {321} (\bibinfo {year} {1966})}\BibitemShut {NoStop}%
\bibitem [{\citenamefont {Fishman}\ \emph {et~al.}(2022{\natexlab{a}})\citenamefont {Fishman}, \citenamefont {White},\ and\ \citenamefont {Stoudenmire}}]{itensor1}%
  \BibitemOpen
  \bibfield  {author} {\bibinfo {author} {\bibfnamefont {M.}~\bibnamefont {Fishman}}, \bibinfo {author} {\bibfnamefont {S.}~\bibnamefont {White}},\ and\ \bibinfo {author} {\bibfnamefont {E.~M.}\ \bibnamefont {Stoudenmire}},\ }\bibfield  {title} {\bibinfo {title} {The itensor software library for tensor network calculations},\ }\href {https://doi.org/10.21468/SciPostPhysCodeb.4} {\bibfield  {journal} {\bibinfo  {journal} {SciPost Phys. Codebases}\ ,\ \bibinfo {pages} {4}} (\bibinfo {year} {2022}{\natexlab{a}})}\BibitemShut {NoStop}%
\bibitem [{\citenamefont {Fishman}\ \emph {et~al.}(2022{\natexlab{b}})\citenamefont {Fishman}, \citenamefont {White},\ and\ \citenamefont {Stoudenmire}}]{itensor2}%
  \BibitemOpen
  \bibfield  {author} {\bibinfo {author} {\bibfnamefont {M.}~\bibnamefont {Fishman}}, \bibinfo {author} {\bibfnamefont {S.}~\bibnamefont {White}},\ and\ \bibinfo {author} {\bibfnamefont {E.~M.}\ \bibnamefont {Stoudenmire}},\ }\bibfield  {title} {\bibinfo {title} {Codebase release 0.3 for itensor},\ }\href {https://doi.org/10.21468/SciPostPhysCodeb.4-r0.3} {\bibfield  {journal} {\bibinfo  {journal} {SciPost Phys. Codebases}\ ,\ \bibinfo {pages} {4}} (\bibinfo {year} {2022}{\natexlab{b}})}\BibitemShut {NoStop}%
\bibitem [{\citenamefont {Giuliani}\ and\ \citenamefont {Vignale}(2005)}]{Giuliani2005}%
  \BibitemOpen
  \bibfield  {author} {\bibinfo {author} {\bibfnamefont {G.}~\bibnamefont {Giuliani}}\ and\ \bibinfo {author} {\bibfnamefont {G.}~\bibnamefont {Vignale}},\ }\href {https://doi.org/10.1017/CBO9780511619915} {\emph {\bibinfo {title} {Quantum {Theory} of the {Electron} {Liquid}}}}\ (\bibinfo  {publisher} {Cambridge University Press},\ \bibinfo {address} {Cambridge},\ \bibinfo {year} {2005})\BibitemShut {NoStop}%
\bibitem [{\citenamefont {Carlen}\ \emph {et~al.}(2016)\citenamefont {Carlen}, \citenamefont {Lieb},\ and\ \citenamefont {Reuvers}}]{Carlen:2016fv}%
  \BibitemOpen
  \bibfield  {author} {\bibinfo {author} {\bibfnamefont {E.~A.}\ \bibnamefont {Carlen}}, \bibinfo {author} {\bibfnamefont {E.~H.}\ \bibnamefont {Lieb}},\ and\ \bibinfo {author} {\bibfnamefont {R.}~\bibnamefont {Reuvers}},\ }\bibfield  {title} {\bibinfo {title} {{E}ntropy and {E}ntanglement {B}ounds for {R}educed {D}ensity {M}atrices of {F}ermionic {S}tates},\ }\href {https://doi.org/10.1007/s00220-016-2651-6} {\bibfield  {journal} {\bibinfo  {journal} {Commun. Math. Phys.}\ }\textbf {\bibinfo {volume} {344}},\ \bibinfo {pages} {655} (\bibinfo {year} {2016})}\BibitemShut {NoStop}%
\bibitem [{\citenamefont {Protopopov}\ \emph {et~al.}(2011)\citenamefont {Protopopov}, \citenamefont {Gutman},\ and\ \citenamefont {Mirlin}}]{Protopopov2011}%
  \BibitemOpen
  \bibfield  {author} {\bibinfo {author} {\bibfnamefont {I.~V.}\ \bibnamefont {Protopopov}}, \bibinfo {author} {\bibfnamefont {D.~B.}\ \bibnamefont {Gutman}},\ and\ \bibinfo {author} {\bibfnamefont {A.~D.}\ \bibnamefont {Mirlin}},\ }\bibfield  {title} {\bibinfo {title} {Many-particle correlations in a non-equilibrium luttinger liquid},\ }\href {https://doi.org/10.1088/1742-5468/2011/11/p11001} {\bibfield  {journal} {\bibinfo  {journal} {J. Stat. Mech.: Theory Exp.}\ }\textbf {\bibinfo {volume} {2011}}\bibinfo  {number} { (11)},\ \bibinfo {pages} {P11001}}\BibitemShut {NoStop}%
\bibitem [{\citenamefont {Rosina}(1968)}]{Rosina1968}%
  \BibitemOpen
\bibfield  {number} {  }\bibfield  {author} {\bibinfo {author} {\bibfnamefont {M.}~\bibnamefont {Rosina}},\ }\bibfield  {title} {\bibinfo {title} {Transition amplitudes as ground-state variational parameters},\ }in\ \href@noop {} {\emph {\bibinfo {booktitle} {Reduced Density Matrices with Applications to Physical and Chemical Systems}}},\ \bibinfo {series and number} {\bibinfo {series} {Queen's Papers in Pure and Applied Mathematics}\ No.~\bibinfo {number} {11}},\ \bibinfo {editor} {edited by\ \bibinfo {editor} {\bibfnamefont {A.~J.}\ \bibnamefont {Coleman}}\ and\ \bibinfo {editor} {\bibfnamefont {R.~M.}\ \bibnamefont {Erdahl}}}\ (\bibinfo  {publisher} {Queen's University},\ \bibinfo {address} {Kingston, Ontario},\ \bibinfo {year} {1968})\ p.\ \bibinfo {pages} {369}\BibitemShut {NoStop}%
\bibitem [{\citenamefont {Mazziotti}(2000)}]{Mazziotti2000Reconstruction}%
  \BibitemOpen
  \bibfield  {author} {\bibinfo {author} {\bibfnamefont {D.~A.}\ \bibnamefont {Mazziotti}},\ }\bibfield  {title} {\bibinfo {title} {Complete reconstruction of reduced density matrices},\ }\href {https://doi.org/10.1016/S0009-2614(00)00773-9} {\bibfield  {journal} {\bibinfo  {journal} {Chem. Phys. Lett.}\ }\textbf {\bibinfo {volume} {326}},\ \bibinfo {pages} {212} (\bibinfo {year} {2000})}\BibitemShut {NoStop}%
\bibitem [{\citenamefont {Massaccesi}\ \emph {et~al.}(2026)\citenamefont {Massaccesi}, \citenamefont {O{\~n}a}, \citenamefont {Lain}, \citenamefont {Torre}, \citenamefont {Peralta}, \citenamefont {Alcoba},\ and\ \citenamefont {Scuseria}}]{Massaccesi2026}%
  \BibitemOpen
  \bibfield  {author} {\bibinfo {author} {\bibfnamefont {G.~E.}\ \bibnamefont {Massaccesi}}, \bibinfo {author} {\bibfnamefont {O.~B.}\ \bibnamefont {O{\~n}a}}, \bibinfo {author} {\bibfnamefont {L.}~\bibnamefont {Lain}}, \bibinfo {author} {\bibfnamefont {A.}~\bibnamefont {Torre}}, \bibinfo {author} {\bibfnamefont {J.~E.}\ \bibnamefont {Peralta}}, \bibinfo {author} {\bibfnamefont {D.~R.}\ \bibnamefont {Alcoba}},\ and\ \bibinfo {author} {\bibfnamefont {G.~E.}\ \bibnamefont {Scuseria}},\ }\bibfield  {title} {\bibinfo {title} {Is the matrix completion of reduced density matrices unique?},\ }\href {https://doi.org/10.1021/acs.jpclett.6c00296} {\bibfield  {journal} {\bibinfo  {journal} {J. Phys. Chem. Lett.}\ }\textbf {\bibinfo {volume} {17}},\ \bibinfo {pages} {3430} (\bibinfo {year} {2026})}\BibitemShut {NoStop}%
\bibitem [{\citenamefont {Thamm}\ \emph {et~al.}(2026)\citenamefont {Thamm}, \citenamefont {Radhakrishnan},\ and\ \citenamefont {{Del Maestro}}}]{repo}%
  \BibitemOpen
  \bibfield  {author} {\bibinfo {author} {\bibfnamefont {M.}~\bibnamefont {Thamm}}, \bibinfo {author} {\bibfnamefont {H.}~\bibnamefont {Radhakrishnan}},\ and\ \bibinfo {author} {\bibfnamefont {A.}~\bibnamefont {{Del Maestro}}},\ }\href {https://doi.org/10.5281/zenodo.21384640} {\bibinfo {title} {{Git{H}ub Repository}}},\ \bibinfo {howpublished} {Zenodo doi:10.5281/zenodo.21384640} (\bibinfo {year} {2026})\BibitemShut {NoStop}%
\end{thebibliography}%

\end{document}